\documentclass[preprint,12pt,3p]{elsarticle}
 \pdfoutput=1
\usepackage{graphicx}
\usepackage{amsmath}
\usepackage{amssymb}
\usepackage{amsthm}
\usepackage{bm}
\usepackage{color}
\usepackage{mathrsfs}
\usepackage{multirow}
\usepackage{booktabs}
\newtheorem{remark}{Remark}

\journal{}

\def\rme{\mathrm e}
\def\rmd{\mathrm d}
\def\rmT{\mathrm T}

\def\bfP{\mathbf \Psi}
\def\rma{\mathrm {Ma}}

\def\bfW{\mathbf W}
\def\bfF{\mathbf F}

\begin{document}

\begin{frontmatter}

\title{An Efficient High-Order Gas-Kinetic Scheme (I): Euler equations}

\author[ia1]{Shiyi Li}
\ead{lishiyi14@tsinghua.org.cn}
\author[ia1]{Yibing Chen \corref{mycorrespondingauthor}}
\cortext[mycorrespondingauthor]{Corresponding author}
\ead{chen\_yibing@iapcm.ac.cn}
\author[ia1]{Song Jiang}
\ead{jiang@iapcm.ac.cn}

\address[ia1]{Institute of Applied Physics and Computational Mathematics, Beijing 100191, China}

\begin{abstract}
In this paper, an efficient high-order gas-kinetic scheme (EHGKS) is proposed to solve the Euler equations for compressible flows.
We re-investigate the underlying mechanism of the high-order gas-kinetic scheme (HGKS) and find a new strategy to improve its efficiency.
The main idea of the new scheme contains two parts.
Firstly, inspired by the state-of-art simplifications on the third-order HGKS, we extend the HGKS to the case of arbitrary high-order accuracy and eliminate its unnecessary high-order dissipation terms.
Secondly, instead of computing the derivatives of particle distribution function and their complex moments, we introduce a Lax-Wendroff procedure to compute the high-order derivatives of macroscopic quantities directly.
The new scheme takes advantage of both HGKS and the Lax-Wendroff procedure, so that it can be easily extended to the case of arbitrary high-order accuracy with practical significance.
Typical numerical tests are carried out by EHGKS, with the third, fifth and seventh-order accuracy.
The presence of good resolution on the discontinuities and flow details, together with the optimal CFL numbers, validates the high accuracy and strong robustness of EHGKS.
To compare the efficiency, we present the results computed by the EHGKS, the original HGKS and Runge-Kutta-WENO-GKS.
This further demonstrates the advantages of EHGKS.
\end{abstract}


\end{frontmatter}

\section{\label{sec:int}Introduction}

In the past decades, a multitude of high-order schemes have been well developed and become very popular in solving the Euler equations for compressible flows, because high-order schemes generally use much less CPU time than low-order schemes to approach the solutions.
There are a series of successful numerical schemes, such as ENO, WENO \cite{eno1996,shu1998}, DG \cite{cockburn1998}, RD \cite{abgrall2006},
SV \cite{wang2002}, SD \cite{liu2006}, which can achieve arbitrary high-order accuracy in space.
Meanwhile, they often need to employ a multi-stage Runge-Kutta method in time to preserve the stability of these schemes.
In practice, the third-order TVD Runge-Kutta method \cite{eno1996} has been most widely utilized for its simplicity.
However, there exist two drawbacks in the multi-stage schemes.
Firstly, as pointed out by Toro et. al. \cite{toro2001,titarev2002}, the accuracy of these schemes can not exceed the time accuracy,
which is called "accuracy barrier".
To obtain the designed accuracy as in space, the CFL condition number should be reduced, which will increase the computational cost.
Secondly, the high-order reconstruction techniques, which are required to be implemented for several times in a single time step in the multi-stage schemes, are often expensive.

To overcome these two drawbacks, it has attracted much attention to develop the one-step schemes with consistent high-order accuracy
in both space and time \cite{Artzi1984,ljq2016,toro2001,titarev2005,lqb2010,pan2016,cyb2016}.
GRP, which solves the generalized Riemann problem with initial piecewise smooth data \cite{Artzi1984}, is one of the earliest in this category.
Limited by the sophisticated wave structures from the high-order piecewise polynomials, it becomes tedious to construct a higher than third-order GRP unless introducing the idea of the multi-stage method \cite{ljq2016}.
By introducing a linearization technique, the ADER (arbitrary derivative in space and time) scheme \cite{toro2001}
simplifies the computation process of the original GRP scheme.
ADER is a one-step and fully discrete Godunov approach with arbitrary high-order accuracy in both space and time.
There are two types of ADER: the state-expansion version and the flux-expansion version \cite{toro2005}.
The previous one is easier to be implemented although not all Riemann solvers are applicable, while the latter is suitable
for any Riemann solver but introduces more complexities.

Among the ways of constructing high-order schemes,
the high-order gas-kinetic scheme (HGKS) has been developed systematically in the recent years \cite{lqb2010,luo2013,liu2014,pan2016,ji2018}.
Different from solving the hydrodynamic wave structures in the traditional Riemann problem \cite{toro2013},
HGKS utilizes the time integral solution of the Bhatnagar-Gross-Krook (BGK) equation as its the evolution model \cite{shen2006,xu1998}.
The integral solution describes the particle free transport and collisions \cite{shen2006}.
It builds up a multi-scale scheme applicable in the whole flow regimes \cite{xu2001,xu2010}.
When simulating the hydrodynamic flows, in accordance with the Euler and Navier-Stokes (NS) equations, the hydrodynamic part in the evolution model is dominant in the smooth regions.
Meanwhile, the kinetic part provides the significant shock-capturing capability \cite{luo22013,zc2018}.
This underlying mechanism ensures HGKS as an accurate and robust scheme for various unsteady compressible flows \cite{lqb2010,liu2014,pan2016,ji2017}.
The straightforward way to construct a high-order HGKS is based on the high-order Taylor extension of the integral solution in both space and time.
The third and fourth-order HGKS have been developed adherently \cite{lqb2010,luo2013,liu2014}.
However, because of the time-consuming computation process of the sophisticated extension, the state-of-art fourth and fifth-order HGKS is compromised to the multi-stage framework in practice \cite{pan2016,ji2017}.

Considering the convenience of the straightforward high-order extensions in HGKS, efforts have been paid to make the existing third-order HGKS simpler
and more efficient \cite{zhou2017,luo22013,ji2018,zc2018}.
HGKS is originally developed to solve the NS equations \cite{xu1998,xu2001}.
Luo distinguished the physical and numerical dissipation parts in the third-order HGKS and eliminated the physical dissipation part to solve the Euler equations \cite{luo22013}.
Zhou simplified appropriately the numerical dissipation part and kept its primary terms to provide necessary numerical dissipation \cite{zhou2017}.
The validity of this simplification to preserve the high accuracy and strong robustness as the original third-order HGKS has been illustrated theoretically and numerically in typical compressible flows \cite{zhou2017}.
The existing simplifications are also inspiring to extend HGKS for the Euler equations with arbitrary high-order accuracy.
It is further expected the high-order terms after these simplifications contain only the time derivatives of the particle distribution function \cite{zhou2017,luo22013}.
Meanwhile, the procedure to obtain the corresponding high-order terms in flux evaluation is consistent with the traditional Lax-Wendroff procedure \cite{titarev2005}.
The Lax-Wendroff procedure is more efficient than computing the derivatives of the particle distribution function.
Consequently, it can be utilized as an alternative way in the simplified HGKS flux evaluation to solve the Euler equations with more practical significance.

In this paper, taking advantage of both HGKS and the Lax-Wendroff procedure, a more efficient one-step EHGKS will be proposed to solve the Euler equations
with arbitrary high-order accuracy in both space and time.
The new scheme is based on the extensions and modifications on the original HGKS and the introduction of an alternative Lax-Wendroff procedure.
This paper is organized as follows.
In Section~\ref{sec:rev}, the construction of HGKS and the existing simplifications are reviewed firstly.
Sections~\ref{sec:hgks} and \ref{sec:ehgks} illustrate the construction of EHGKS, including the extension of HGKS to arbitrary
high-order accuracy and its simplifications and modifications, together with the introduction of Lax-Wendroff procedure in flux evaluation.
In Section~\ref{sec:nta}, the new scheme is tested in several typical examples, demonstrating its high accuracy, robustness and efficiency.
The conclusion is made in the last section.

\section{\label{sec:rev} Review}

\subsection{\label{sec:rev-gkt} Gas-kinetic theory}

\subsubsection{\label{sec:rev-gkt-bgk} BGK model}

The evolution model in GKS is based on the BGK equation and its time integral solution \cite{xu1998,xu2001,lqb2010}.
The BGK equation is a widely used simplified model of the Boltzmann equation \cite{xu1998}.
In the one-dimensional (1D) case, the BGK equation can be written as \cite{xu1998}
\begin{eqnarray}
\label{eq:bgk}
   \frac{\partial f}{\partial t}
+u \frac{\partial f}{\partial x}
 = \frac{g - f}{\tau }.
\end{eqnarray}
Here $f=f({ x},t,{ u},{\bm \xi})$ is the particle distribution function at space ${ x}$, time $t$, particle velocity ${ u}$ and
the internal variables $\bm \xi$ \cite{shen2006}.
$g$ is the Maxwellian equilibrium state
\begin{eqnarray}
\label{eq:bgkg}
g =g\left( \bfW \right) =\rho
\left( \frac{\lambda}{\pi } \right)^{\frac{K+1}{2}}
\rme^{ - \lambda  \left( ({ u} - { U})^2+ {\bm \xi}^2 \right)} ,
\end{eqnarray}
which can be determined by the macroscopic conservative variables $\bfW = ( \rho ,\rho { U},\rho E )^{\rmT}$.
$\rho$ is the local density, ${ U} $ is the velocity, the total energy $\rho E = \frac{1}{2}\rho U^2  + {p}/(\gamma  - 1)$.
The pressure $p=\rho RT$. ${R}$ is the gas constant. The temperature $T=1/(2R \lambda)$.
$K$ is the degree of freedom in $\bm \xi$, such as the particle motion in the $y,z$ direction, molecular rotation or vibration \cite{shen2006,xu1998}.
In the 1D case, $K=(3-\gamma)/(\gamma-1)$, where $\gamma$ is the specific heat ratio.
$\tau=\mu/p$ is the mean collision time \cite{shen2006}.
$\mu$ is the dynamical viscosity.

The time integral solution of the BGK equation is given as \cite{xu2001}
\begin{eqnarray}
f({ x},t,{ u},{\bm \xi}) =
\frac{1}{\tau} \int_0^t
{g({ x}',t',{ u},{\bm \xi}) \rme^{-(t - t')/\tau} {\rmd t'}}
+ \rme^{-t/\tau} f({ x} - { u} t,0,{ u},{\bm \xi}),
\label{eq:ints}
\end{eqnarray}
where ${ x}'= { x} - { u} (t-t')$.
In this integral solution, the terms related to the equilibrium state and the initial distribution function are weighted roughly through $\rme^{-t/\tau}$.
Under different situations of $t/\tau$, the integral solution recurs to different mechanisms inherently \cite{xu2001}.
When $t \ll \tau$,
\begin{eqnarray}
f({ x},t,{ u},{\bm \xi}) \approx f({ x} - { u} t,0,{ u},{\bm \xi}),
\label{eq:ints0}
\end{eqnarray}
where the particle collisions are seldom encountered.
It corresponds to the free molecular flow.
In practical simulations, this situation also corresponds to the unresolved numerical discontinuities \cite{xu2017}.
On the other side, when $t \gg \tau$,
\begin{eqnarray}
f({ x},t,{ u},{\bm \xi}) \approx g({ x},t,{ u},{\bm \xi}).
\label{eq:intse}
\end{eqnarray}
It corresponds to the situation with adequate relaxation to the local equilibrium state.

The BGK equation and its integral solution present the mesoscopic description of the gas dynamics \cite{shen2006}.
The statistics over the particle distribution function provides the macroscopic description of flow structures.

\subsubsection{\label{sec:rev-gkt-mm} Relationship between mesoscopic and macroscopic descriptions}

The relationship between $f$ and $\bfW $ is (see \cite{shen2006} for example)
\begin{eqnarray}
\bfW =  \int {f{\bfP}{\rmd \Xi}},
\label{eq:rew}
\end{eqnarray}
where the moment vector ${\bfP} = ({1,{ u},({ u}^2+{\bm \xi}^2 )/2})^{\rmT}$ and $\rmd \Xi  = {\rmd { u}}{\rmd {\bm \xi}}$.
The macroscopic conservative equations can be obtained by taking the moments of the BGK equation on ${\bfP} $, i.e.
\begin{eqnarray}
\label{eq:ree}
\frac{\partial \bfW}{\partial t}
+ \frac{\partial \bfF}{\partial x} = {\bm 0},
\end{eqnarray}
where the relationship between $f$ and the macroscopic flux $\bfF$ is
\begin{eqnarray}
\label{eq:reflux}
\bfF=\int {f { u} {\bfP}{\rmd \Xi}}.
\end{eqnarray}
This derivation is based on the conservative constraint \cite{xu1998,lqb2010}
\begin{eqnarray}
\int {\left( g - f \right){\bfP}{\rmd \Xi}} = {\bm 0}.
\label{eq:conc}
\end{eqnarray}
More specific macroscopic equations, such as the Euler equations or NS equations, can be derived from the BGK equation according
to the Chapman-Enskog expansion on $\tau$ \cite{xu1998}.
The Euler equations correspond to the zeroth-order Chapman-Enskog expansion with adequate relaxation to the local Maxwellian equilibrium state
\begin{eqnarray}
f=g.
\label{eq:ce0}
\end{eqnarray}
According to the moments of the Maxwellian equilibrium state \cite{xu1998}, the macroscopic equations can be derived as \cite{shen2006}
\begin{eqnarray}
\label{eq:weuler}
\frac{\partial }
{{\partial t}}\left( \begin{gathered}
  \rho   \\
  \rho U  \\
  \rho E  \\
\end{gathered}  \right) + \frac{\partial }
{{\partial x}}\left( \begin{gathered}
  \rho U  \\
  \rho U^2  + p  \\
  \rho EU + pU  \\
\end{gathered}  \right) = {\bm 0}.
\end{eqnarray}
The flux $\bfF$ can be determined by $\bfW$ as
\begin{eqnarray}
\label{eq:fluxeuler0}
{\bfF}=\int {g { u} {\bfP}{\rmd \Xi}}
={\bfF}_{Eu}(\bfW)=\left( \begin{gathered}
  \rho U  \\
  \rho U^2  + p  \\
  \rho EU + pU  \\
\end{gathered}  \right) .
\end{eqnarray}
The first-order Chapman-Enskog expansion
\begin{eqnarray}
f= f_{NS}=g-\tau(\frac{\partial g}{\partial t}+u\frac{\partial g}{\partial x}),
\label{eq:ce1}
\end{eqnarray}
yields the NS equations:
\begin{eqnarray}
\label{eq:wns}
\frac{\partial }
{{\partial t}}\left( \begin{gathered}
  \rho   \\
  \rho U  \\
  \rho E  \\
\end{gathered}  \right) + \frac{\partial }
{{\partial x}}\left( \begin{gathered}
  \rho U  \\
  \rho U^2  + p  \\
  \rho EU + pU \\
\end{gathered}  \right) =
\frac{\partial }
{{\partial x}}\left( {\begin{array}{*{20}c}
   0  \\
   {\frac{{2K}}
{{K + 1}}\mu \frac{{\partial U}}
{{\partial x}}}  \\[2mm]
   {\frac{{K + 3}}
{4}\mu \frac{\partial }
{{\partial x}}\left( {\frac{1}
{\lambda }} \right) + \frac{{2K}}
{{K + 1}}\mu U\frac{{\partial U}}
{{\partial x}}\,}  \\
 \end{array} } \right).
\end{eqnarray}

The relationship between the partial derivative $\frac{\partial g}{\partial x}$ and $\frac{\partial \bfW}{\partial x}$ is
\begin{eqnarray}
\frac{{\partial \bfW}}{{\partial x}}  =  \int {\frac{{\partial g}}{{\partial x}}{\bfP}{\rmd \Xi}}.
\label{eq:wxgx}
\end{eqnarray}
Because of the typical form of the Maxwellian equilibrium state, $\frac{\partial g}{\partial x}$ can be written as
\begin{eqnarray}
\frac{{\partial g}}{{\partial x}} = g \left( \bfP ^{\rmT} \cdot {\bf a}\right),
\label{eq:gx1}
\end{eqnarray}
where ${\bf a}=\left( {a_1 ,a_2 ,a_2 } \right)^{\rmT}$ is independent of $u$ and ${\bm \xi}$.
Define
\begin{eqnarray}
{\bf M}^1({m_1},{m_2})=\int\limits {gu^{m_1}{\bm \xi}^{m_2}{\bfP}^{\rmT} {\bfP}{\rmd}\Xi }.\nonumber
\label{eq:moment1}
\end{eqnarray}
The moments in ${\bf M}^1({m_1},{m_2})$ can be determined completely by $\bfW$.
According to the moments of the Maxwellian distribution function \cite{xu1998},
\begin{eqnarray}
\label{eq:pp}
{\bf M}^1({0},{0})
&=&\int\limits {gu^{0}{\bm \xi}^{0}{\bfP}^{\rmT} {\bfP}{\rmd}\Xi }\nonumber\\
&=& \rho \left[
  {\begin{array}{*{20}c}
   1 & U & {\frac{1}
{2}\left( {U^2  + \frac{{K + 1}}
{{2\lambda }}} \right)}  \\[2mm]
   U & {U^2  + \frac{1}
{{2\lambda }}} & {\frac{1}
{2}\left( {U^3  + \frac{{K + 3}}
{{2\lambda }}U} \right)}  \\[2mm]
   {\frac{1}
{2}\left( {U^2  + \frac{{K + 1}}
{{2\lambda }}} \right)} & {\frac{1}
{2}\left( {U^3  + \frac{{K + 3}}
{{2\lambda }}U} \right)} & {\frac{1}
{4}\left( {U^4  + \frac{{K + 3}}
{\lambda }U^2  + \frac{{K^2  + 4K + 3}}
{{4\lambda ^2 }}} \right)}  \\
 \end{array} }
 \right].\nonumber
\end{eqnarray}
Combining Eq.(\ref{eq:wxgx}) with Eq.(\ref{eq:gx1}) yields
\begin{eqnarray}
{\bf M}^1({0},{0}) \cdot {\bf{a}}
= \frac{\partial {\bfW}}{\partial x}.
\label{eq:gxa}
\end{eqnarray}
Consequently, {\bf{a}} and $\frac{\partial g}{\partial x}$ can be determined completely by $\bfW$ and $\frac{\partial \bfW}{\partial x}$.

Furthermore, the second-order derivative $\frac{\partial ^2 g}{\partial x^2}$ can be determined by $\frac{\partial ^2 \bfW}{\partial x^2}$ from more complex derivations.
Considering the cross terms between the lower-order derivatives,
\begin{eqnarray}
\frac{\partial ^2 g}{\partial x^2} =
g\left( \bfP ^{\rmT} \cdot {\bf a}\right)\left( \bfP ^{\rmT} \cdot {\bf a}\right)
+g\left( \bfP ^{\rmT} \cdot {\bf b}\right),
\label{eq:gxb0}
\end{eqnarray}
where ${\bf b}=\left( {b_1 ,b_2 ,b_2 } \right)^{\rmT}$ is also independent of $u$ and ${\bm \xi}$.
Define
\begin{eqnarray}
&&{\mathbf{M}}^2 \left( {m_1 ,m_2 ,{\mathbf{a}}} \right)\nonumber\\
&=&\int\limits {g\left( \bfP ^{\rmT} \cdot {\bf a}\right)u^{m_1}{\bm \xi}^{m_2}{\bfP}^{\rmT} {\bfP}{\rmd}\Xi } \nonumber\\
&=& \left[ {{\mathbf{M}}^1 \left( {m_1 ,m_2 } \right) \cdot {\mathbf{a}},{\mathbf{M}}^1 \left( {m_1  + 1,m_2 } \right) \cdot {\mathbf{a}},\frac{{{\mathbf{M}}^1 \left( {m_1  + 2,m_2 } \right) + {\mathbf{M}}^1 \left( {m_1 ,m_2  + 2} \right)}}
{2} \cdot {\mathbf{a}}} \right].\nonumber
\label{eq:moment2}
\end{eqnarray}
${\bf b}$ can be determined by taking the moments of Eq.(\ref{eq:gxb0}),
\begin{eqnarray}
{\bf M}^1({0},{0}) \cdot {\bf{b}}
= \frac{\partial^2 {\bfW}}{\partial x^2}
-{\mathbf{M}}^2 \left( {0 ,0 ,{\mathbf{a}}} \right) \cdot {\mathbf{a}}.
\label{eq:gxb}
\end{eqnarray}
See \cite{xu1998} for more specific expressions of the complex moments.
To obtain the higher-order derivatives, more complex moments are required.

\begin{remark}
\label{re:gxgt}
It occupies large amounts of floating point operations to determine the high-order partial derivative of g from $\bfW$ and its derivatives \cite{pan2016,liu2014}.
The moments from ${\bf M}^1({m_1},{m_2})$ to ${\mathbf{M}}^2 \left( {m_1 ,m_2 ,{\mathbf{a}}} \right)$ indicate these floating point operations grow up nearly exponentially as the order increases.
\end{remark}

\subsection{\label{sec:gks} HGKS }

\subsubsection{\label{sec:gks-fv} Finite volume framework }

HGKS is based on the finite volume framework.
The space domain is discretized into $N$ computational cells, indexed as $i$.
The cell size is $\Delta x$.
The time domain is discretized into $0,\Delta t^0,...,t^n,t^{n+1}=t^n+\Delta t^n,...$.
The cell-averaged conservative variables at time $t^n$ are recorded as $\overline {\bfW} _i^{n}$.
In the following discussions, $\Delta t^n$ is abbreviate as $\Delta t$ in the absence of ambiguity.
$t^n=0$ is adopted without loss of generality.
Under the one-step fully discrete finite volume framework, the cell-averaged conservative variables are updated by
\begin{eqnarray}
\label{eq:flux}
\overline {\bfW} _i^{n+1}  = \overline {\bfW} _i^n  - \frac{1}
{{\Delta x}}\left( {\int_0^{\Delta t} {{\bfF}_{i + \frac{1}{2}}\left( {t} \right){\rmd}t}  - \int_0^{\Delta t} {{\bfF}_{i - \frac{1}
{2}}\left( {t} \right){\text{d}}t} } \right).
\end{eqnarray}
In HGKS, the integral solution Eq.(\ref{eq:ints}) is utilized as its evolution model in flux evaluation
\begin{eqnarray}
\label{eq:flux}
\bfF_{i + \frac{1}{2}}(t) = \int {f({ x_{i + \frac{1}{2}}},t,{ u},{\bm \xi}) { u} {\bfP}{\rmd \Xi}}.
\end{eqnarray}
HGKS is originally developed to solve the NS equations from the Taylor expansion and first-order Chapman-Enskog expansion on the integral solution \cite{xu2001}.
Through the modifications on $\tau$ to replace the physical dissipation by the numerical dissipation, HGKS is also applicable to the Euler equations \cite{luo22013}.
Because the original HGKS provides the basis for higher-order extensions and simplifications, the state-of-art third-order HGKS \cite{luo2013,lqb2010} is reviewed firstly, followed by the existing simplifications \cite{zhou2017} for a more efficient NS solver and an Euler solver \cite{luo2013,luo22013} ultimately.

\subsubsection{\label{sec:gks-3d} Third-order HGKS }

To simplify the presentation, $f({ x},t)=f({ x},t,{ u},{\bm \xi})$ is adopted in the following discussions.
At the cell interface, $x_{i + \frac{1}{2}}=0$ without loss of generality.
To construct the third-order HGKS, the second-order Taylor expansion is implemented on the integral solution Eq.(\ref{eq:ints}), specifically on $f({ x} - { u} t,0)$ and $g({ x}',t')$ respectively.

Firstly, the second-order Taylor expansion on $f({ x}_{i+\frac{1}{2}} - { u} t,0)$ is given as
\begin{eqnarray}
f({ x}_{i+\frac{1}{2}} - { u} t,0) =
    \sum\limits_{l = 0}^2 {\frac{\left( {- {{u}}t} \right)^l}{{l!}}
                        \frac{\partial ^l  f^0 }{{\partial x^l}}},
\label{eq:taylord1f00}
\end{eqnarray}
where $f^0=f(x_{i+\frac{1}{2}} ,0)$.
Originally devoted for the NS equations, $f^0$ is given the first-order Chapman-Enskog expansion \cite{xu2001}
\begin{eqnarray}
f^0= f^0_{NS}
=g^0-\tau(\frac{\partial g^0}{\partial t}+u\frac{\partial g^0}{\partial x}).
\label{eq:ce1}
\end{eqnarray}
$g^0$ is constructed as
\begin{eqnarray}
g^{0} =
\left\{
\begin{array}{l}
 g^{\rm L}=g\left( \bfW^{\rm L} \right) ,u \ge 0, \\
 g^{\rm R}=g\left( \bfW^{\rm R} \right) ,u < 0,
 \end{array}
 \right.
\label{eq:consf0}
\end{eqnarray}
where $g^{\rm L}$ and $g^{\rm R}$ are determined by the point-wise values $\bfW^{\rm L}$ and $\bfW^{\rm R}$ respectively.
After the third-order reconstruction at the cell interface $x_{i+\frac{1}{2}}$,
\begin{eqnarray}
&& \bfW^{\rm L} = \mathop {\lim }\limits_{{x}  \to {x_{i+\frac{1}{2}}}^ - }
                    \bfW_i(x,0) ,\\
&& \bfW^{\rm R} = \mathop {\lim }\limits_{{x}  \to {x_{i+\frac{1}{2}}}^ + }
                    \bfW_{i+1}(x,0) ,\nonumber
\label{eq:consf01}
\end{eqnarray}
where $\bfW_i({x},0)$ is the reconstructed quadratic polynomials over the $i$-th cell.
A typical choice of reconstruction is given in ~\ref{app:rec}.
After the replacement by $f^0= f^0_{NS}$, the expansion yields \cite{luo22013}
\begin{eqnarray}
f({ x}_{i+\frac{1}{2}} - { u} t,0) &=&
    \sum\limits_{l = 0}^2 {\frac{\left( {- {{u}}t} \right)^l}{{l!}}
                        \frac{\partial ^l  f^0_{NS} }{{\partial x^l}}} \\
&=&g^0  - \left( {t + \tau } \right)u\frac{{\partial g^0 }}
{{\partial x}} - \tau \frac{{\partial g^0 }}
{{\partial t}} + \left( {\tau tu^2  + \frac{1}
{2}u^2 t^2 } \right)\frac{{\partial ^2 g^0 }}
{{\partial x^2 }} + \tau tu\frac{{\partial ^2 g^0 }}
{{\partial x\partial t}}
-\tau^0*,\nonumber
\label{eq:taylord1f0luo}
\end{eqnarray}
where $\tau^0*=\tau t^2  \frac{u^2{\partial ^2 }}{{2\partial x^2 }}\left( {\frac{{\partial g_{}^0 }}{{\partial t}} + u\frac{{\partial g_{}^0 }}{{\partial x}}} \right)$ are the third-order partial derivatives and should be eliminated in the third-order evolution model.
To build up a simpler expression of the expansion, a more intensive summation formula is introduced into Eq.(\ref{eq:taylord1f0luo}).
The expansion is then re-written as
\begin{eqnarray}
\label{eq:taylord1f0}
f({ x}_{i+\frac{1}{2}} - { u} t,0) =
    \sum\limits_{l = 0}^2 {\frac{\left( {- {{u}}t} \right)^l}{{l!}}
                        \frac{\partial ^l  f_{NS}^0 }{{\partial x^l}}}
+ \tau^0*.
\end{eqnarray}
The extra term $ \tau^0*$ is present for the establishment of the intensive summation formula in Eq.(\ref{eq:taylord1f0}).

Secondly, the same-order Taylor expansion on $g({ x}',t')$ in both space and time yields
\begin{eqnarray}
g(x',t') = \sum\limits_{l = 0}^2 {\frac{1}{{{l!}}}
\left( - {{u\left( {t - t'} \right)}}
\frac{\partial }{{\partial x}}+
{t'\frac{\partial }{{\partial t}} }
\right)^l g^e } ,
\label{eq:taylord1ge}
\end{eqnarray}
where $g^e $ can be determined by
\begin{eqnarray}
g^e=g\left( \bfW^{e} \right).
\label{eq:consge}
\end{eqnarray}
$\bfW ^e$ is obtained according to the conservation constraint \cite{xu2001}
\begin{eqnarray}
\bfW ^e = \int_{u \ge 0} { g^{\rm L} {\bfP}{\rmd \Xi}} + \int_{u<0} { g^{\rm R} {\bfP}{\rmd \Xi}}.
\label{eq:compg}
\end{eqnarray}
After the integration on $t'$, the high-order terms related to $\tau$, such as $\tau^2$ and $\tau^3$, are introduced.
To be consistent with the first-order Chapman-Enskog expansion on $\tau$, the high-order terms related to $\tau^2$, $\tau^3$ and so on, should be eliminated \cite{luo2013}, which yields
\begin{eqnarray}
\label{eq:taylord1fe}
\frac{1}{\tau} \int_0^t {g({ x}',t',{ u},{\bm \xi}) \rme^{-(t - t')/\tau} {\rmd t'}}
=
\sum\limits_{l = 0}^2 {\frac{ t^l}{{l!}}
                         {\frac{\partial ^l f_{\rm{NS}}^e}{{\partial t^l}} }  }
- {\rm{e}}^{ - t/\tau }
    \sum\limits_{l = 0}^2 {\frac{\left( {- {{u}}t} \right)^l}{{l!}}
                        \frac{\partial ^l  {f_{\rm{NS}}^e } }{{\partial x^l}}}+\tau^e*,
\end{eqnarray}
where
\begin{eqnarray}
\label{eq:taylord1fetau}
&&f^e_{NS}
=g^e-\tau(\frac{\partial g^e}{\partial t}+u\frac{\partial g^e}{\partial x}),\nonumber\\
&&\tau^e*= \tau \frac{{t^2 }}
{2}\frac{{\partial ^2 }}
{{\partial t^2 }}\left( {\frac{{\partial g_{}^e }}
{{\partial t}} + u\frac{{\partial g_{}^e }}
{{\partial x}}} \right)
- \tau {\text{e}}^{ - t/\tau } \frac{{u^2 t^2 }}
{2}\frac{{\partial ^2 }}
{{\partial x^2 }}\left( {\frac{{\partial g_{}^e }}
{{\partial t}} + u\frac{{\partial g_{}^e }}
{{\partial x}}} \right).\nonumber
\end{eqnarray}

Combining Eq.(\ref{eq:taylord1f0}) and Eq.(\ref{eq:taylord1fe}), the evolution model for the third-order HGKS can be written as
\begin{eqnarray}
\label{eq:taylor3}
f(x_{i+\frac{1}{2}} ,t)
=
\sum\limits_{l = 0}^2 {\frac{ t^l}{{l!}}
                         {\frac{\partial ^l f_{\rm{NS}}^e}{{\partial t^l}} }  }
+ {\rm{e}}^{ - t/\tau }
    \sum\limits_{l = 0}^2 {\frac{\left( {- {{u}}t} \right)^l}{{l!}}
                        \frac{\partial ^l  \left( {f_{\rm{NS}}^{0}  - f_{\rm{NS}}^e } \right)}{{\partial x^l}}}+ \tau^*,
\end{eqnarray}
where
\begin{eqnarray}
\label{eq:taylor3tau}
\tau^*= \tau \frac{{t^2 }}
{2}\frac{{\partial ^2 }}
{{\partial t^2 }}\left( {\frac{{\partial g_{}^e }}
{{\partial t}} + u\frac{{\partial g_{}^e }}
{{\partial x}}} \right)
+ \tau {\text{e}}^{ - t/\tau } \frac{{u^2 t^2 }}
{2}\frac{{\partial ^2 }}
{{\partial x^2 }}\left( {\frac{{\partial  }}
{{\partial t}} + u\frac{{\partial }}
{{\partial x}}} \right)\left(g^0-g^e\right).\nonumber
\end{eqnarray}
\begin{remark}
\label{re:ft3}
The evolution model Eq.(\ref{eq:taylor3}) for the third-order HGKS is written in a different way from the existing work \cite{luo2013,luo22013}.
With a more intensive summation formula, we find it more convenient for higher-order extensions based on this way.
\end{remark}

To construct the space derivatives of $g^{e}$, $g^0=g^{\rm L}$ or $ g^{\rm R}$ in Eq.(\ref{eq:taylor3}), the third-order reconstruction is utilized firstly to obtain the quadratic polynomials of the conservative variables.
A typical choice is given in ~\ref{app:rec}.
Based on the point-wise derivatives of $\bfW^{e}$, $\bfW^{\rm L}$ and $ \bfW^{\rm R}$, the derivatives of $g^{e}$, $g^{\rm L}$ and $ g^{\rm R}$ can be solved from Eq.(\ref{eq:gxa}) and Eq.(\ref{eq:gxb}) respectively.

For the time derivatives, we denote $\bf a$ in Eq.(\ref{eq:gxa}) as ${\bf g}_x$ and $\bf b$ in Eq.(\ref{eq:gxb}) as ${\bf g}_{xx}$.
Similar definition can be given to introduce ${\bf g}_t,{\bf g}_{xt}$ and ${\bf g}_{tt}$.
The time derivative $ \frac{\partial g}{{\partial t}}$ for $g=g^{{\rm L}},g^{{\rm R}}$ or $g^{e}$ is determined from the compatibility conditions \cite{lqb2010,luo2013}
\begin{eqnarray}
{\int \left( {\frac{{\partial g}}{{\partial t}}}
 + u\frac{\partial g}{\partial x}  \right){\bfP}{\rmd \Xi}} = {\bm 0}.
\label{eq:compacond}
\end{eqnarray}
Specifically, ${\bf g}_{t}$ is obtained from ${\bf g}_{x}$ by solving
\begin{eqnarray}
{\bf M}({0},{0}) \cdot {\bf{g}}_{{t}}=-{\bf M}({1},{0}) \cdot {\bf{g}}_{{x}}.
\label{eq:gxgt}
\end{eqnarray}
To obtain the second-order time derivative, the compatibility condition \cite{luo2013,liu2014} is further derived to solve ${\bf g}_{xt} $ firstly from
\begin{eqnarray}
{\int \frac{\partial }{\partial x} \left( {\frac{{\partial g}}{{\partial t}}}
 + u\frac{\partial g}{\partial x}  \right){\bfP}{\rmd \Xi}} = {\bm 0},
\label{eq:compacond2}
\end{eqnarray}
and followed by
\begin{eqnarray}
{\int \frac{\partial }{\partial t} \left( {\frac{{\partial g}}{{\partial t}}}
 + u\frac{\partial g}{\partial x}  \right){\bfP}{\rmd \Xi}} = {\bm 0}.
\label{eq:compacond3}
\end{eqnarray}
to solve ${\bf g}_{tt} $.

Finally, ${\bfF}_{i + \frac{1}{2}}(t)$ can be obtained by taking the moments on $g^{\rm{L}}$, $g^{\rm{R}}$, $g^{e}$ and their space and time derivatives\cite{xu1998}.
For a better illustration, the procedure for the flux evaluation in the state-of-art third-order HGKS is plotted as follows
\begin{eqnarray}
\left. {\begin{array}{*{20}c}
   {}  \\
   {}  \\
   {\boxed{{\mathbf{W}}^{\text{L}} }\;\;\; \to \;\;}  \\
   \begin{gathered}
   \hfill \\
   \hfill \\
   \hfill \\
   \hfill \\
   \hfill \\
   \hfill \\
   \hfill \\
  \boxed{{\mathbf{W}}^{\text{R}} }\;\;\; \to \;\; \hfill \\
   \hfill \\
   \hfill \\
\end{gathered}   \\
 \end{array} \begin{array}{*{20}c}
   {}  \\
   {}  \\
   {g^{\text{L}} \;\;\;\;\;\; \to \;\;\;\;}  \\
   \begin{gathered}
   \hfill \\
   \downarrow  \hfill \\
   \hfill \\
  {\mathbf{W}}^e  \to g^e  \to  \hfill \\
   \hfill \\
   \uparrow  \hfill \\
   \hfill \\
  g^{\text{R}} \;\;\;\;\;\; \to \;\;\;\; \hfill \\
   \hfill \\
   \hfill \\
\end{gathered}   \\
 \end{array} \begin{array}{*{20}c}
   {\left\{ \begin{gathered}
  \boxed{\frac{{\partial {\mathbf{W}}^{\text{L}} }}
{{\partial x}}} \to \frac{{\partial g^{\text{L}} }}
{{\partial x}} \to \frac{{\partial g^{\text{L}} }}
{{\partial t}} \hfill \\
  \boxed{\frac{{\partial ^2 {\mathbf{W}}^{\text{L}} }}
{{\partial x^2 }}} \to \frac{{\partial ^2 g^{\text{L}} }}
{{\partial x^2 }} \to \frac{{\partial ^2 g^{\text{L}} }}
{{\partial x\partial t}}\;\;\;\;\;\;\;\;\;\;\; \hfill \\
\end{gathered}  \right.}  \\
   {\left\{ \begin{gathered}
  \boxed{\frac{{\partial {\mathbf{W}}^e }}
{{\partial x}}} \to \frac{{\partial g^e }}
{{\partial x}} \to \frac{{\partial g^e }}
{{\partial t}} \hfill \\
  \boxed{\frac{{\partial ^2 {\mathbf{W}}^e }}
{{\partial x^2 }}} \to \frac{{\partial ^2 g^e }}
{{\partial x^2 }} \to \frac{{\partial ^2 g^e }}
{{\partial x\partial t}} \to \frac{{\partial ^2 g^e }}
{{\partial t^2 }} \hfill \\
\end{gathered}  \right.}  \\
   {\left\{ \begin{gathered}
  \boxed{\frac{{\partial {\mathbf{W}}^{\text{R}} }}
{{\partial x}}} \to \frac{{\partial g^{\text{R}} }}
{{\partial x}} \to \frac{{\partial g^{\text{R}} }}
{{\partial t}} \hfill \\
  \boxed{\frac{{\partial ^2 {\mathbf{W}}^{\text{R}} }}
{{\partial x^2 }}} \to \frac{{\partial ^2 g^{\text{R}} }}
{{\partial x^2 }} \to \frac{{\partial ^2 g^{\text{R}} }}
{{\partial x\partial t}}\;\;\;\;\;\;\;\;\;\; \hfill \\
\end{gathered}  \right.}  \\
 \end{array} } \right\} \to {\mathbf{F}}_{i + \frac{1}
{2}} (t)\nonumber
\label{eq:compacond}
\end{eqnarray}
The quantities in the box are obtained from the reconstructions.
\begin{remark}
A number of second-order derivatives and their moments are required to be computed in the third-order HGKS.
They occupy high proportion of the total computation cost \cite{zc2018,liu2014}.
\end{remark}

\subsubsection{\label{sec:simp0} Simplifications on HGKS }

The original HGKS possesses the high accuracy and strong robustness to solve the NS and Euler equations for compressible flows \cite{xu2001,lqb2010,luo22013}.
But the process to compute a number of high-order derivatives of particle distribution function holds back HGKS to higher-order extensions and their practical simulations.
In the meantime, researches on the simplifications of the third-order HGKS have been carried out for better efficiency \cite{zhou2017,luo22013}.
The existing simplifications mainly include two types.
One is implemented on the numerical dissipation part to solve NS equations more efficiently \cite{zhou2017}.
The other is implemented on the physical dissipation part to solve the Euler equations \cite{luo22013}.

Zhou introduced two simplifications on the third-order HGKS \cite{zhou2017}.
One is the linearization on the high-order derivatives, as $\frac{\partial ^2 g}{\partial x^2} =g\left( {\bf b} \cdot \bfP \right)$ for example \cite{zhou2017}.
Its validation is still doubtable in non-linear cases.
The other simplification is to eliminate all the high-order terms in the numerical dissipation part in the evolution model \cite{zhou2017}.
Based on the mechanism analysis \cite{zhou2017}, in the smooth regions where $g^{\rm{L}} = g^{\rm{R}} = g^e$, only the physical dissipation part works \cite{zhou2017}.
The evolution model remains as
\begin{eqnarray}
\label{eq:taylor3s1}
f(x_{i+\frac{1}{2}} ,t)
=
\sum\limits_{l = 0}^2 {\frac{ t^l}{{l!}}
                         {\frac{\partial ^l f_{\rm{NS}}^e}{{\partial t^l}} }  }
+ \tau \frac{{t^2 }}
{2}\frac{{\partial ^2 }}
{{\partial t^2 }}\left( {\frac{{\partial g_{}^e }}
{{\partial t}} + u\frac{{\partial g_{}^e }}
{{\partial x}}} \right).
\end{eqnarray}
The extra term
\begin{eqnarray}
\label{eq:taylor3s2}
{\rm{e}}^{ - t/\tau }
    \sum\limits_{l = 0}^2 {\frac{\left( {- {{u}}t} \right)^l}{{l!}}
                        \frac{\partial ^l  \left( {f_{\rm{NS}}^{0}  - f_{\rm{NS}}^e } \right)}{{\partial x^l}}}+ \tau {\text{e}}^{ - t/\tau } \frac{{u^2 t^2 }}
{2}\frac{{\partial ^2 }}
{{\partial x^2 }}\left( {\frac{{\partial  }}
{{\partial t}} + u\frac{{\partial }}
{{\partial x}}} \right)\left(g^0-g^e\right),
\end{eqnarray}
is the numerical dissipation part.
According to the analysis by Zhou, only the primary terms related to $t$ in the 1D cases are required to provide necessary numerical dissipation \cite{zhou2017}.
The numerical dissipation part remains as
\begin{eqnarray}
\label{eq:taylor3s2}
{\rm{e}}^{ - t/\tau }
    \sum\limits_{l = 0}^1 {\frac{\left( {- {{u}}t} \right)^l}{{l!}}
                        \frac{\partial ^l  \left( {g^{0}  - g^e } \right)}{{\partial x^l}}}.
\end{eqnarray}
As a summary, from the mechanism analysis on the evolution model \cite{zhou2017}, the third-order HGKS evolution model is simplified as
\begin{eqnarray}
\label{eq:taylor3s3}
f(x_{i + \frac{1}{2}} ,t) = \sum\limits_{l = 0}^2 {\frac{{t^l }}
{{l!}}\frac{{\partial ^l f_{{\text{NS}}}^e }}
{{\partial t^l }}}  + {\text{e}}^{ - t/\tau } \sum\limits_{l = 0}^1 {\frac{{\left( { - ut} \right)^l }}
{{l!}}\frac{{\partial ^l \left( {g}^0  - g^e  \right)}}
{{\partial x^l }}}  \\
+\tau \frac{{t^2 }}
{2}\frac{{\partial ^2 }}
{{\partial t^2 }}\left( {\frac{{\partial g_{}^e }}
{{\partial t}} + u\frac{{\partial g_{}^e }}
{{\partial x}}} \right).\nonumber
\end{eqnarray}
The high-order derivatives $\frac{\partial ^2 g^{\rm{L}}}{\partial x^2},\frac{\partial ^2 g^{\rm{L}}}{{\partial x \partial t}}$,
$\frac{\partial ^2 g^{\rm{R}}}{\partial x^2},\frac{\partial ^2 g^{\rm{R}}}{{\partial x \partial t}}$ are eliminated.
Consequently, the computation cost to solve these derivatives and take their moments in flux evaluation is saved.
This simplification is also inspiring to extend HGKS to achieve arbitrary high-order accuracy.

What is more, since the original HGKS is devoted to solve the NS equations, there still exists the physical dissipation in the evolution model.
To solve the Euler equations, the physical dissipation should be eliminated by approaching $\tau$ to $0$ \cite{luo2013},
with the numerical dissipation terms ruled out.
Based on the original third-order HGKS, the simplified evolution model to solve the Euler equations is given as \cite{luo2013}
\begin{eqnarray}
\label{eq:taylor3s4}
f(x_{i + \frac{1}{2}} ,t) = \sum\limits_{l = 0}^2 {\frac{{t^l }}
{{l!}}\frac{{\partial ^l g^e }}
{{\partial t^l }}}  +
 {\rm{e}}^{ - t/\tau }
    \sum\limits_{l = 0}^2 {\frac{\left( {- {{u}}t} \right)^l}{{l!}}
                        \frac{\partial ^l  \left( {f_{\rm{NS}}^{0}  - f_{\rm{NS}}^e } \right)}{{\partial x^l}}}\\
 + \tau {\text{e}}^{ - t/\tau } \frac{{u^2 t^2 }}
{2}\frac{{\partial ^2 }}
{{\partial x^2 }}\left( {\frac{{\partial  }}
{{\partial t}} + u\frac{{\partial }}
{{\partial x}}} \right)\left(g^0-g^e\right),\nonumber
\end{eqnarray}
where the moments on $\frac{\partial ^2 g^{e}}{{\partial x\partial t}}$ are no longer included in the flux evaluation.

\begin{remark}
The existing simplifications on the third-order HGKS are mostly from the reduction of the derivatives of particle distribution functions and their moments to be computed.
However, no more than half of them are able to be reduced.
Consequently, the improved efficiency is limited.
The difficulty still exists in extending the state-of-art HGKS to arbitrary high-order accuracy with practical significance even to solve the Euler equations.
Not to mention the multi-dimensional cases where much more and complex derivatives are included \cite{zhou2017}.
\end{remark}

\section{\label{sec:hgks} HGKS with arbitrary high-order accuracy: Euler equations}

\subsection{\label{sec:hgks-e} HGKS with arbitrary high-order accuracy}

In this section, the state-of-art HGKS is extended mandatorily to arbitrary high-order accuracy, to provide the basis of the following simplifications and modifications for a more efficient Euler solver.

From the arbitrary $r$-th-order Taylor expansion on the integral solution, the evolution model to build up HGKS with $(r+1)$th-order accuracy is given as
\begin{eqnarray}
f(x_{i+\frac{1}{2}},t) =
\sum\limits_{l = 0}^r {\frac{ t^l}{{l!}}
                         {\frac{\partial ^l f_{\rm{NS}}^e}{{\partial t^l}} }  }
+ {\rm{e}}^{ - t/\tau }
    \sum\limits_{l = 0}^r {\frac{\left( {- {{u}}t} \right)^l}{{l!}}
                        \frac{\partial ^l \left( {f_{\rm{NS}}^{0}  - f_{\rm{NS}}^e } \right)}{{\partial x^l}}}+ \tau^*,
\label{eq:taylord1}
\end{eqnarray}
where
\begin{eqnarray}
\tau^*= \tau \frac{{t^r }}{r!}\frac{{\partial ^r }}{{\partial t^r }}\left( {\frac{{\partial g_{}^e }}{{\partial t}} + u\frac{{\partial g_{}^e }}
{{\partial x}}} \right)
+ \tau {{\rme}}^{ - t/\tau }
\frac{{u^r t^r }}{r!}\frac{{\partial ^r }}{{\partial x^r }}\left( {\frac{{\partial }}{{\partial t}}
+ u\frac{{\partial }}{{\partial x}}} \right)\left(g^0-g^e\right).\nonumber
\label{eq:taylord1tau}
\end{eqnarray}
Compared with the third-order HGKS Eq.(\ref{eq:taylor3}), the higher-order derivatives of $g^0=g^{{\rm L}}$ or $g^{{\rm R}}$ and $g^{e}$ are required
to be computed for $r>2$.
The corresponding $(r+1)$th-order reconstructions are implemented to obtain the polynomials of $\bfW_i({x},0)$ and $\bfW^e$ firstly.
By further taking derivations on Eqs.(\ref{eq:gxa}) and (\ref{eq:gxb}), the higher-order space derivatives of $g=g^{{\rm L}},g^{{\rm R}}$ or $g^{e}$
can be solved from the point-wise derivatives of $\bfW^{{\rm L}},\bfW^{{\rm R}}$ or $\bfW^{e}$ respectively.
Meanwhile, the high-order time related derivatives $\frac{\partial ^{m+q+1} g}{\partial x^m  t^{q+1}} $ are obtained according to the compatibility conditions \cite{luo2013,liu2014}
\begin{eqnarray}
{\frac{\partial ^{m+q}}{\partial x^m  t^q}
\int \left( {\frac{{\partial g}}{{\partial t}}}
 + { u}{\frac{{\partial g}}{{\partial x}}} \right){\bfP}{\rmd \Xi}} = {\bm 0}.
\label{eq:hgxgt}
\end{eqnarray}

In the fifth-order HGKS, computations for the derivatives of $g^{e}$ follow from this sequence:
\begin{eqnarray}
&&\frac{\partial g^{e}}{\partial x} \to \frac{\partial g^{e}}{\partial t}; \nonumber\\
&&\frac{\partial ^2 g^{e}}{\partial x^2} \to \frac{\partial ^2 g^{e}}{\partial x \partial t} \to \frac{\partial ^2 g^{e}}{\partial t^2};\nonumber\\
&&\frac{\partial ^3 g^{e}}{\partial x^3}  \to \frac{\partial ^3 g^{e}}{\partial x^2 \partial t}  \to \frac{\partial ^3 g^{e}}{\partial x \partial t^2} \to \frac{\partial ^3 g^{e}}{\partial t^3 } ;\nonumber\\
&&\frac{\partial ^4 g^{e}}{\partial x^4} \to \frac{\partial ^4 g^{e}}{\partial x^3 \partial t} \to \frac{\partial ^4 g^{e}}{\partial x^2 \partial t^2} \to \frac{\partial ^4 g^{e}}{\partial x \partial t^3} \to {\frac{\partial ^4 g^{e}}{\partial t^4}}.\nonumber
\label{eq:gxs1}
\end{eqnarray}

\begin{remark}
The sequence to translate the space and time derivatives of $g^e$ in HGKS is in accord with the traditional Lax-Wendroff procedure \cite{titarev2005},
except that  the space-time translations in the Lax-Wendroff procedure are based on the macroscopic variables \cite{xu1998,luo22013,titarev2005}.
The requirement to solve the space and time derivatives of particle distribution functions in HGKS makes it more complicated.
It brings in much rapider increase of computation cost as the order of accuracy increases \cite{pan2016,luo2013,ji2017,liu2014}.
It even occupies most of the computation cost in the fourth-order HGKS \cite{liu2014} to
calculate the third-order derivatives of particle distribution functions and take their moments.
Consequently, the state-of-art HGKS with higher than third-order accuracy is compromised to the multi-stage framework where only
low-order HGKS evolution model is utilized\cite{pan2016,ji2017,ji2018}.
\end{remark}

\subsection{\label{sec:hgks-s} Simplification and modification on HGKS}
In this section, the existing simplifications in \cite{zhou2017,luo22013} are extended and modified to develop a relatively more applicable HGKS with arbitrary high-order accuracy to solve the Euler equations.

Similar to the notation by Zhou \cite{zhou2017}, the evolution model of the $(r+1)$th-order HGKS Eq.(\ref{eq:taylord1}) is again written as
\begin{eqnarray}
{ f}(x_{i+\frac{1}{2}},t) = {\mathscr P_r} f_{\rm{NS}}^e
+ {\rm{e}}^{ - t/\tau } {\mathscr N_r}\left( {f_{\rm{NS}}^{0}  - f_{\rm{NS}}^e } \right)+\tau*,
\label{eq:taylor2}
\end{eqnarray}
where
\begin{eqnarray}
{\mathscr P_r}
    =\sum\limits_{l = 0}^r {\frac{ t^l}{{l!}}
                         {\frac{\partial ^l }{{\partial t^l}} }  },
{\mathscr N_r}=
    \sum\limits_{l = 0}^r {\frac{\left( {- {{u}}t} \right)^l}{{l!}}
                        \frac{\partial ^l }{{\partial x^l}}}.\nonumber
\label{eq:taylor20}
\end{eqnarray}
From the previous discussions in Section~\ref{sec:simp0}, it is able to get rid of the high-order terms $\tau*$ in the analysis of Zhou \cite{zhou2017} to simplify the evolution model.
Based on the analysis \cite{zhou2017}, the numerical dissipation terms in Eq.(\ref{eq:taylor2}) are ${\rm{e}}^{ - t/\tau }{\mathscr N_r} ({ f_{\rm{NS}}^{0}- f_{\rm{NS}}^e } )$.
Since only the primary terms related to $t$ are required to provide necessary numerical dissipation \cite{zhou2017}, the numerical dissipation part remains as ${\rm{e}}^{ - t/\tau } {\mathscr N_1} ( g^{0}- g^e )$.
Only the first and second derivatives of distribution functions are reserved in the numerical dissipation part.
After eliminating the physical dissipation \cite{luo2013}, the HGKS evolution model to solve the Euler equations remains as
\begin{eqnarray}
{ f}(x_{i+\frac{1}{2}},t)={\mathscr P_r} g^e
+ {\rm{e}}^{ - t/\tau } {\mathscr N_1} \left( g^{0}- g^e \right).
\label{eq:is0}
\end{eqnarray}

\begin{remark}
There exist drawbacks in the straightforward extension Eq.(\ref{eq:is0}).
It is inevitable to encounter serious instability problem if without elaborate reconstructions on $\bfW ^e$.
While in fact, considering the dominant role of $ {\mathscr P_r} g^e $ in the smooth regions, the derivatives of $ \bfW ^e $ are uniformly obtained based on the continuous flow distribution hypothesis, by linear reconstruction for example \cite{zc2018,lqb2010,liu2014,pan2016,ji2017}, wherever a discontinuity exists.
Near the discontinuity, the numerical oscillations generated by the high-order terms in $ {\mathscr P_r} g^e $ cannot be balanced sufficiently by the low-order numerical dissipation part.
This problem becomes more serious theoretically as $r$ increases.
\end{remark}

To overcome the above-mentioned problem without elaborate reconstructions, the evolution model Eq.(\ref{eq:is0}) is further modified.
Firstly, ${\rm{e}}^{ -  t/\tau }  {\mathscr N_1}g^e $ in the numerical dissipation part is replaced by ${\rm{e}}^{ -  t/\tau } {\mathscr P_r} g^e$.
Since all the terms related to $g^e$ are vanished near the discontinuity where $\tau \gg t$, no elaborate reconstruction on $\bfW ^e$ is required.
At the same time, the weight ${\rm{e}}^{ -  t/\tau }$ is replaced by ${\rm{e}}^{ - \Delta t/\tau }$.
This modification preserves the evolution model the underlying idea of HGKS.
And it also makes the following flux evaluation much simpler.
Under these modifications, the evolution model Eq.(\ref{eq:is0}) yields
\begin{eqnarray}
{ f}(x_{i+\frac{1}{2}},t)=
\left(1- {\rm{e}}^{ -  \Delta t/\tau }\right) {\mathscr P_r} g^e
+ {\rm{e}}^{ - \Delta t/\tau }  {\mathscr N_1}g^{0},
\label{eq:iss}
\end{eqnarray}
for arbitrary time in the interval $0 \le t \le \Delta t$ and $\tau$ is given by \cite{xu2001}
\begin{eqnarray}
\tau  =\epsilon _1 \Delta t
+ \epsilon_2 \frac{{\left|p^{\rm L}-p^{\rm R} \right|}}{p^{\rm L}+p^{\rm R} }\Delta t ,
\label{eq:tau}
\end{eqnarray}
where $\epsilon _1$ and $\epsilon _2$ are two constant parameters.
${\mathscr P_r} g^e$ and ${\mathscr N_1}g^{0}$ are weighted by a factor related to $p^{\rm L}$ and $p^{\rm R}$.
In the smooth regions where $\tau$ is small, the modified evolution model Eq.(\ref{eq:iss}) recurs to ${ f}(x_{i+\frac{1}{2}},t)={\mathscr P_r} g^e$.
While near the discontinuity, it also approaches to ${ f}(x_{i+\frac{1}{2}},t)= {\mathscr N_1}g^{0}$ which builds up the second-order KFVS flux \cite{xu1998}, the typical Euler solver with strong robustness \cite{estivalezes1996}.
Consequently, the modified evolution model still preserves the high-order accuracy in the smooth regions and the strong robustness near the discontinuity.
In fact, a similar weighting has already been implemented in a simplified version of low-order GKS evolution model \cite{xu1998,xu1996}.

After taking the moments of the evolution model Eq.(\ref{eq:iss}) on ${u}{\bfP}$, the flux is evaluated as
\begin{eqnarray}
{\bfF}_{i+\frac{1}{2}}({t})=
\left( 1- {\rm{e}}^{ - \Delta t/\tau } \right)
      {\mathscr P_r}  {\bfF}^e
+   {\rm{e}}^{ - \Delta t/\tau }  {\bfF}^k ,
\label{eq:ft0}
\end{eqnarray}
where
\begin{eqnarray}
\label{eq:ft0fefk}
&&{\bfF}^e = \int { g^{e} {u}{\bfP}{\rmd \Xi}},\nonumber\\
&&{\bfF}^k = \int_{u \ge 0} {\left( g^{\rm{L}}  - ut\frac{\partial g^{\rm{L}}  }{{\partial x}}\right) {u} {\bfP}{\rmd \Xi}}  +
\int_{u>0} {\left( g^{\rm{R}}  - ut\frac{\partial g^{\rm{R}}  }{{\partial x}}\right) {u} {\bfP}{\rmd \Xi}} .
\end{eqnarray}

\begin{remark}
To solve the compressible Euler equations, the evolution model Eq.(\ref{eq:iss}) and flux evaluation Eq.(\ref{eq:ft0}) build up a more practical HGKS than the version based on Eq.(\ref{eq:is0}).
If one has to solve the compressible Euler equations based on HGKS with arbitrary high-order accuracy, Eq.(\ref{eq:iss}) and Eq.(\ref{eq:ft0}) are recommended.
To provide a fair comparison to solve the Euler equations, the scheme version based on Eq.(\ref{eq:iss}) and Eq.(\ref{eq:ft0}) is utilized as a reference in the following numerical tests, which is called the original HGKS briefly.
Since the higher than first-order derivatives of $g^{\rm{L}}$ and $g^{\rm{R}}$ are eliminated in this version, its computation cost is less than the complete version of HGKS for the NS equations.
\end{remark}

\section{\label{sec:ehgks} Efficient HGKS with arbitrary high-order accuracy: Euler equations}

Although HGKS based on Eq.(\ref{eq:iss}) and Eq.(\ref{eq:ft0}) gets rid of all the high-order space and time derivatives of $g^{\rm{L}}$ and $g^{\rm{R}}$, those of $g^e$ still exist and are required to be solved following the whole space-time translation sequence.
In this section, a much more efficient flux evaluation based on the same evolution model Eq.(\ref{eq:iss}) is introduced.
Since computing the high-order terms ${\mathscr P_r}  {\bfF}^e$ in Eq.(\ref{eq:ft0}) occupies most of the computation cost for $r > 2$, our focus for a more efficient HGKS flux evaluation is on ${\mathscr P_r}  {\bfF}^e$.

\subsection{\label{sec:ehgks-lw} Efficient HGKS flux evaluation: Lax-Wendroff procedure}

In the complete version of HGKS \cite{xu2001,lqb2010,luo2013,liu2014}, it is relatively more convenient to solve the derivatives of particle distribution functions firstly and take the moments of them afterwards for flux evaluation.
However, based on the simplified evolution model Eq.~(\ref{eq:iss}), it is able to bypass the computations of the derivatives of $g^{e}$ in ${\mathscr P_r}  {\bfF}^e$ evaluations.
In fact, after taking the moments in $\bfF^e$, $\bfF^e$ has the same form of flux as in the Euler equations
\begin{eqnarray}
{\bfF}^e={\bfF}_{Eu}(\bfW ^e)=
\left( \begin{gathered}
  \rho^e U^e  \\
  \rho^e U^{e2}  + p^e  \\
  \rho^e E^eU^e + p^eU^e  \\
\end{gathered}  \right) ,
\label{eq:fluxeuler}
\end{eqnarray}
and consequently,
\begin{eqnarray}
{\mathscr P_r} {\bfF}^e={\mathscr P_r} {\bfF}_{Eu}(\bfW ^e).
\label{eq:fluxeuler}
\end{eqnarray}
Meanwhile, taking the moments of the compatibility conditions in Eq.(\ref{eq:hgxgt})  for $g=g^e$ yields
\begin{eqnarray}
\frac{\partial ^{m+q}}{\partial x^m t^q}
 \left({\frac{{\partial {\bfW}^{e}}}{{\partial t}}}
 + {\frac{{\partial  {\bfF}_{Eu}(\bfW ^e)}}{{\partial x}}} \right)   = {\bm 0}.
\label{eq:wxwt}
\end{eqnarray}
Since ${\bfF}_{Eu}(\bfW ^e)$ can be determined completely by $\bfW ^e$, $\frac{\partial^{m+q+1} }{ \partial x^m t^{q+1}}{\bfW}^{e}$ can be
translated from $ \frac{\partial^{m+q+1} }{ \partial x^{m+1} t^{q}} {\bfW}^{e}$.
Following the space-time translation sequence, the time related derivatives of $\bfW ^e$ can be obtained from the corresponding space derivatives
of $\bfW ^e$ step by step.
No derivatives of $g^e$ are included in the space-time translation and in the final flux evaluation of ${\mathscr P_r} {\bfF}_{Eu}(\bfW ^e)$.
It is consequently more convenient to compute ${\mathscr P_r} {\bfF}_{Eu}(\bfW ^e)$ directly from the space-time translation sequence based on Eq.(\ref{eq:wxwt}).
The computation cost to solve the high-order derivatives of $g^e$ which occupies mostly in the original HGKS flux evaluation Eq.(\ref{eq:ft0}) for $r >2$
can be saved contemporarily.

In fact, the space-time translation based on the macroscopic variables is what the traditional Lax-Wendroff procedure always means to do \cite{titarev2005}.
The Lax-Wendroff procedure has been widely used in ADER \cite{toro2001,titarev2005}.
The flexible applications of the Lax-Wendroff procedure in ADER \cite{harten1987,toro2005} are also available here.
Compared with the flux-expansion version, the state-expansion is computationally cheaper \cite{toro2001,toro2005}.
The introduction the Lax-Wendroff procedure into the HGKS flux evaluation is given as follows.

Define the time integrated ${\mathscr P_r} {\bfF}_{Eu}(\bfW ^e)$ by
\begin{eqnarray}
{{\mathbb F} ^e}_{i+\frac{1}{2}}({\Delta t})=\int_0^{\Delta t} {\mathscr P_r}  {\bfF}_{Eu}(\bfW ^e){\rmd t}.
\label{eq:timefluxe}
\end{eqnarray}
The Gaussian rule is adopted in the integration \cite{toro2005}:
\begin{eqnarray}
\label{eq:gaussw1}
{\mathbb F}^e_{i+\frac{1}{2}}({\Delta t}) =
\sum\limits_{\alpha  = 0}^{K  } {{\bf{F}}_{Eu}^{}
\left( {\mathcal Q}^e \left( {\kappa _\alpha  \Delta t} \right) \right)
\omega _\alpha  },\\
{\mathcal Q}^e \left( {t} \right) =
  \sum\limits_{l = 0}^r {{\frac{t^l}{l!}}{\frac{\partial ^l{\bf Q}^e}{{\partial t^l}}}   } ,\nonumber
\end{eqnarray}
where ${\bf Q}=(\rho,U,p)$, $\kappa _\alpha$ is the scaled node, $\omega _\alpha$ is the weight,
$K+1 $ is the number of the Gauss nodes.
The order of accuracy of the Gauss rule satisfies $2K \ge r$.
The space-time translation Eq.(\ref{eq:wxwt}) is replaced equivalently by \cite{harten1987}
\begin{eqnarray}
\frac{\partial ^{m+q}}{\partial x^m  t^q}
 \left({\frac{\partial {\bf Q}^{e}}{\partial t}}
 + {\bf A}^e \cdot {\frac{\partial {\bf Q}^{e}}{\partial x}}
  \right)= {\bm 0},\quad
{\bf A}=
\left[
{\begin{array}{*{20}c}
   U & \rho   & {0}  \\
   {0} & U  & {1/\rho }  \\
   {0} & {\gamma p}  & U  \\
\end{array}}
\right],
\label{eq:qxqta1}
\end{eqnarray}
for a simple ${\bf A}$ and thus better efficiency.

As a summary, the procedure for the time integrated flux evaluation ${{\mathbb F} }_{i+\frac{1}{2}}({\Delta t})=\int_0^{\Delta t} {\bfF}_{i+\frac{1}{2}}(t){\rmd t}$ is plotted as follows.
\begin{eqnarray}
\begin{array}{*{20}c}
   {}  \\
   \begin{gathered}
   \hfill \\
  \boxed{{\mathbf{W}}^{\text{L}} }\;\;\; \to \;\;\; \hfill \\
\end{gathered}   \\
   {}  \\
   \begin{gathered}
   \hfill \\
   \hfill \\
   \hfill \\
   \hfill \\
   \hfill \\
   \hfill \\
   \hfill \\
  \boxed{{\mathbf{W}}^{\text{R}} }\;\;\; \to \;\;\; \hfill \\
   \hfill \\
   \hfill \\
\end{gathered}   \\
 \end{array} \begin{array}{*{20}c}
   {}  \\
   \begin{gathered}
   \hfill \\
  g^{\text{L}} \;\;\;\;\;\; \to \;\;\;\; \hfill \\
\end{gathered}   \\
   {}  \\
   \begin{gathered}
   \hfill \\
   \downarrow  \hfill \\
   \hfill \\
  {\mathbf{W}}^e  \to g^e  \to  \hfill \\
   \hfill \\
   \uparrow  \hfill \\
   \hfill \\
  g^{\text{R}} \;\;\;\;\;\; \to \;\;\;\; \hfill \\
   \hfill \\
   \hfill \\
\end{gathered}   \\
 \end{array} \left. {\begin{array}{*{20}c}
   {\left\{ {\boxed{\frac{{\partial {\mathbf{W}}^{\text{L}} }}
{{\partial x}}} \to \frac{{\partial g^{\text{L}} }}
{{\partial x}} \to \frac{{\partial g^{\text{L}} }}
{{\partial t}}\;\;\;\;\;\;\;\;\;\;\;\;\;\;\;\;\;\;\;\;} \right.}  \\
   {\left\{ \begin{gathered}
  \boxed{\frac{{\partial {\mathbf{W}}^e }}
{{\partial x}}} \to \frac{{\partial {\mathbf{Q}}^e }}
{{\partial t}} \hfill \\
  \boxed{\frac{{\partial ^2 {\mathbf{W}}^e }}
{{\partial x^2 }}} \to \frac{{\partial ^2 {\mathbf{Q}}^e }}
{{\partial x\partial t}} \to \frac{{\partial ^2 {\mathbf{Q}}^e }}
{{\partial t^2 }} \hfill \\
  ... \hfill \\
  \boxed{\frac{{\partial ^r {\mathbf{W}}^e }}
{{\partial x^r }}} \to \frac{{\partial ^r {\mathbf{Q}}^e }}
{{\partial x^{r - 1} \partial t}} \to ... \to \frac{{\partial ^r {\mathbf{Q}}^e }}
{{\partial t^r }} \hfill \\
\end{gathered}  \right.}  \\
   {\left\{ {\boxed{\frac{{\partial {\mathbf{W}}^{\text{R}} }}
{{\partial x}}} \to \frac{{\partial g^{\text{R}} }}
{{\partial x}} \to \frac{{\partial g^{\text{R}} }}
{{\partial t}}\;\;\;\;\;\;\;\;\;\;\;\;\;\;\;\;\;\;\;} \right.}  \\
 \end{array} } \right\} \to {{\mathbb F} }_{i+\frac{1}{2}}({\Delta t})\nonumber
\label{eq:qxqta2}
\end{eqnarray}

\subsection{\label{sec:muld} Multi-dimensional extensions}

Both 1D evolution model and flux evaluation introduced in the former sections can be extended straightforward to the multi-dimensional cases.
The existing simplifications and modifications are also applicable for the multi-dimensional cases \cite{zhou2017}.

In the two-dimensional (2D) case, further considering the Taylor expansion in the tangential direction, the evolution model with $(r+1)$th-order accuracy is expanded and modified as
\begin{eqnarray}
f({\bf{x}},t,{\bf u},{\bm \xi}) =
\left(1- {\rm{e}}^{ -  \Delta t/\tau }\right) {\mathscr P_r} g^e
+ {\rm{e}}^{ - \Delta t/\tau }  {\mathscr N_1}g^{0},
\label{eq:taylord2}
\end{eqnarray}
where
\begin{eqnarray}
{\mathscr P_r}
    =\sum\limits_{l = 0}^r {\frac{1}{{l!}}
                        \left( {t\frac{\partial }{{\partial t}} + {\bf{x}} \cdot \nabla } \right)^l },
{\mathscr N_r}=
    \sum\limits_{l = 0}^r {\frac{1}{{l!}}
                        \left[ {\left( {{\bf{x}} - {\bf{u}}t} \right) \cdot \nabla } \right]^l } ,\nonumber
\label{eq:taylor20}
\end{eqnarray}
and ${\bf{x}}=(x,y)$, ${\bf{u}}=(u,v)$ and $\nabla=(\frac{\partial}{\partial x},\frac{\partial}{\partial y})$.

Under the finite volume framework, the discretized cell is indexed as $(i,j)$.
Assuming the space domain is discretized uniformly,
the time integrated flux along the cell interface $(i+\frac{1}{2},j)$ is evaluated as
\begin{eqnarray}
{\mathbb F }_{i+\frac{1}{2},j}({\Delta t})&=&\int_{-\Delta y/2}^{\Delta y/2} \int_0^{\Delta t}
\int {f({ \bf x},t,{\bf u},{\bm \xi}) { u} {\bfP}{\rmd \Xi}} {\rmd y}{\rmd t}\\
&=&\left(1- {\rm{e}}^{ -  \Delta t/\tau }\right)
   {\mathbb F }^e_{i+\frac{1}{2},j}({\Delta t}) +
   {\rm{e}}^{ - \Delta t/\tau }
   {\mathbb F }^k_{i+\frac{1}{2},j}({\Delta t}),\nonumber
\label{eq:flux2d}
\end{eqnarray}
where the high-order terms exist in
\begin{eqnarray}
{\mathbb F }^e_{i+\frac{1}{2},j}({\Delta t})
=\int_{-\Delta y/2}^{\Delta y/2} \int_0^{\Delta t} {\mathscr P_r}  {\bfF}_{Eu}(\bfW ^e){\rmd y}{\rmd t},
\label{eq:fluxe2d}
\end{eqnarray}
and ${\mathbb F }^k_{i+\frac{1}{2},j}({\Delta t})$ is the second-order KFVS flux.

The high-order terms ${\mathbb F }^e_{i+\frac{1}{2},j}({\Delta t})$ are approximated by the Gaussian rule
\begin{eqnarray}
\label{eq:gaussw2}
{\mathbb F }^e_{i+\frac{1}{2},j}({\Delta t})=
\sum\limits_{\alpha  = 0}^{K  }
\sum\limits_{\beta  = 0}^{K  }
{{\bf{F}}_{Eu}^{}
\left( {\mathcal Q}^e \left( {\kappa _\alpha  \Delta y},
                             {\kappa _\beta   \Delta t} \right) \right)
\omega _\alpha \omega _\beta  },\\
{\mathcal Q}^e \left( {y,t} \right)
    =\sum\limits_{l = 0}^r {\frac{1}{{l!}}
                        \left( {t\frac{\partial }{{\partial t}}}
                              +{y\frac{\partial }{{\partial y}}}  \right)^l }{\bf Q}^e.\nonumber
\end{eqnarray}
The time related derivatives of ${\bf Q}^e$ are obtained based on \cite{harten1987}
\begin{eqnarray}
\frac{\partial ^{m+n+q}}{\partial x^m y^n t^q}
 \left({\frac{\partial {\bf Q}^{e}}{\partial t}}
 + {\bf A}^e \cdot {\frac{\partial {\bf Q}^{e}}{\partial x}}
 + {\bf B}^e \cdot {\frac{\partial {\bf Q}^{e}}{\partial y}} \right)= {\bm 0},
\label{eq:qxqt2}
\end{eqnarray}
where
\begin{eqnarray}
{\bf A}=
\left[
{\begin{array}{*{20}c}
   U & \rho  & {0} & {0}  \\
   {0} & U & {0} & {1/\rho }  \\
   {0} & {0} & U & {0}  \\
   {0} & {\gamma p} & {0} & U  \\
\end{array}}
\right],
{\bf B}=
\left[
{\begin{array}{*{20}c}
   V & {0} & \rho  & {0}  \\
   {0} & V & {0} & {0}  \\
   {0} & {0} & V & {1/\rho }  \\
   {0} & {0} & {\gamma p} & V  \\
\end{array}}
\right].\nonumber
\label{eq:qxqta2}
\end{eqnarray}

In the 2D case, the reconstruction is implemented direction by direction \cite{eno1996}.
The linear reconstruction is utilized in the tangential direction on $\bfW^e$.
It is worthy of mention that ${\mathcal Q}^e \left( {y,t} \right)$ has already taken the tangential derivatives into consideration.
The reconstruction is implemented only once at the central point of the cell interface.
No more reconstructions are required at each of the gauss points in the tangential direction.
This implement initiates from the multi-dimensional effect of the HGKS evolution model, which has been widely used under many other frameworks \cite{zc2018}.

It's easy to further extend the current scheme to the three-dimensional case by the similar way.
According to the traditional three-dimensional Lax-Wendroff procedure, the derivatives in the $z$ direction should also be included in the space-time translation \cite{titarev2005}.

\subsection{\label{sec:rem} Efficient HGKS with arbitrary high-order accuracy: Euler equations}

As a summary, the time integrated flux along the cell interface is evaluated as
\begin{eqnarray}
{\mathbb F}_{i+\frac{1}{2}}({\Delta t})=
\left( 1- {\rm{e}}^{ - \Delta t/\tau } \right)
        {\mathbb F}^e_{i+\frac{1}{2}}({\Delta t})
+   {\rm{e}}^{ - \Delta t/\tau } {\mathbb F}_{i+\frac{1}{2}}^k({\Delta t}) ,
\label{eq:ffdt}
\end{eqnarray}
where ${\mathbb F}^e_{i+\frac{1}{2}}({\Delta t})$ is given by Eqs.~(\ref{eq:gaussw1}) and (\ref{eq:gaussw2}).
${\mathbb F}^k_{i+\frac{1}{2}}({\Delta t})$ is the time integrated KFVS flux Eq.~(\ref{eq:ft0fefk}) along the cell interface.
The time related derivatives of ${\bf Q}^e$ are obtained from the space derivatives according to the Lax-Wendroff procedure Eqs.(\ref{eq:qxqta1})
and (\ref{eq:qxqt2}). The scheme based on Eq.(\ref{eq:ffdt}) has the same evolution model as the original HGKS Eq.(\ref{eq:ft0}).
But the computation cost is much less with the same-order magnitude as the traditional Lax-Wendroff procedure.
We call the currently constructed scheme in this paper EHGKS.

The high efficiency of EHGKS initiates from two aspects.
The first and basic is the simplification on the numerical dissipation part.
It brings in the feasibility of the second aspect that the remaining high-order derivatives of particle distribution function and their moments in flux evaluation are replaced by the more efficient Lax-Wendroff procedure.
Taking the advantage of both HGKS and Lax-Wendroff procedure, EHGKS preserves not only the high-order accuracy and strong robustness, but also the high efficiency.

In practical simulations, $\tau$ is given as
\begin{eqnarray}
\tau  =\epsilon _1 \Delta t
+ \epsilon_2 \frac{{p_{max}-p_{min} }}{p_{max}+p_{min} }\Delta t ,
\label{eq:tau}
\end{eqnarray}
where $p_{max}$ and $p_{min}$ are the maximum and minimum pressure respectively among $p^{\rm L}$, $p^{\rm R}$ and those in the nearest-layer stencils in the tangential direction of the 2D case.
Instead of only introducing $p^{\rm L}$ and $p^{\rm R}$ \cite{xu2001,lqb2010,liu2014,cyb2016}, those in the tangential direction are included with consideration of the multi-directional effect. We shall take
$\epsilon_1=0.02,\epsilon_2=2$ in the following numerical tests.

\section{\label{sec:nta}Numerical test}

In order to validate the accuracy, efficiency and robustness of the newly developed EHGKS in this paper, numerical test results for the Euler equations of compressible flows are presented in this section.
The direct comparisons of the $r$-th-order EHGKS (EHGKS-$r$), the original HGKS (HGKS-$r$) based on Eq.(\ref{eq:ft0}) and the third-order TVD Runge-Kutta-WENO-GKS scheme (RK3-WENO$r$-GKS) are also performed in the 1D cases.
In all comparisons, the HGKS models (only the first-order GKS is required in RK3-WENO$r$-GKS), the WENO reconstruction techniques and the CFL number are the same.

The time step $\Delta t$ is determined by the CFL condition
\begin{eqnarray}
\Delta t=
CFL\times \min \left\{\frac{\Delta x}{\left| \overline U \right| + \overline c_s },
                      \frac{\Delta y}{\left| \overline V \right| + \overline c_s } \right\},
\label{eq:dt}
\end{eqnarray}
where $c_s$ is the speed of sound.
The CFL condition number $CFL=0.95$ in the 1D cases and $CFL=0.45$ in the 2D cases if without specification.

The validation of EHGKS is based on a series of 1D and 2D benchmark test cases, namely:
\begin{enumerate}[(1)]
  \item
  1D linear advection of the density perturbation \cite{pan2016}{--–} the Euler equations are reduced to the linear advection equation problem with smooth solution. This case is to assess the accuracy and efficiency of the numerical schemes.
  \item
  1D Shu-Osher shock acoustic wave interaction \cite{titarev2002,cyb2016}{--–} the solution includes the small-scale smooth features while the low-order schemes often introduce over diffusion \cite{dumbser2014}.
  \item
  1D Woodward-Colella blast wave \cite{shu1998}{--–} the solution contains complex interactions of the shock waves and contact discontinuities. We use this test to demonstrate the robustness of EHGKS.
  \item
  2D linear advection of the density perturbation \cite{zhong2013}{--–} the 2D Euler equations are reduced to the linear advection equation problem with smooth solution. This case is to assess the accuracy and efficiency of EHGKS in the 2D case.
  \item
  2D isotropic vortex propagation \cite{ljq2016}{--–}the exact solution of 2D Euler equations is smooth to assess the high-order accuracy of the numerical schemes.
  \item
  2D Riemann problems \cite{pan2016,cyb2016}{--–}we use two of these problems to demonstrate the ability of the high-order schemes to solve 2D Riemann problems genuinely.
  \item
  Double Mach reflection problem \cite{cyb2016,lqb2010}{--–}we use this classical test problem to investigate the ability of the high-order schemes to capture the details of complex flows produced by the interaction of shock waves.
\end{enumerate}

\subsection{\label{sec:sna}1D linear advection of the density perturbation}

Here we assess the accuracy and the efficiency of EHGKS when the solution is linear and smooth in the 1D case.
And the comparisons of EHGKS-$r$, the original HGKS-$r$ and RK3-WENO$r$-GKS schemes are presented.

The initial condition is given by
\begin{eqnarray}
 \left( \rho,U,p \right) = \left( 1 + 0.2\sin (x),1,1 \right).\nonumber
\label{eq:ac10}
\end{eqnarray}
Under the periodic boundary condition, the analytic solution is
\begin{eqnarray}
 \left( \rho,U,p \right) = \left( 1 + 0.2\sin (x - t),1,1 \right).\nonumber
\label{eq:ac1}
\end{eqnarray}
The computational domain is $[ {0,2\pi } ]$ divided by $N$ uniform cells.
The output time is 400 periods.
$CFL = 0.95$ is adopted.

The results of the errors and the accuracy order computed by EHGKS-$3$, EHGKS-$5$ and EHGKS-$7$ are shown in Table~\ref{tab:ac1}.
All the EHGKS-$r$ schemes can achieve the designed order of accuracy in both space and time.

\begin{table}[htb!]
\centering
\begin{tabular}{c|cc|cc|cc}
\hline
\multicolumn{7}{c}{EHGKS-3}  \\
\hline
$N$  &$L_1$ error &Order &$L_2$ error &Order &$L_{\infty}$ error   & Order  \\
\hline
    20    & 1.20E-01 &      & 1.33E-01 &      & 1.86E-01 &      \\
    40    & 3.90E-02 & 1.62  & 4.33E-02 & 1.62  & 6.17E-02 & 1.59 \\
    80    & 5.74E-03 & 2.76  & 6.38E-03 & 2.76  & 9.45E-03 & 2.71 \\
    160   & 7.35E-04 & 2.97  & 8.15E-04 & 2.97  & 1.23E-03 & 2.95 \\
\hline
\multicolumn{7}{c}{EHGKS-5}  \\
\hline
$N$  &$L_1$ error &Order &$L_2$ error &Order &$L_{\infty}$ error   & Order  \\
\hline
    20    & 7.67E-03 &   & 8.51E-03 &   & 1.21E-02 &  \\
    40    & 2.50E-04 & 4.94  & 2.77E-04 & 4.94  & 4.13E-04 & 4.88 \\
    80    & 7.84E-06 & 4.99  & 8.69E-06 & 5.00  & 1.30E-05 & 4.99 \\
    160   & 2.45E-07 & 5.00  & 2.72E-07 & 5.00  & 4.07E-07 & 5.00 \\
\hline
\multicolumn{7}{c}{EHGKS-7}  \\
\hline
$N$  &$L_1$ error &Order &$L_2$ error &Order &$L_{\infty}$ error   & Order  \\
\hline
    20    & 1.72E-04 &   & 1.91E-04 &   & 2.80E-04 &  \\
    40    & 1.37E-06 & 6.98  & 1.52E-06 & 6.98  & 2.26E-06 & 6.96 \\
    80    & 1.07E-08 & 6.99  & 1.19E-08 & 6.99  & 1.78E-08 & 6.99 \\
    160   & 9.71E-11 & 6.79  & 1.08E-10 & 6.78  & 1.57E-10 & 6.82 \\
\hline
\end{tabular}
\caption{\label{tab:ac1} Accuracy test for EHGKS in the 1D advection of the density perturbation.}
\end{table}

\begin{figure*}[htb!]
\centering
\begin{minipage}{0.49\linewidth}
  \centerline{\includegraphics[width=\linewidth]{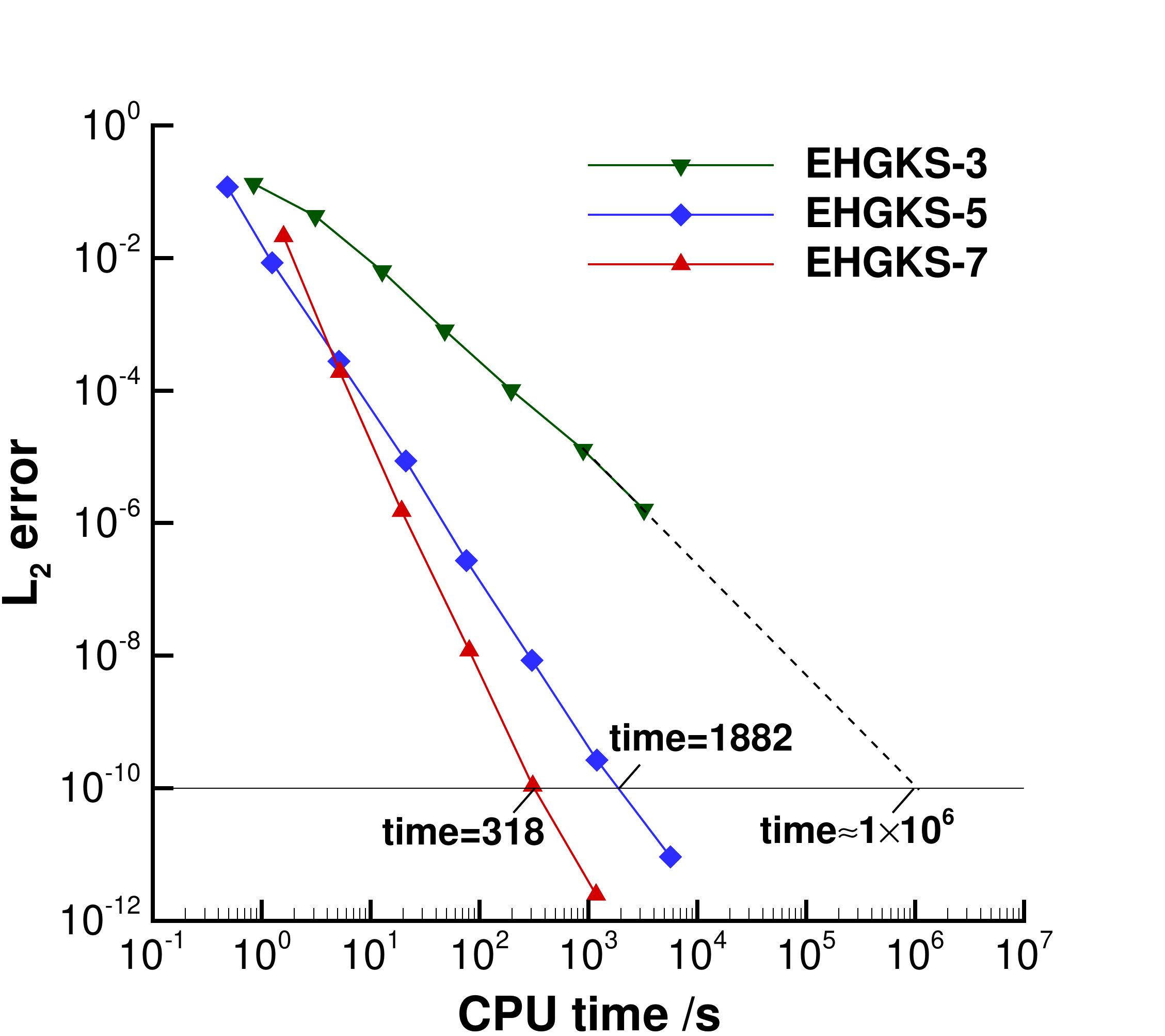}}
\end{minipage}
\caption{\label{fig:cputime} The CPU time vs $L_2$ error for EHGKS. The test case is the 1D advection of the density perturbation.}
\end{figure*}

We further investigate the efficiency of EHGKS-$r$.
The comparison of the CPU time averaged in one single time step under $80$ uniform grids are presented in Table~\ref{tab:actime}.
The errors and the corresponding CPU times of EHGKS-$r$ are shown in Fig.~\ref{fig:cputime}.
The results show that the higher-order schemes use more CPU time than the lower-order schemes in a single time step.
However, given the computational errors which are small enough, the higher-order schemes use less CPU time than the lower-order schemes.

\begin{table}[htb!]
\centering
\begin{tabular}{c|c}
\hline
                &CPU time per step   \\
\hline
    EHGKS-3     & 1.64E-04 \\
    EHGKS-5     & 2.70E-04 \\
    EHGKS-7     & 1.03E-03 \\
\hline
\end{tabular}
\caption{\label{tab:actime} CPU time(seconds) per time step for EHGKS in the 1D advection of the density perturbation.}
\end{table}

\begin{figure*}[htb!]
\centering
\begin{minipage}{0.49\linewidth}
  \centerline{\includegraphics[width=\linewidth]{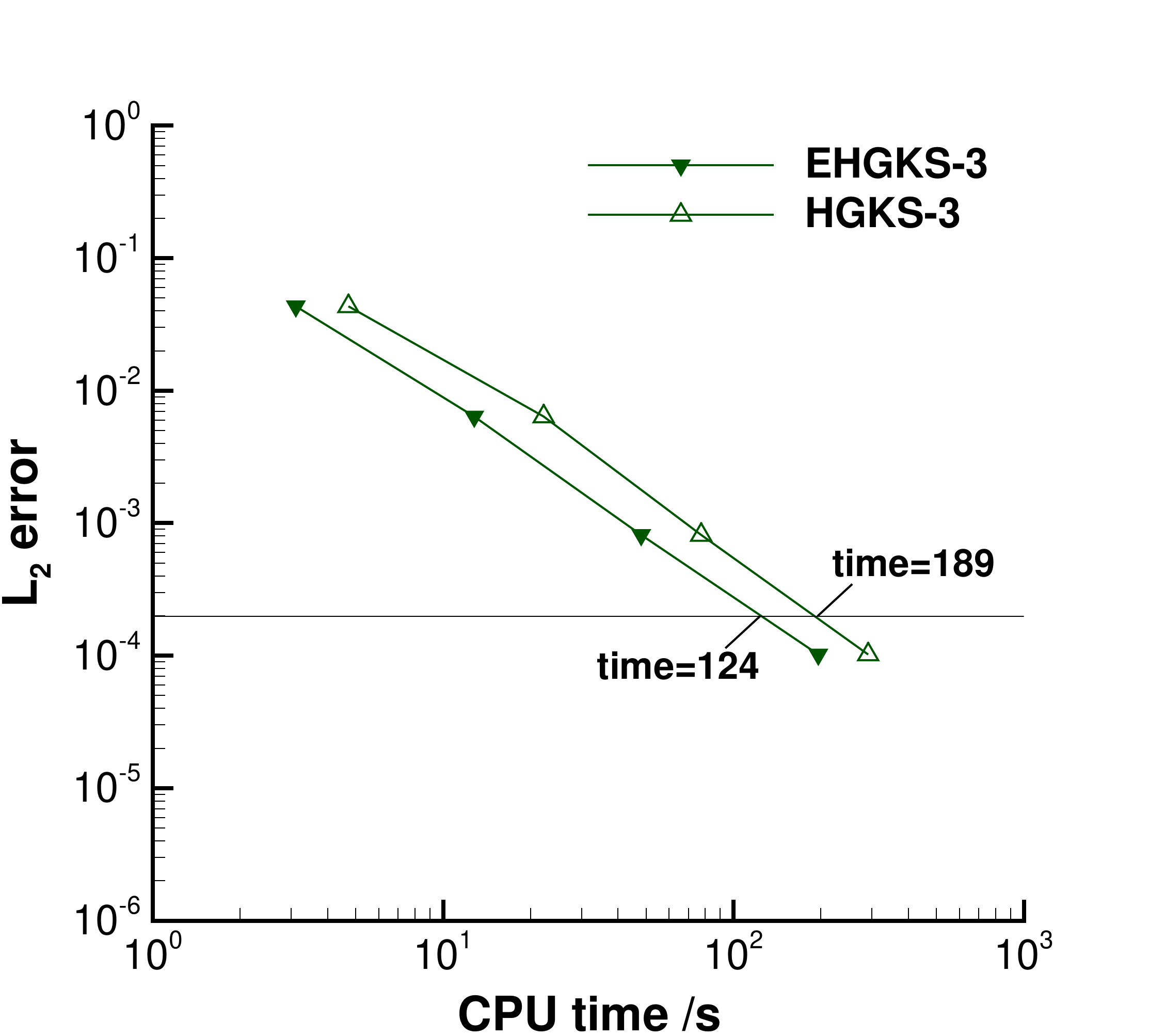}}
\end{minipage}
\begin{minipage}{0.49\linewidth}
  \centerline{\includegraphics[width=\linewidth]{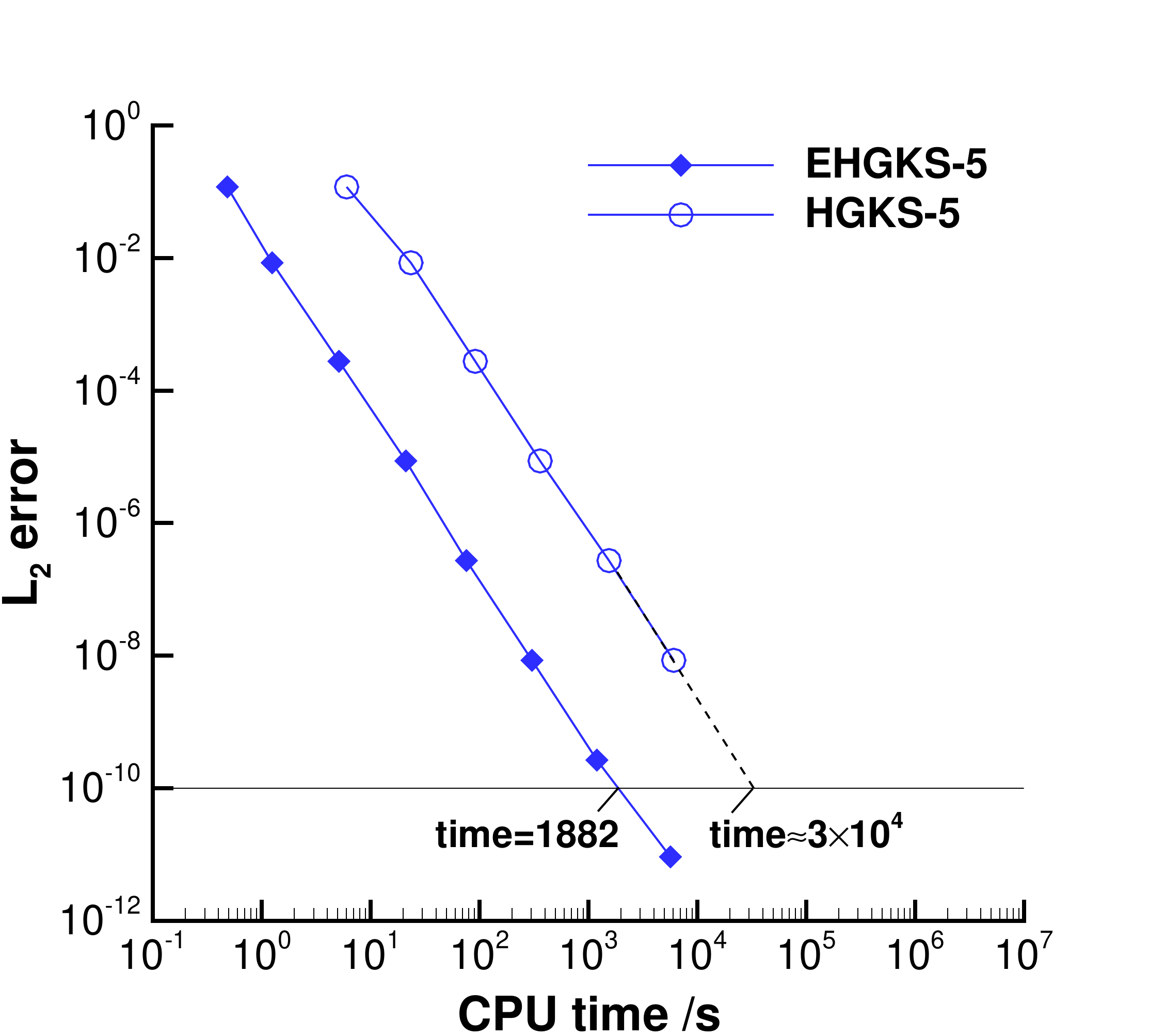}}
\end{minipage}
\caption{\label{fig:actime-hgks} The CPU time vs $L_2$ error for HGKS. The test case is the 1D advection of the density perturbation.}
\end{figure*}

Under the same computational conditions, the results of the errors and the accuracy order of HGKS-$r$ are shown in Table~\ref{tab:ac1-hgks}.
We only display the results of HGKS-$3$ and HGKS-$5$ since it is really complex to implement HGKS-$7$.
It is clear that HGKS-$r$ can achieve the designed order of accuracy in both space and time.
The CPU times averaged in one single time step under $80$ uniform grids are presented in Table~\ref{tab:actime-hgks}.
Comparisons on the errors and the corresponding CPU times between EHGKS-$r$ and HGKS-$r$ are shown in Fig.~\ref{fig:actime-hgks}.
The results demonstrate that EHGKS-$r$ is more efficient than the original HGKS-$r$.

\begin{table}[htb!]
\centering
\begin{tabular}{c|cc|cc|cc}
\hline
\multicolumn{7}{c}{HGKS-3}  \\
\hline
$N$  &$L_1$ error &Order &$L_2$ error &Order &$L_{\infty}$ error   & Order  \\
\hline
    20    & 1.20E-01 &   & 1.33E-01 &   & 1.86E-01 &  \\
    40    & 3.90E-02 & 1.62  & 4.33E-02 & 1.62  & 6.17E-02 & 1.59 \\
    80    & 5.74E-03 & 2.76  & 6.38E-03 & 2.76  & 9.45E-03 & 2.71 \\
    160   & 7.35E-04 & 2.97  & 8.15E-04 & 2.97  & 1.23E-03 & 2.95 \\
\hline
\multicolumn{7}{c}{HGKS-5}  \\
\hline
$N$  &$L_1$ error &Order &$L_2$ error &Order &$L_{\infty}$ error   & Order  \\
\hline
    20    & 7.67E-03 &   & 8.51E-03 &   & 1.21E-02 &  \\
    40    & 2.50E-04 & 4.94  & 2.77E-04 & 4.94  & 4.13E-04 & 4.88 \\
    80    & 7.84E-06 & 4.99  & 8.69E-06 & 5.00  & 1.30E-05 & 4.99 \\
    160   & 2.45E-07 & 5.00  & 2.72E-07 & 5.00  & 4.07E-07 & 5.00 \\
\hline
\end{tabular}
\caption{\label{tab:ac1-hgks} Accuracy test for HGKS in the 1D advection of the density perturbation.}
\end{table}

\begin{table}[htb!]
\centering
\begin{tabular}{c|c|c}
\hline
         &CPU time per step &HGKS-$r$/EHGKS-$r$ \\
\hline
    HGKS-3     & 2.85E-04 & 1.74  \\
    HGKS-5     & 4.61E-03 & 17.1  \\
\hline
\end{tabular}
\caption{\label{tab:actime-hgks} CPU time(seconds) per time step for HGKS in the 1D advection of the density perturbation.}
\end{table}

\begin{figure*}[htb!]
\centering
\begin{minipage}{0.49\linewidth}
  \centerline{\includegraphics[width=\linewidth]{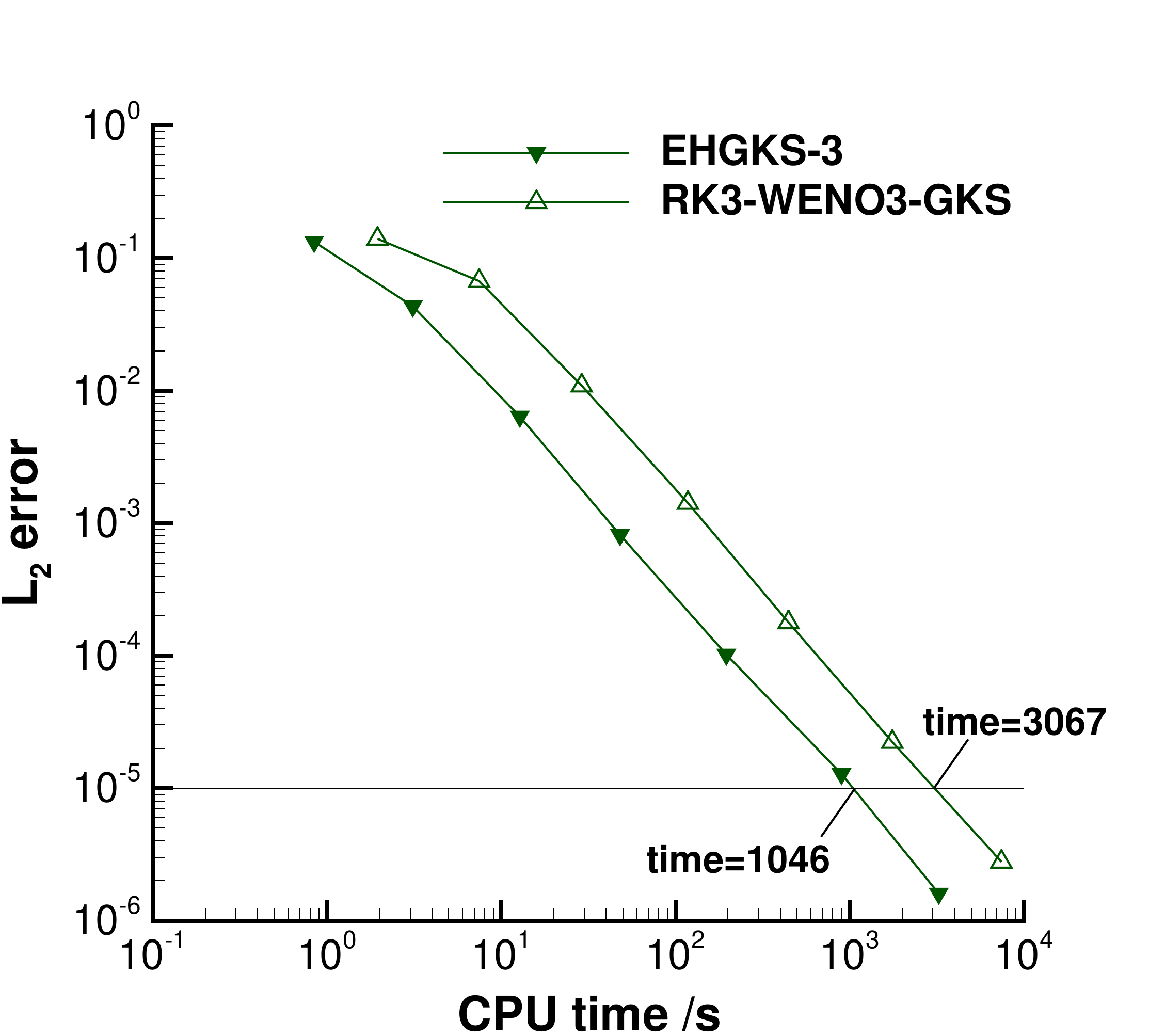}}
\end{minipage}
\begin{minipage}{0.49\linewidth}
  \centerline{\includegraphics[width=\linewidth]{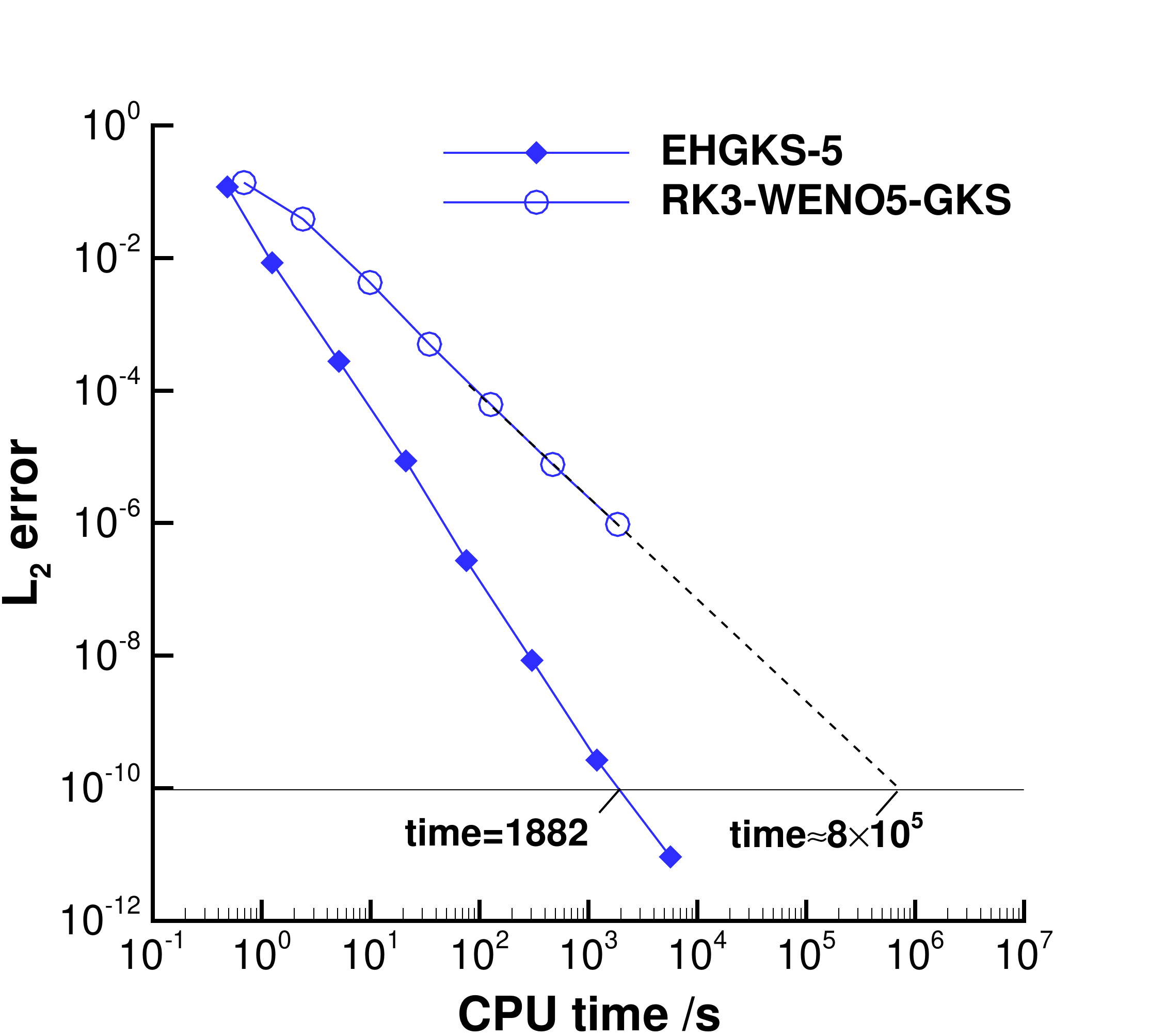}}
\end{minipage}
\begin{minipage}{0.49\linewidth}
  \centerline{\includegraphics[width=\linewidth]{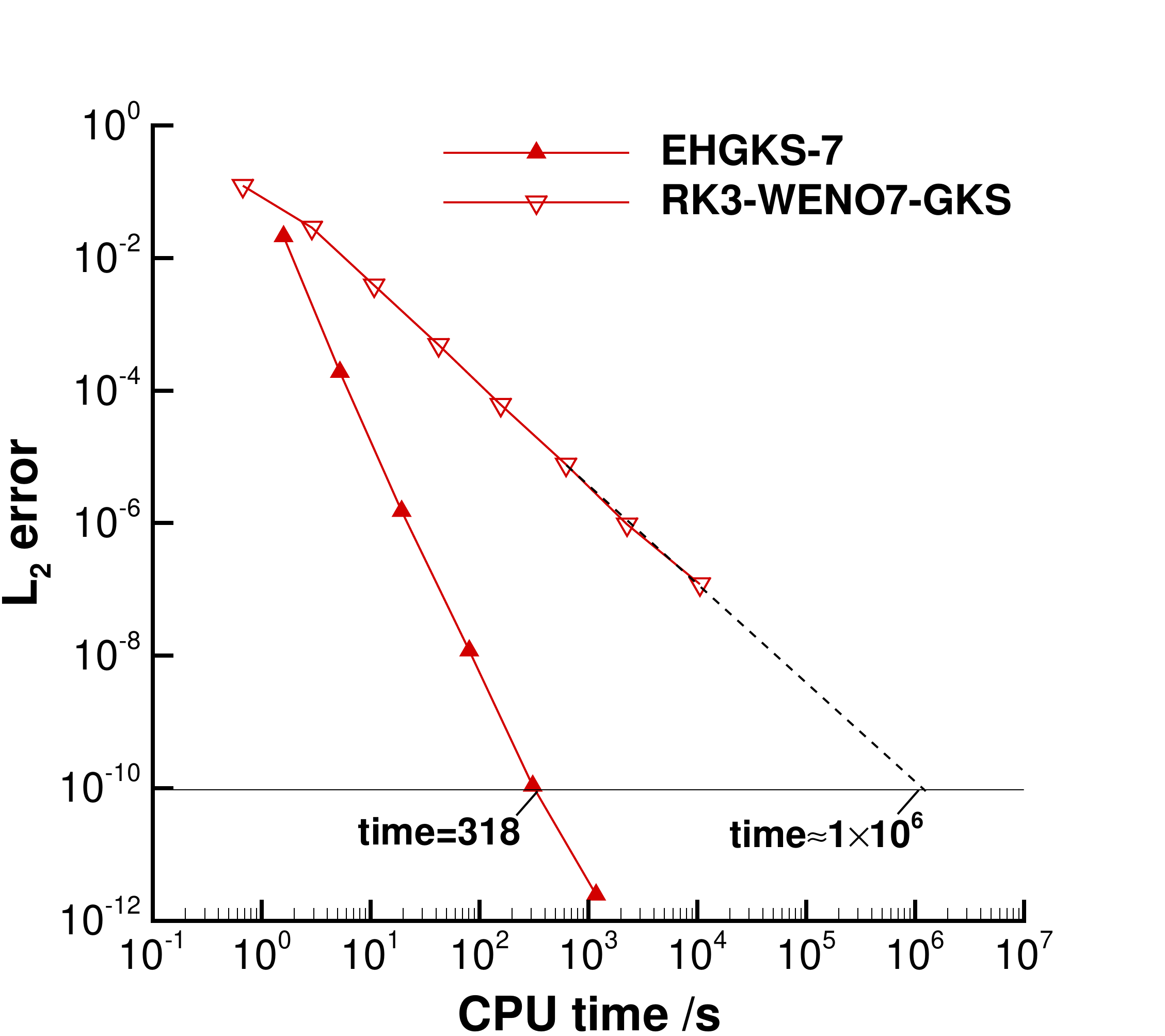}}
\end{minipage}
\caption{\label{fig:actime-rk} The CPU time vs $L_2$ error for RK3-WENO-GKS. The test case is the 1D advection of the density perturbation.}
\end{figure*}

The results of the errors and the accuracy order of RK3-WENO$r$-GKS are also shown in Table~\ref{tab:ac1-rk}.
Limited by the accuracy in time, all these schemes can only achieve third-order accuracy.
To improve the accuracy in time, more complex Runge-Kutta methods are required.
Obviously, it is unfair to compare the accuracy between EHGKS-$r$  and RK3-WENO$r$-GKS since the designed accuracy order in time is different.
To investigate the difference in efficiency between EHGKS-$r$ and RK3-WENO$r$-GKS, we give the errors and the corresponding CPU times in Fig.~\ref{fig:actime-rk}.
The numerical results show that EHGKS-$r$ can significantly reduce the computation cost of  RK3-WENO$r$-GKS.
Consequently, the necessity to preserve the high-order accuracy in both space and time is further confirmed.

\begin{table}[htb!]
\centering
\begin{tabular}{c|cc|cc|cc}
\hline
\multicolumn{7}{c}{RK3-WENO3-GKS}  \\
\hline
$N$  &$L_1$ error &Order &$L_2$ error &Order &$L_{\infty}$ error   & Order  \\
\hline
    20    & 1.27E-01 &   & 1.40E-01 &   & 1.96E-11 &  \\
    40    & 6.05E-02 & 1.07  & 6.72E-02 & 1.06  & 9.52E-02 & 1.04 \\
    80    & 9.85E-03 & 2.62  & 1.09E-02 & 2.62  & 1.59E-02 & 2.59 \\
    160   & 1.28E-03 & 2.95  & 1.42E-03 & 2.95  & 2.09E-03 & 2.93 \\
\hline
\multicolumn{7}{c}{RK3-WENO5-GKS}  \\
\hline
$N$  &$L_1$ error &Order &$L_2$ error &Order &$L_{\infty}$ error   & Order  \\
\hline
    20    & 3.51E-02 &   & 3.89E-02 &   & 5.45E-02 &  \\
    40    & 3.87E-03 & 3.18  & 4.29E-03 & 3.18  & 6.09E-03 & 3.16 \\
    80    & 4.53E-04 & 3.09  & 5.03E-04 & 3.09  & 7.13E-04 & 3.10 \\
    160   & 5.56E-05 & 3.03  & 6.18E-05 & 3.03  & 8.74E-05 & 3.03 \\
\hline
\multicolumn{7}{c}{RK3-WENO7-GKS}  \\
\hline
$N$  &$L_1$ error &Order &$L_2$ error &Order &$L_{\infty}$ error   & Order  \\
\hline
    20    & 2.60E-02 &   & 2.89E-02 &   & 4.07E-02 &  \\
    40    & 3.50E-03 & 2.89  & 3.88E-03 & 2.89  & 5.49E-03 & 2.89 \\
    80    & 4.41E-04 & 2.99  & 4.90E-04 & 2.99  & 6.93E-04 & 2.99 \\
    160   & 5.52E-05 & 3.00  & 6.13E-05 & 3.00  & 8.67E-05 & 3.00 \\
\hline
\end{tabular}
\caption{\label{tab:ac1-rk} Accuracy test for RK3-WENO-GKS in the 1D advection of the density perturbation.}
\end{table}

\begin{table}[htb!]
\centering
\begin{tabular}{c|c}
\hline
   &CPU time per step \\
\hline
    RK3-WENO3-GKS    & 3.71E-04 \\
    RK3-WENO5-GKS    & 4.43E-04 \\
    RK3-WENO7-GKS    & 5.39E-04  \\
\hline
\end{tabular}
\caption{\label{tab:actime-rk} CPU time(seconds) per time step for RK3-WENO-GKS in the 1D advection of the density perturbation.}
\end{table}

\subsection{ {Shu-Osher} shock acoustic wave interaction}

This is a well-known 1D shock tube problem which contains small-scale perturbations and shock waves\cite{dumbser2014}.
This case is to assess whether the high-order schemes can capture the small-scale information of the flows exactly.
The flow field is initialized as \cite{titarev2002,cyb2016}
\begin{eqnarray}
\left( \rho,U,p \right)=
\left\{ \begin{array}{l}
 \left( 3.857134,2.629369, 10.33333 \right),x \le -0.8, \\
 \left( 1 + 0.2\sin (5\pi x),0, 1 \right), x > -0.8 .
 \end{array} \right.\nonumber
\label{eq:so1}
\end{eqnarray}
The computational domain is $[-1, 1]$ divided by $200$ uniform cells.
The output time $t = 0.47$.
$CFL = 0.9$ is adopted in this case.
The exact solution is calculated from the refined grids.

\begin{figure*}[htb!]
\centering
\begin{minipage}{0.49\linewidth}
  \centerline{\includegraphics[width=\linewidth]{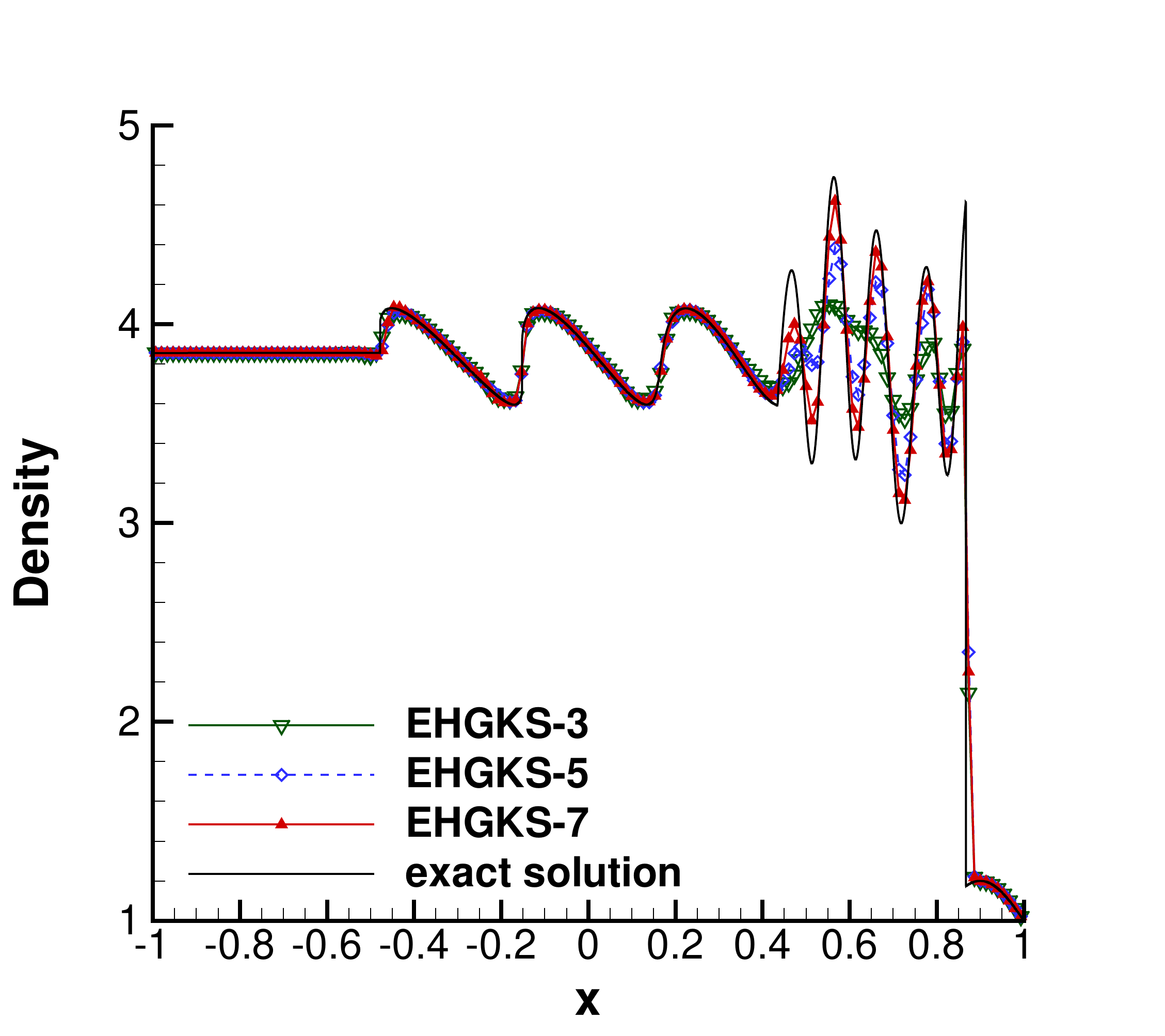}}
\end{minipage}
\begin{minipage}{0.49\linewidth}
  \centerline{\includegraphics[width=\linewidth]{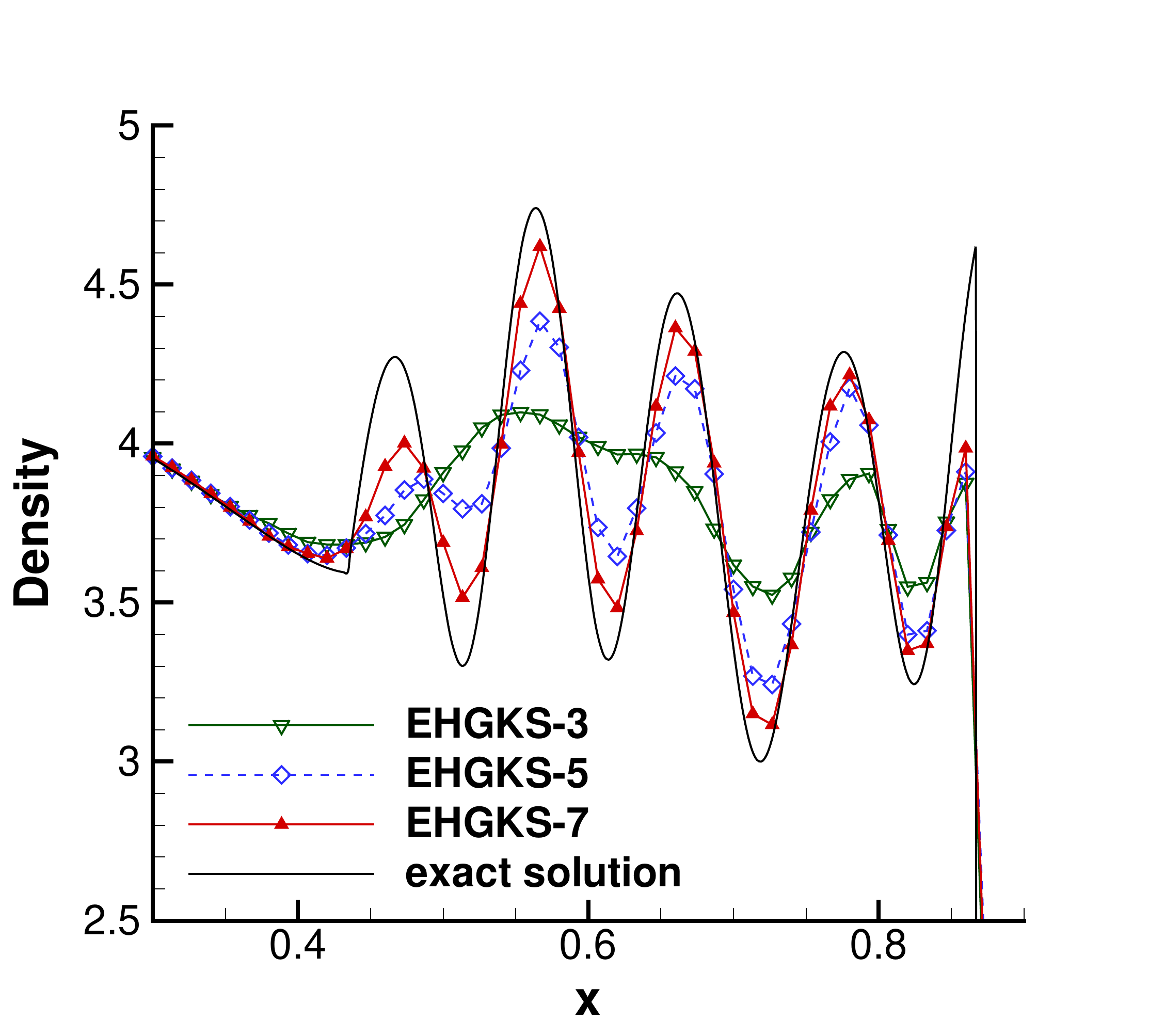}}
\end{minipage}
  \centerline{(a)}
\begin{minipage}{0.49\linewidth}
  \centerline{\includegraphics[width=\linewidth]{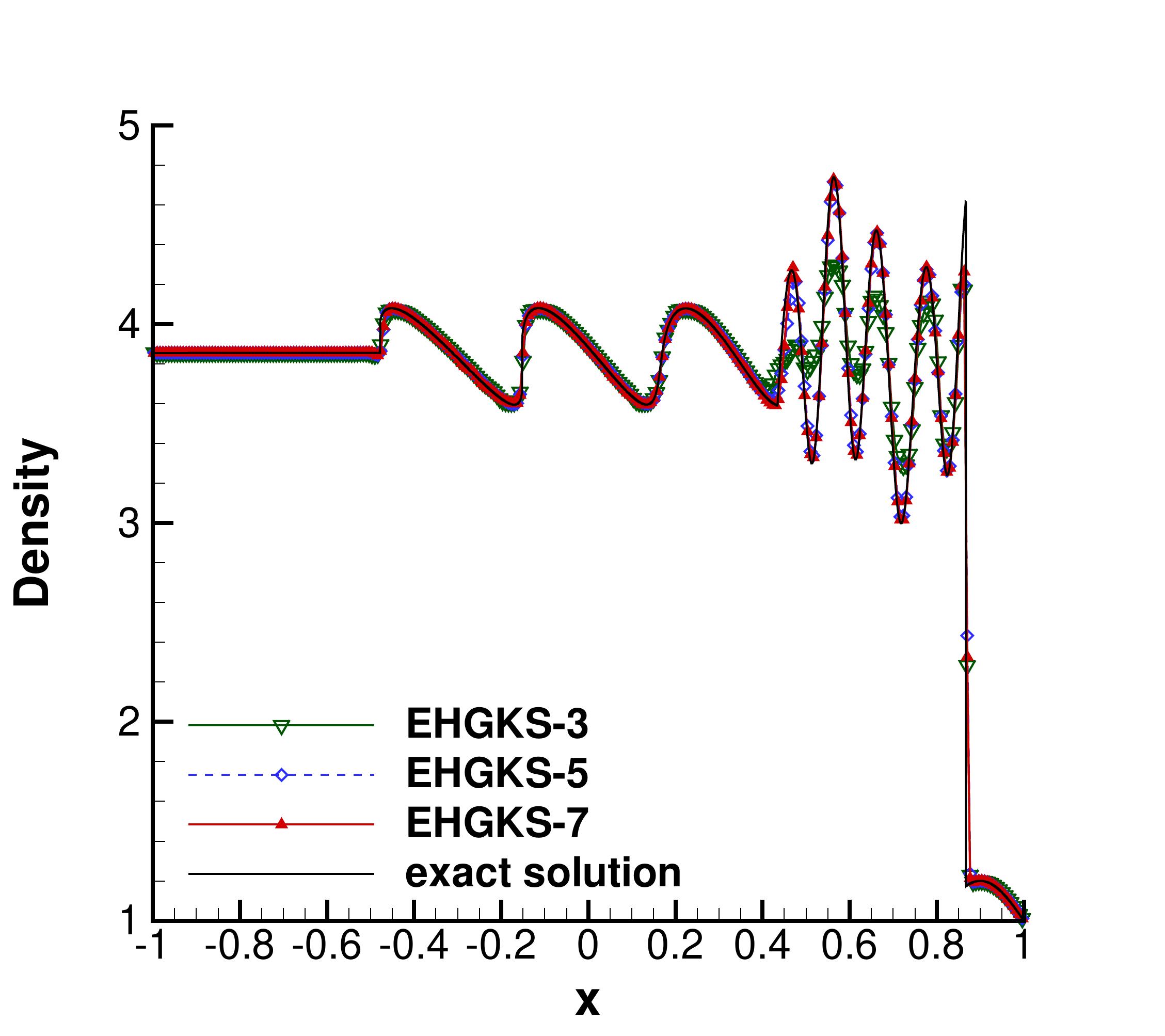}}
\end{minipage}
\begin{minipage}{0.49\linewidth}
  \centerline{\includegraphics[width=\linewidth]{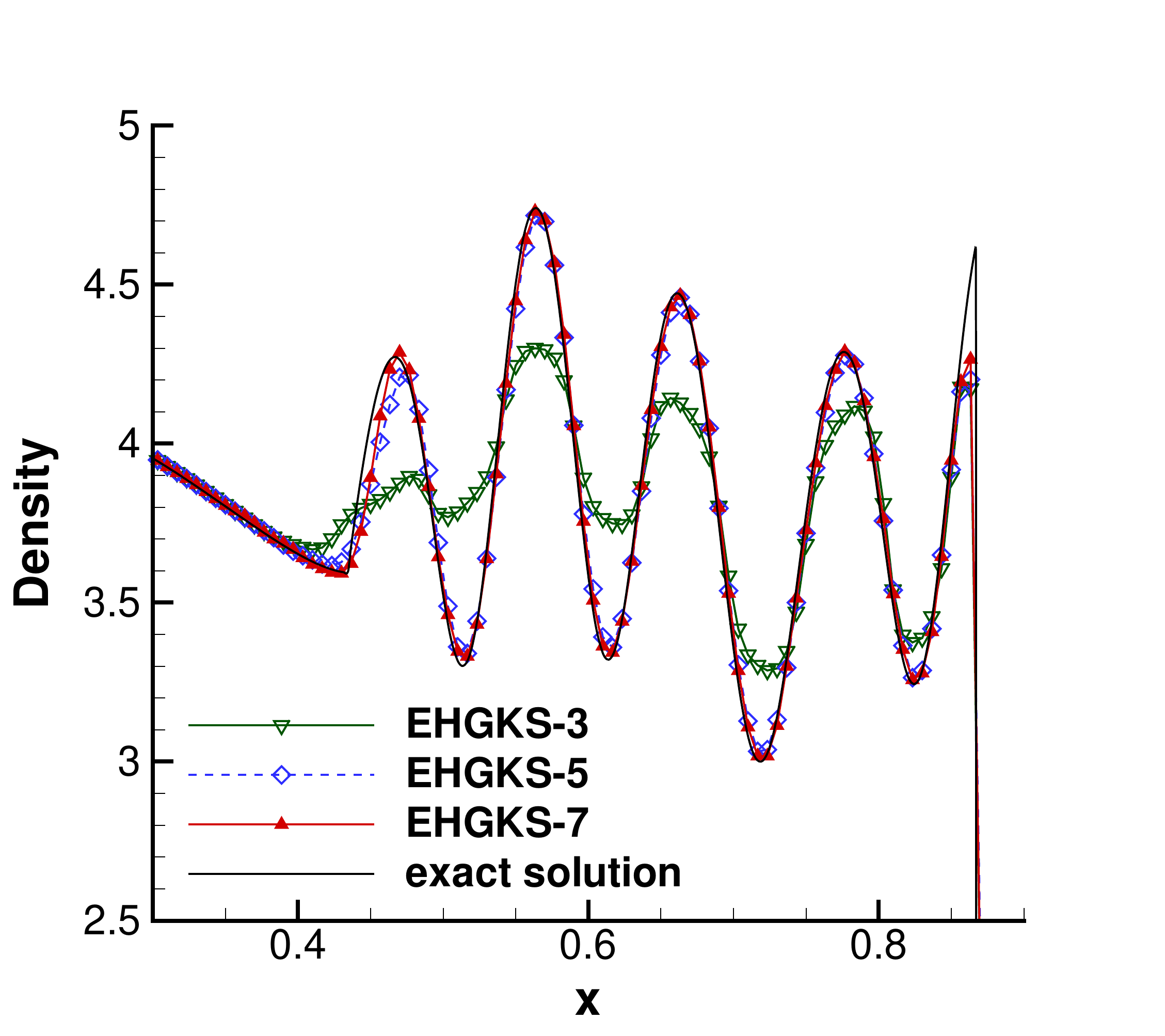}}
\end{minipage}
  \centerline{(b)}
\caption{\label{fig:so} The density distribution of the Shu-Osher shock acoustic interaction with (a) 150 and (b) 300 uniform cells at $t=0.47$.}
\end{figure*}

The density distribution and its local enlargement are shown in Fig.~\ref{fig:so}.
As expected, all the three EHGKS-$r$ schemes can capture the solution well, and higher-order EHGKS-$r$ schemes present better results with sharper shock discontinuity and more details of small-scale perturbations.
This also demonstrates the necessity to use higher-order EHGKS.
Comparisons are also made between EHGKS and HGKS.
The good agreement between EHGKS-$r$ and HGKS-$r$ demonstrates the validity of the Lax-Wendroff procedure introduced into the flux evaluation.
The comparisons between EHGKS and RK3-WENO-GKS illustrate the necessity for the scheme with high-order accuracy in both space and time.

\begin{figure*}[htb!]
\centering
\begin{minipage}{0.49\linewidth}
  \centerline{\includegraphics[width=\linewidth]{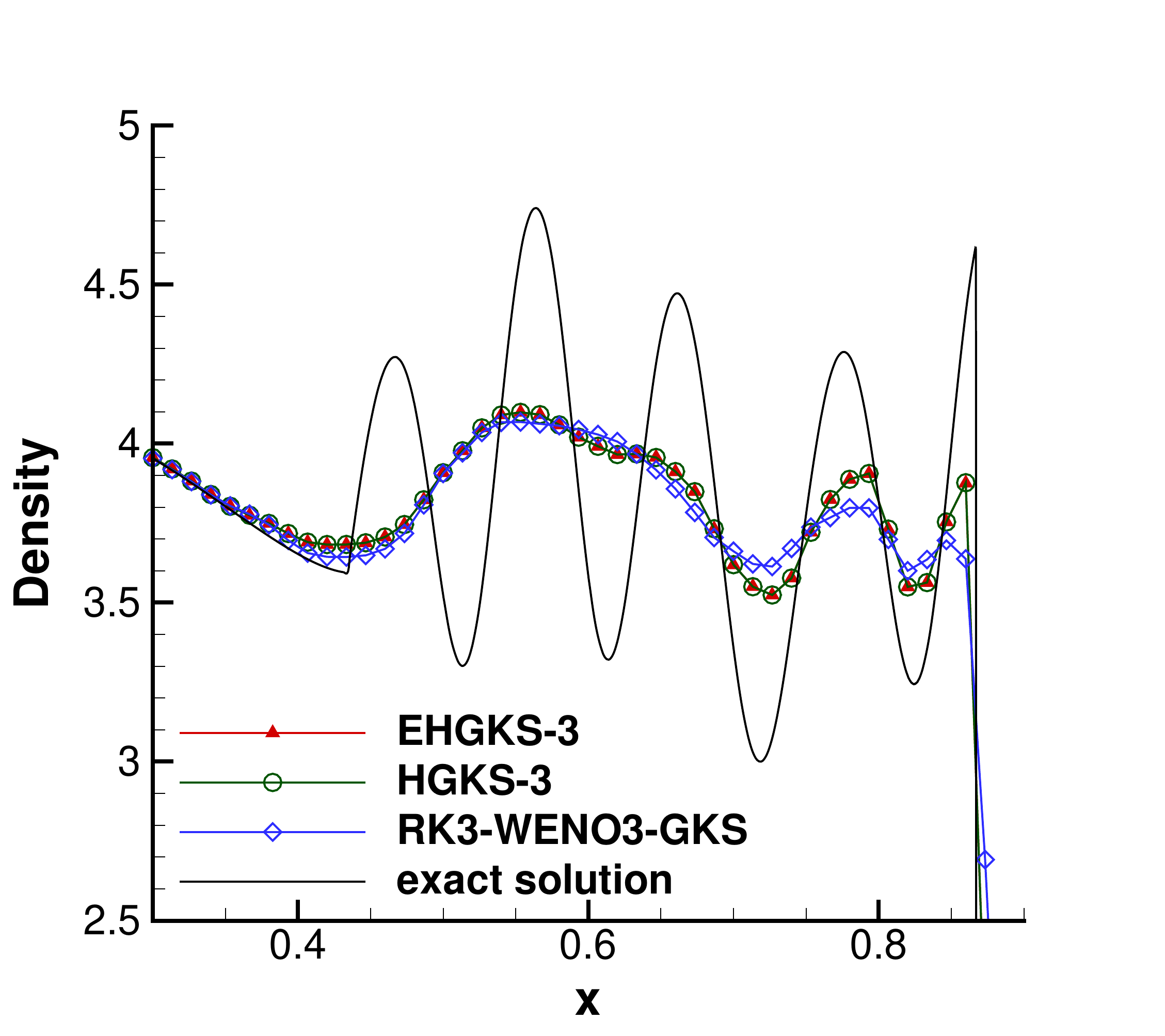}}
\end{minipage}
\begin{minipage}{0.49\linewidth}
  \centerline{\includegraphics[width=\linewidth]{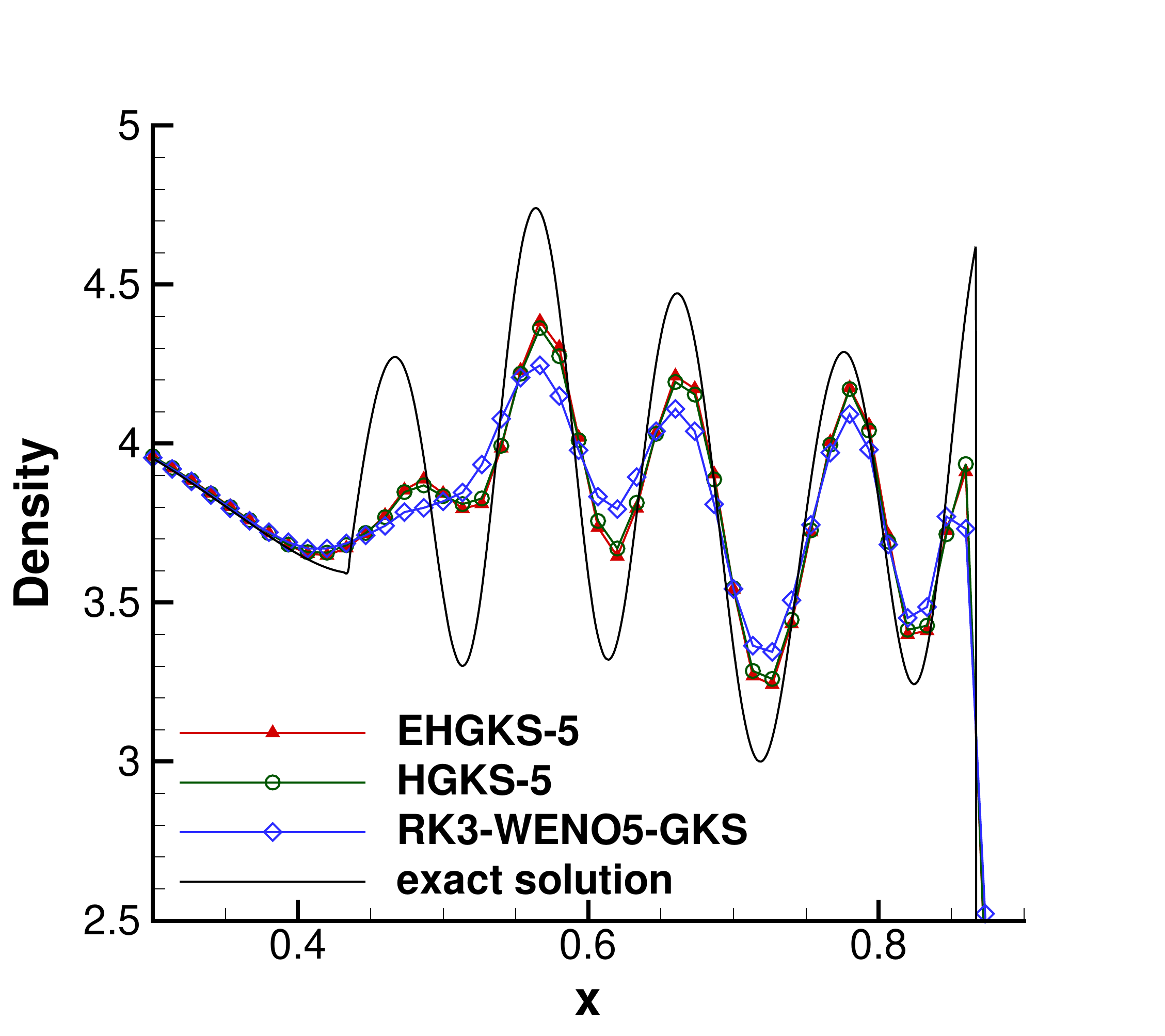}}
\end{minipage}
\begin{minipage}{0.49\linewidth}
  \centerline{\includegraphics[width=\linewidth]{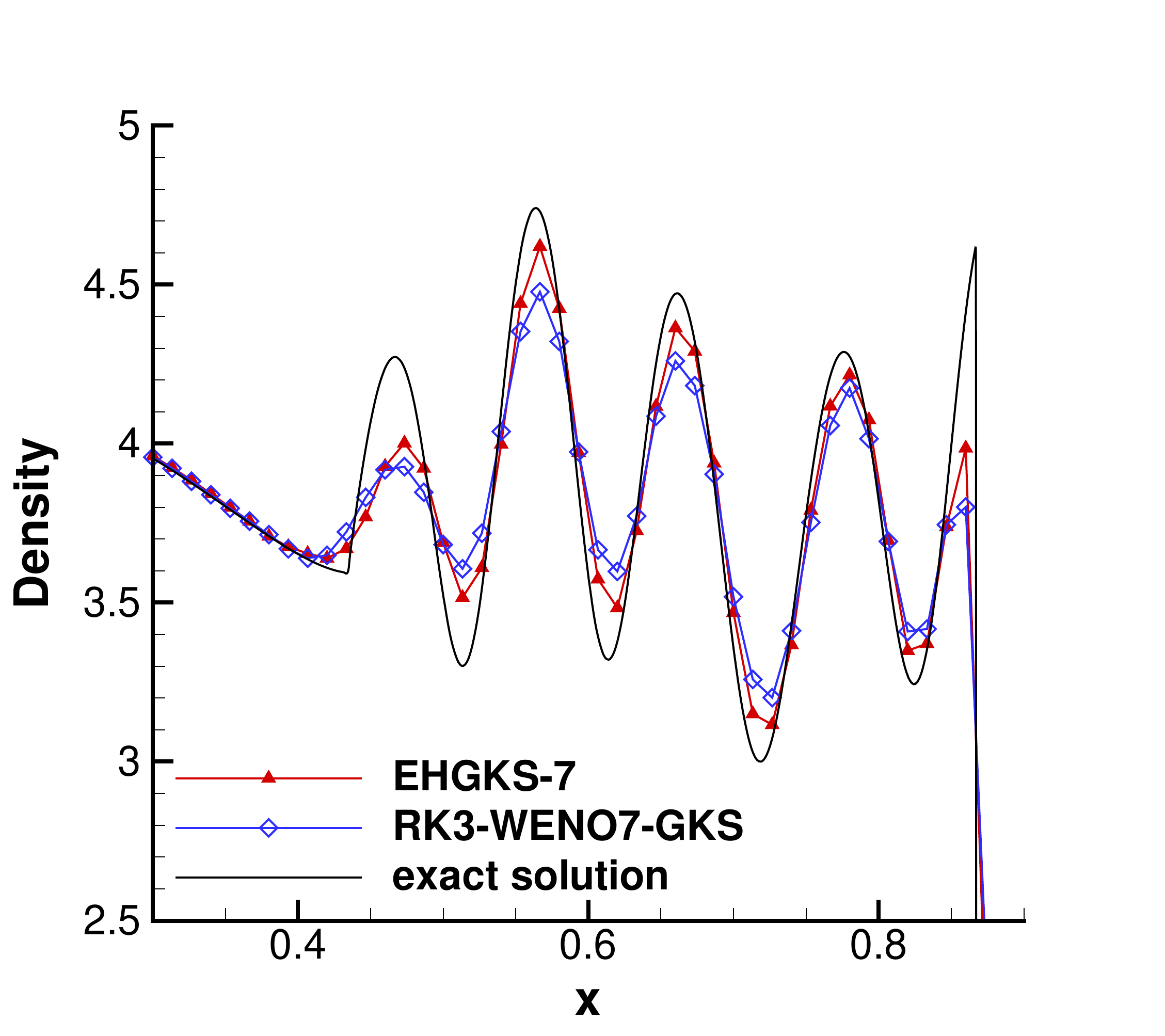}}
\end{minipage}
\caption{\label{fig:so-comp} The density distribution of the Shu-Osher shock acoustic interaction with 150 uniform cells at $t=0.47$.}
\end{figure*}

\subsection{Woodward-Colella blast wave}

\begin{figure*}[htb!]
\centering
\begin{minipage}{0.49\linewidth}
  \centerline{\includegraphics[width=\linewidth]{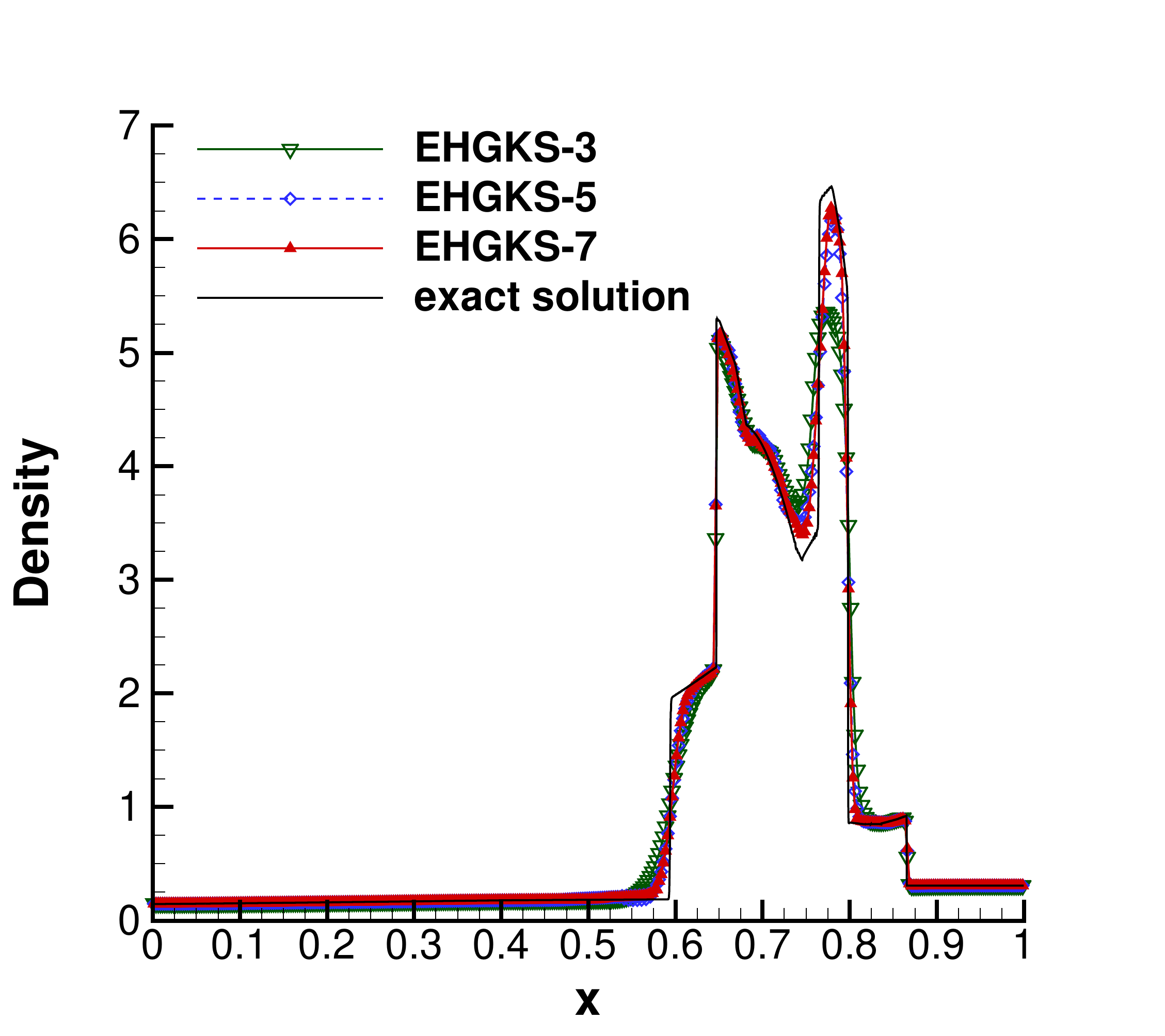}}
\end{minipage}
\begin{minipage}{0.49\linewidth}
  \centerline{\includegraphics[width=\linewidth]{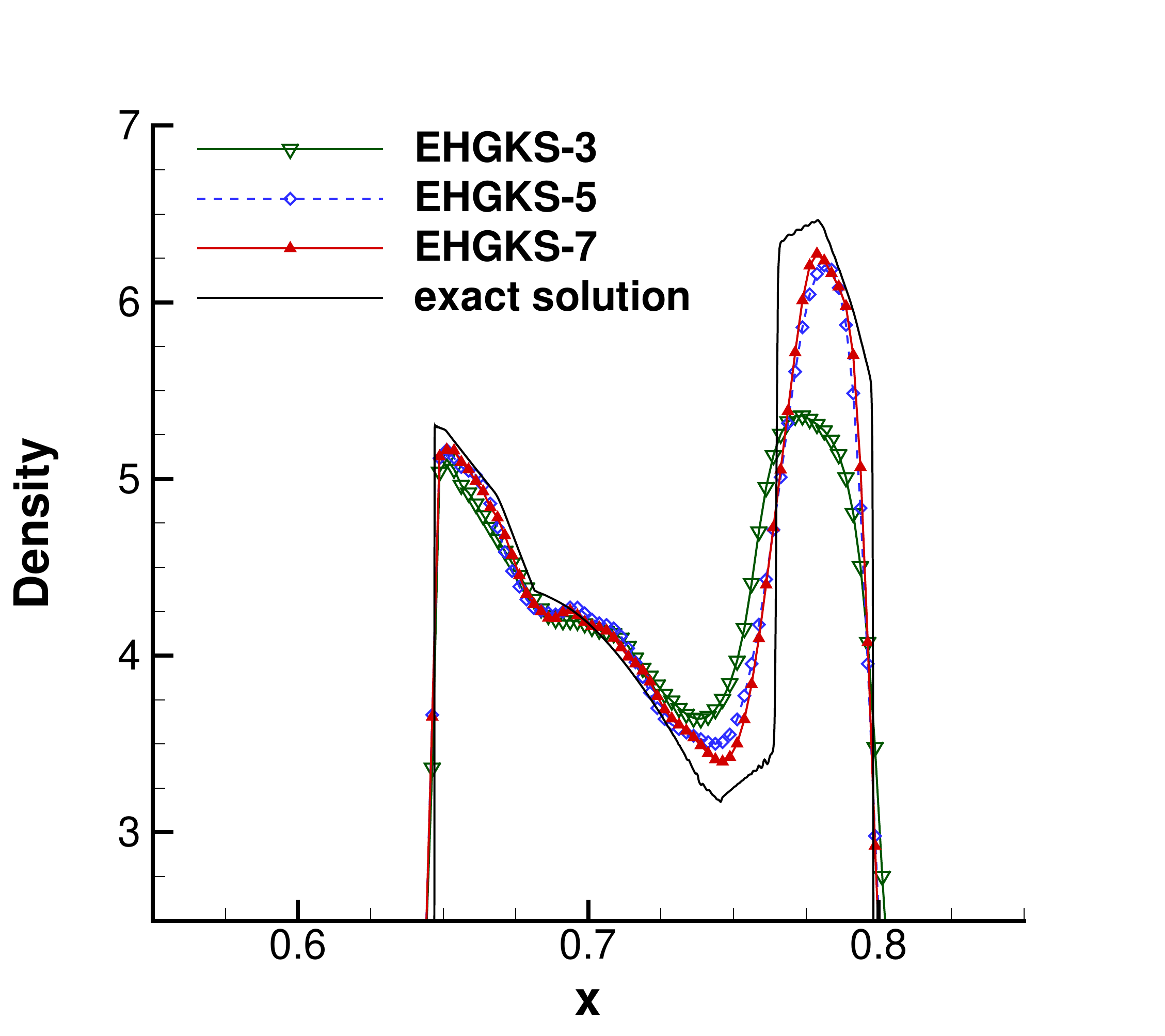}}
\end{minipage}
\caption{\label{fig:bw1} The density distribution of the blast wave problem with 400 uniform cells at $t=0.038$.}
\end{figure*}

This case contains the interactions between strong shock waves and contact discontinuities, which is a very challenging problem to assess the robustness
of a numerical scheme \cite{shu1998}.
The initial conditions are given by
\begin{eqnarray}
\left( \rho ,U,p \right)=
\left\{ \begin{array}{l}
 \left( 1, 0, 1000\right),0 \le x \le 0.1, \\
 \left( 1, 0, 0.01\right),0.1 < x \le 0.9, \\
 \left( 1, 0, 100 \right),0.9 < x \le 1,
 \end{array} \right.\nonumber
\label{eq:bw0}
\end{eqnarray}
with reflective boundary conditions on both sides of the computational domain $[0, 1]$.
$400$ uniform cells are used in the simulation, while the output time $t = 0.038$ and $CFL = 0.9$ are taken in this case
to preserve the stability.
It should be mentioned that $CFL = 0.9$ is still very delighted to be accepted in this case with very strong shocks, contact discontinuities, rarefaction waves and their interactions.
Fig.~\ref{fig:bw1} shows the density distributions given by EHGKS-3, EHGKS-5 and EHGKS-7, which
demonstrates the robustness of EHGKS-$r$.

The results computed by EHGKS are comparable to the existing high-order schemes in other work \cite{lqb2010,cyb2016}.
It illustrates the good resolution of EHGKS.
Similar to the previous tests, higher-order EHGKS-$r$ behaves better on the results with sharper discontinuities and resolved local extrema.
Specifically, the improvement from EHGKS-5 to EHGKS-7 is less significant compared with that from EHGKS-3 to EHGKS-5.
It's worthy of mention that the similar phenomenon has also been observed where the space accuracy is improved from the fourth to eighth order \cite{zhao2017}, or the time accuracy from the fourth to fifth order \cite{ji2017}.
It seems less significant improvement has been made in this case by increasing the very high-order accuracy, which still needs further studies.

\subsection{2D linear advection of the density perturbation}
Here we assess the accuracy and efficiency of EHGKS when the solution is linear and smooth in the 2D advection of the density perturbation.

In this case, the initial condition is given by
\begin{eqnarray}
 \left( \rho,U,V,p \right) = \left( 1 + 0.2\sin (x),0.7,0.3,1 \right),\nonumber
\label{eq:ac10}
\end{eqnarray}
and the analytic solution is
\begin{eqnarray}
 \left( \rho,U,V,p \right) = \left( 1 + 0.2\sin (x - t),0.7,0.3,1 \right).\nonumber
\label{eq:ac1}
\end{eqnarray}
The computational domain is $[ {0,2\pi } ] \times [ {0,2\pi } ]$ divided by $N \times N$ uniform cells.
The output time $t=2\pi$.

\begin{figure*}[htb!]
\centering
\begin{minipage}{0.49\linewidth}
  \centerline{\includegraphics[width=\linewidth]{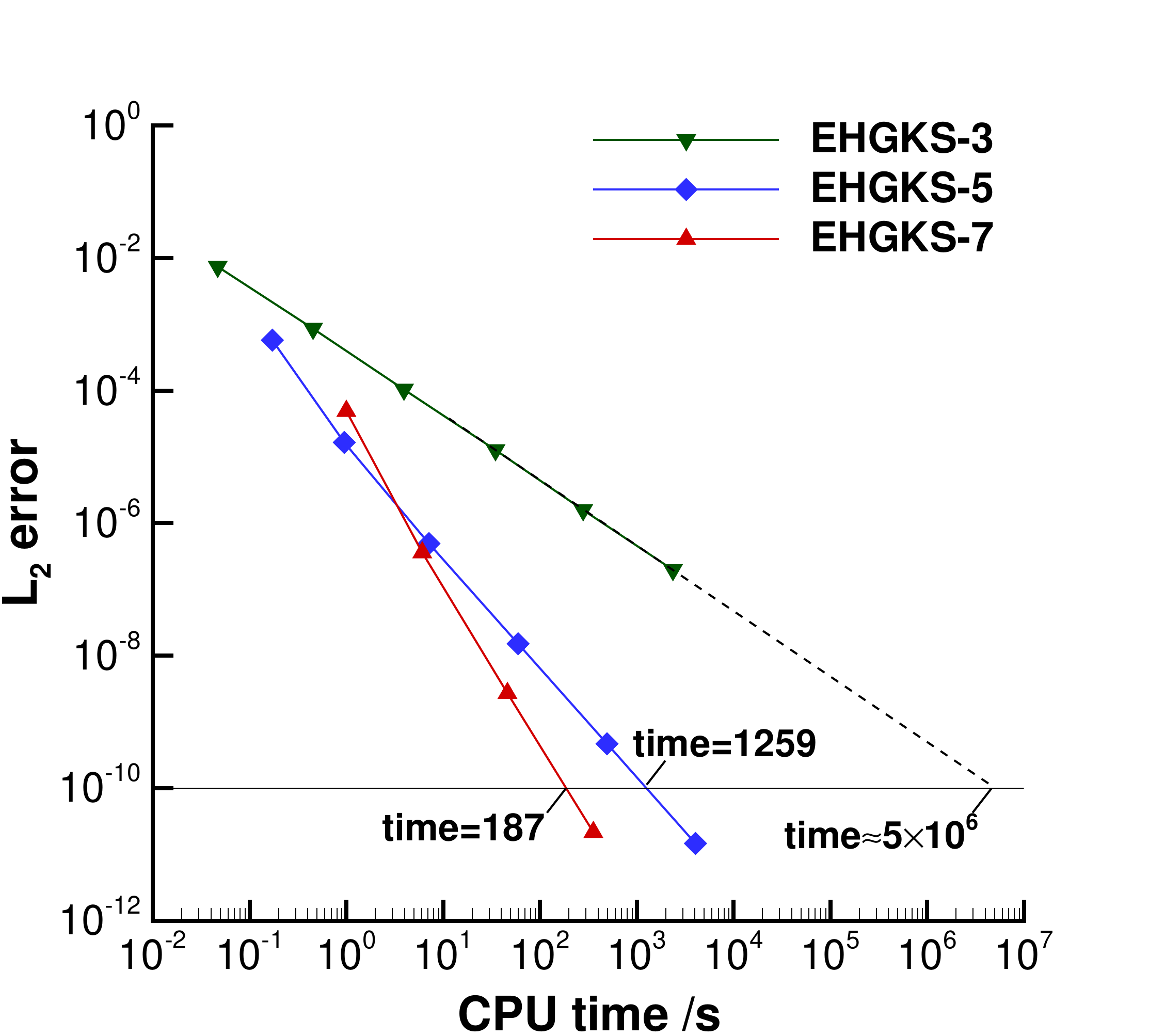}}
\end{minipage}
\caption{\label{fig:cputime2d} The CPU time vs $L_2$ error for EHGKS. The test case is the 2D advection of the density perturbation.}
\end{figure*}

The results of the errors and the accuracy order given by EHGKS-$3$, EHGKS-$5$ and EHGKS-$7$ are shown in Table~\ref{tab:ac2}.
It is observed that all the EHGKS-$r$ schemes can achieve the designed order of accuracy in both space and time in the 2D cases.
The errors and the corresponding CPU times of EHGKS-$r$ are shown in Fig.~\ref{fig:cputime2d}.
The same conclusion can be drawn that given the computational errors, the higher-order EHGKS uses less CPU time than the lower-order scheme.

\begin{table}[htb!]
\centering
\begin{tabular}{c|cc|cc|cc}
\hline
\multicolumn{7}{c}{EHGKS-3}  \\
\hline
$N$  &$L_1$ error &Order &$L_2$ error &Order &$L_{\infty}$ error   & Order  \\
\hline
    10    & 6.59E-03 &       & 7.43E-03 &       & 1.02E-02 &  \\
    20    & 7.75E-04 & 3.09  & 8.63E-04 & 3.11  & 1.25E-03 & 3.03 \\
    40    & 9.35E-05 & 3.05  & 1.04E-04 & 3.06  & 1.51E-04 & 3.05 \\
    80    & 1.15E-05 & 3.02  & 1.27E-05 & 3.02  & 1.87E-05 & 3.02 \\
\hline
\multicolumn{7}{c}{EHGKS-5}  \\
\hline
$N$  &$L_1$ error &Order &$L_2$ error &Order &$L_{\infty}$ error   & Order  \\
\hline
    10    & 5.04E-04 &       & 5.78E-04 &       & 8.04E-04 &  \\
    20    & 1.49E-05 & 5.08  & 1.65E-05 & 5.13  & 2.39E-05 & 5.07 \\
    40    & 4.44E-07 & 5.06  & 4.92E-07 & 5.07  & 7.18E-07 & 5.05 \\
    80    & 1.36E-08 & 5.03  & 1.51E-08 & 5.03  & 2.22E-08 & 5.02 \\
\hline
\multicolumn{7}{c}{EHGKS-7}  \\
\hline
$N$  &$L_1$ error &Order &$L_2$ error &Order &$L_{\infty}$ error   & Order  \\
\hline
    10    & 4.28E-05 &       & 4.92E-05 &       & 6.81E-05 &  \\
    20    & 3.24E-07 & 7.04  & 3.60E-07 & 7.09  & 5.21E-07 & 7.03 \\
    40    & 2.45E-09 & 7.05  & 2.71E-09 & 7.05  & 3.95E-09 & 7.04 \\
    80    & 1.95E-11 & 6.97  & 2.16E-11 & 6.97  & 3.17E-11 & 6.96 \\
\hline
\end{tabular}
\caption{\label{tab:ac2} Accuracy test for EHGKS in the 2D advection of the density perturbation.}
\end{table}

\subsection{2D isotropic vortex propagation problem}
This problem for the 2D compressible Euler equations is to test the accuracy of numerical methods, since the exact solution is smooth and has a simple analytical expression.
Some schemes with attested high-order accuracy in the previous test cases may fail in this test case.
The mean flow is
\begin{eqnarray}
 \left( \rho,U,V,p \right) = \left( 1,1,1,1 \right).\nonumber
\label{eq:ac30}
\end{eqnarray}
An isotropic vortex is added to the mean flow with the perturbations in velocities, temperature and no perturbation in entropy $S = p/\rho ^\gamma$ \cite{ljq2016}, which gives
\begin{eqnarray}
 \left( \delta U,\delta V \right)
 =  \frac{10 \varepsilon }{{2\pi }} \rme^{\frac{{1 - \eta^2 }}{2}}
 \left( -y,x \right),
 \delta T =  - \frac{{\left( {\gamma  - 1} \right)\varepsilon ^2 }}{{8\gamma \pi ^2 }}\rme^{1 - \eta^2 } ,
 \delta S = 0, \nonumber
\label{eq:ac30}
\end{eqnarray}
where $\varepsilon  = 5 $ and $\eta^2=100( x^2 + y^2)$.
The periodic boundary condition is adopted.
The exact solution is the perturbation propagating with $(U, V)=(1,1)$ \cite{ljq2016,pan2016}.
The computational domain is $[ { - 0.5,0.5} ] \times [ { - 0.5,0.5} ]$ divided by $N \times N$ uniform cells.
The output time $t=2$.

The test results based on the density are shown in Table~\ref{tab:ac2i}.
It further illustrates that all the EHGKS-$3$, EHGKS-$5$ and EHGKS-$7$ schemes can achieve their designed order of accuracy in the 2D case.

\begin{table}[htb!]
\centering
\begin{tabular}{c|cc|cc|cc}
\hline
\multicolumn{7}{c}{EHGKS-3}  \\
\hline
$N$  &$L_1$ error &Order &$L_2$ error &Order &$L_{\infty}$ error   & Order  \\
\hline
    25    & 8.59E-03 &       & 2.29E-02 &       & 1.87E-01 &      \\
    50    & 1.86E-03 & 2.21  & 4.57E-03 & 2.32  & 3.16E-02 & 2.56 \\
    100   & 2.87E-04 & 2.69  & 7.24E-04 & 2.66  & 4.62E-03 & 2.78 \\
    200   & 3.76E-05 & 2.93  & 9.66E-05 & 2.91  & 6.05E-04 & 2.93 \\
\hline
\multicolumn{7}{c}{EHGKS-5}  \\
\hline
$N$  &$L_1$ error &Order &$L_2$ error &Order &$L_{\infty}$ error   & Order  \\
\hline
    25    & 2.22E-03 &       & 5.23E-03 &       & 3.85E-02 &      \\
    50    & 1.82E-04 & 3.61  & 4.09E-04 & 3.67  & 2.56E-03 & 3.91 \\
    100   & 6.90E-06 & 4.72  & 1.83E-05 & 4.49  & 1.55E-04 & 4.04 \\
    200   & 2.21E-07 & 4.97  & 6.35E-07 & 4.85  & 3.91E-06 & 5.31 \\
\hline
\multicolumn{7}{c}{EHGKS-7}  \\
\hline
$N$  &$L_1$ error &Order &$L_2$ error &Order &$L_{\infty}$ error   & Order  \\
\hline
    25    & 1.26E-03 &       & 2.97E-03 &       & 2.18E-02 &      \\
    50    & 8.06E-05 & 3.96  & 3.14E-04 & 3.24  & 4.57E-03 & 2.25 \\
    100   & 2.22E-06 & 5.18  & 1.06E-05 & 4.88  & 1.46E-04 & 4.97 \\
    200   & 1.27E-08 & 7.45  & 5.99E-08 & 7.47  & 7.80E-07 & 7.55 \\
\hline
\end{tabular}
\caption{\label{tab:ac2i} Accuracy test for EHGKS in the 2D isotropic vortex propagation.}
\end{table}

\subsection{2D Riemann problems}

Two cases of the 2D Riemann problems are tested to assess the ability of EHGKS to solve the multi-dimensional problems genuinely.

\begin{figure*}[htb!]
\centering
\begin{minipage}{0.49\linewidth}
  \centerline{\includegraphics[width=\linewidth]{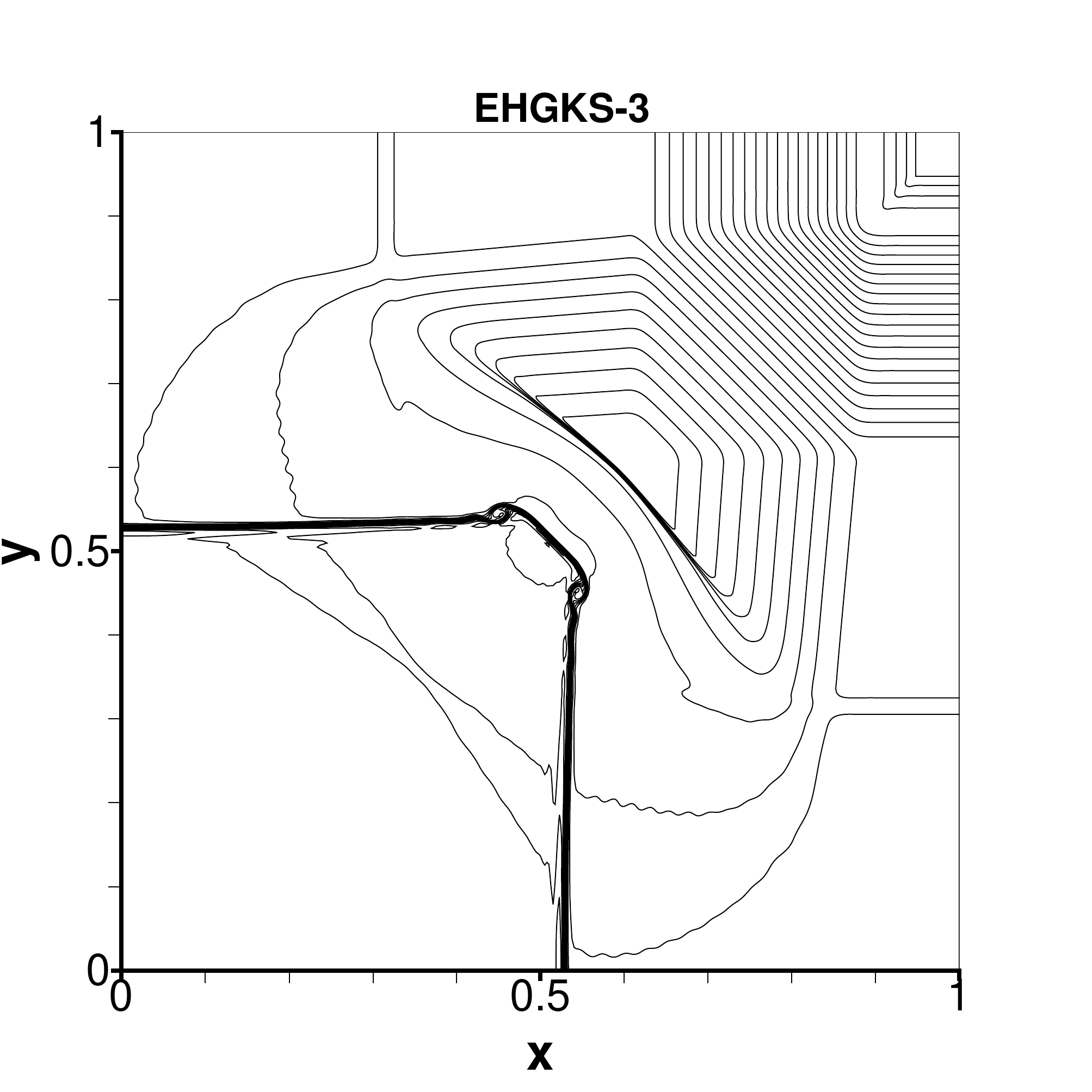}}
\end{minipage}
\begin{minipage}{0.49\linewidth}
  \centerline{\includegraphics[width=\linewidth]{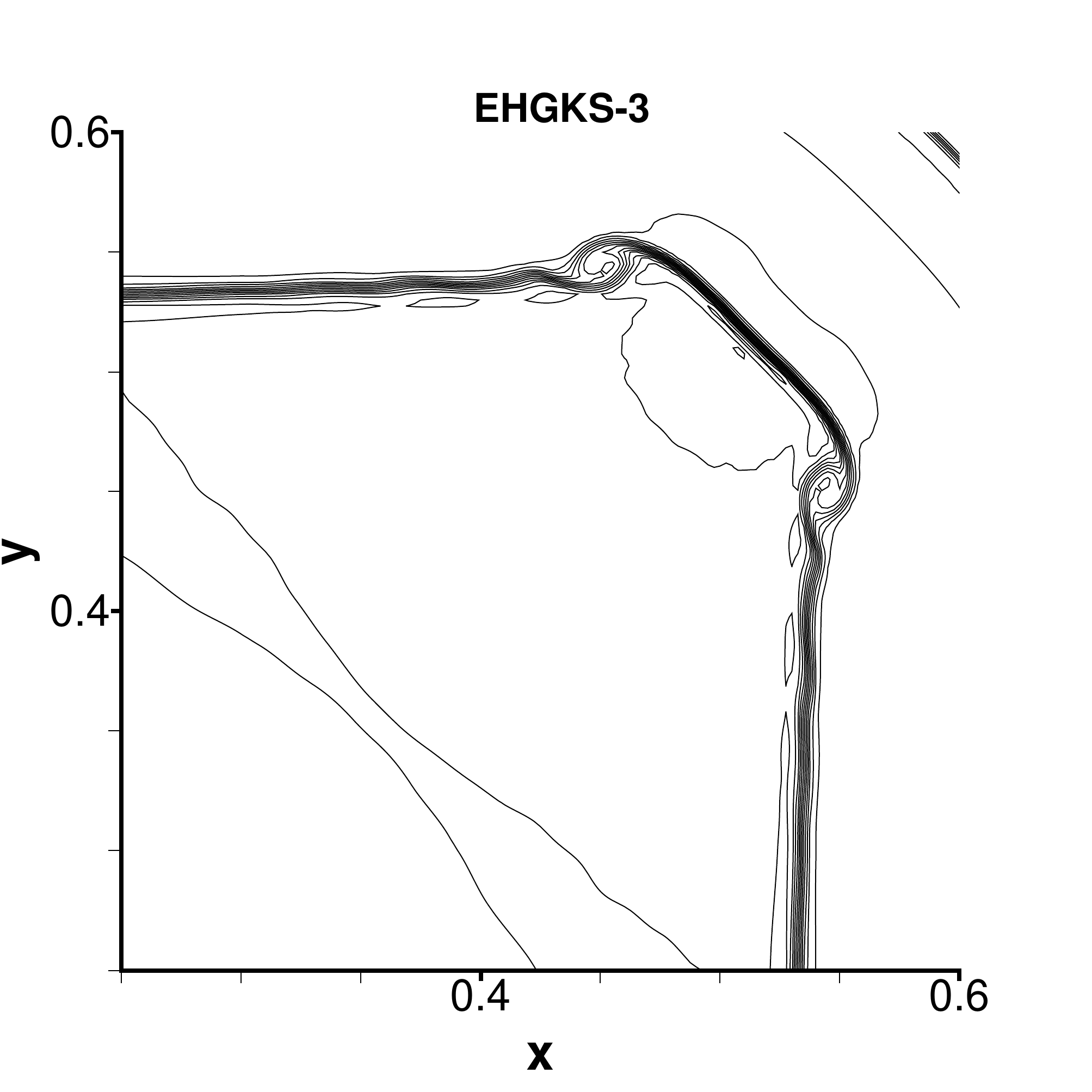}}
\end{minipage}
\begin{minipage}{0.49\linewidth}
  \centerline{\includegraphics[width=\linewidth]{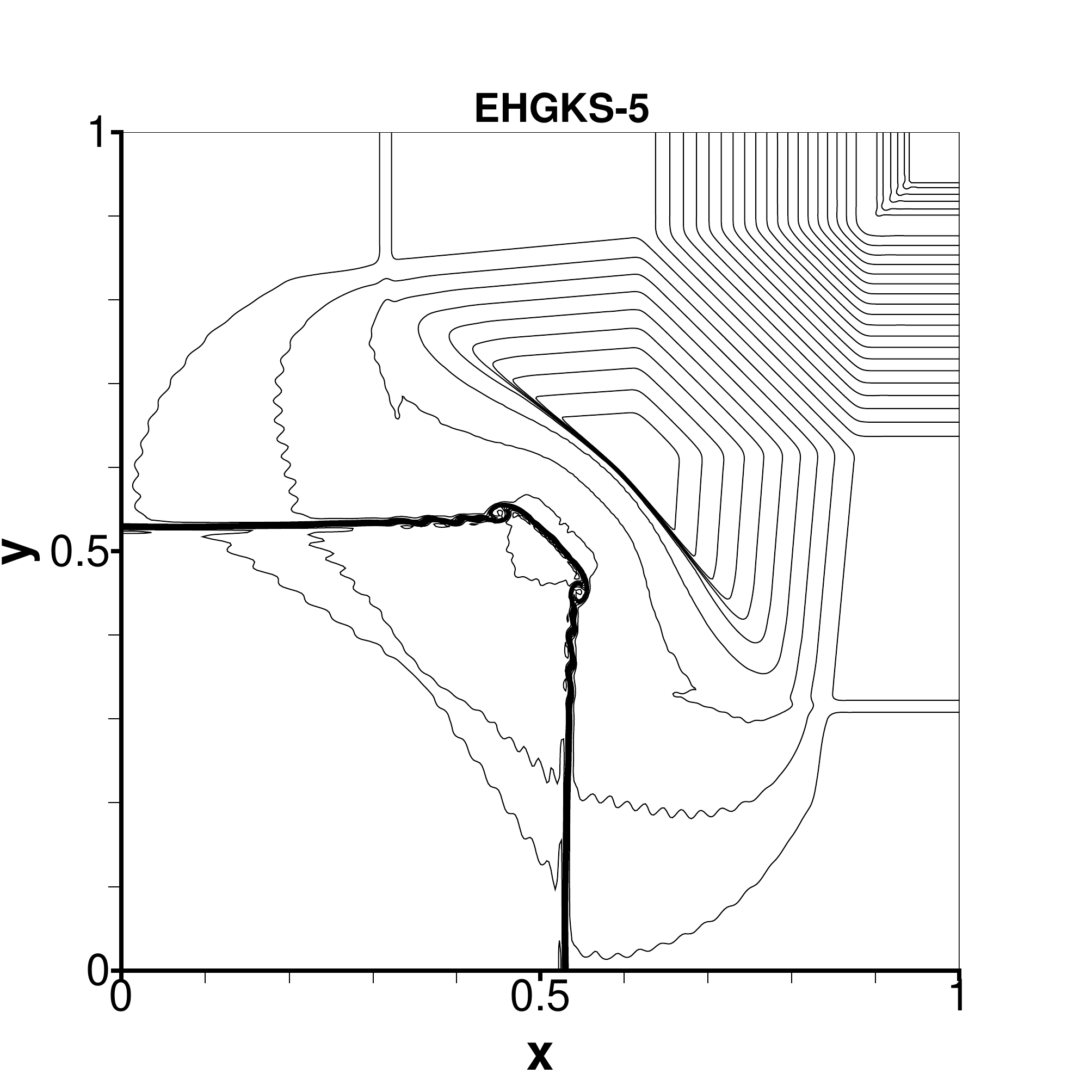}}
\end{minipage}
\begin{minipage}{0.49\linewidth}
  \centerline{\includegraphics[width=\linewidth]{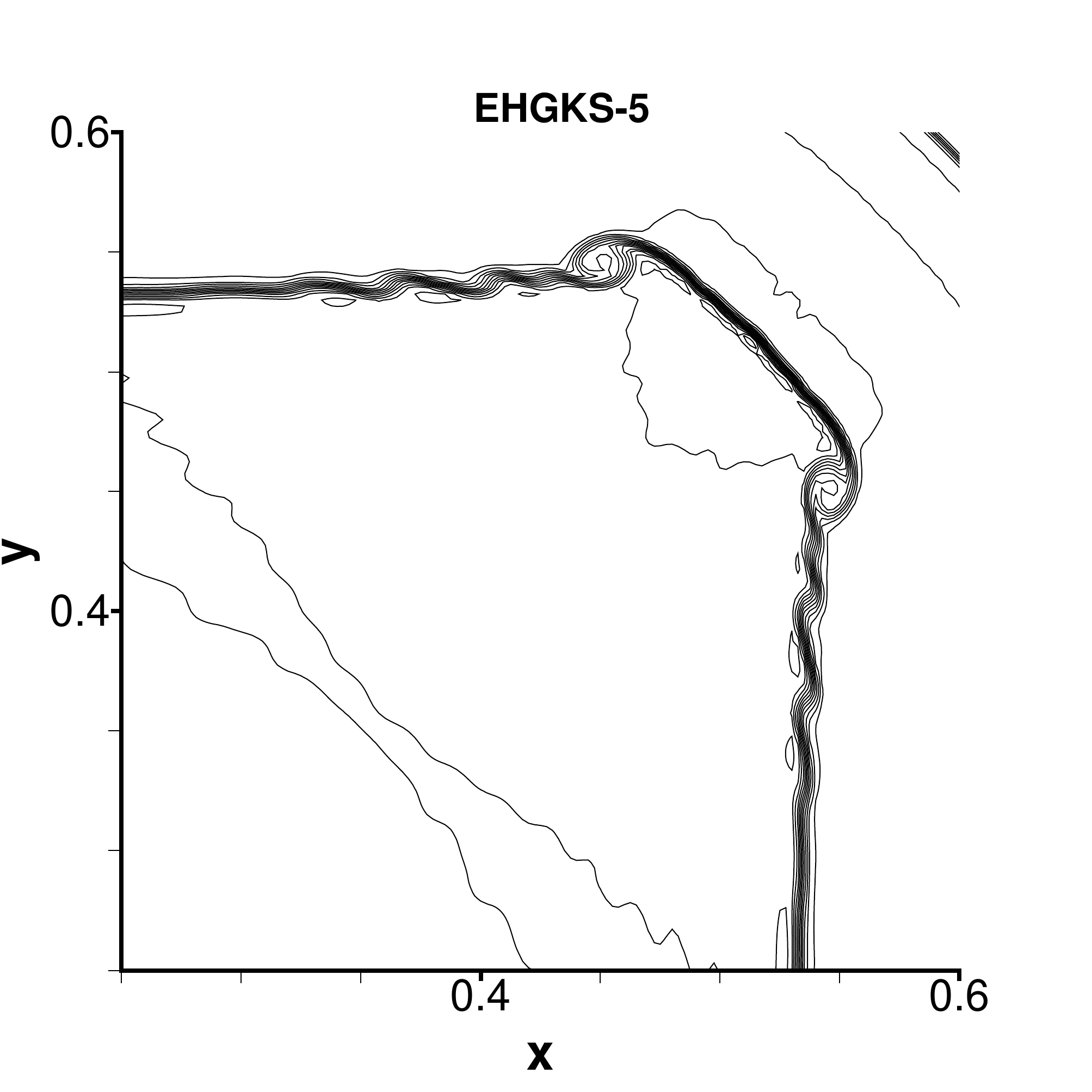}}
\end{minipage}
\begin{minipage}{0.49\linewidth}
  \centerline{\includegraphics[width=\linewidth]{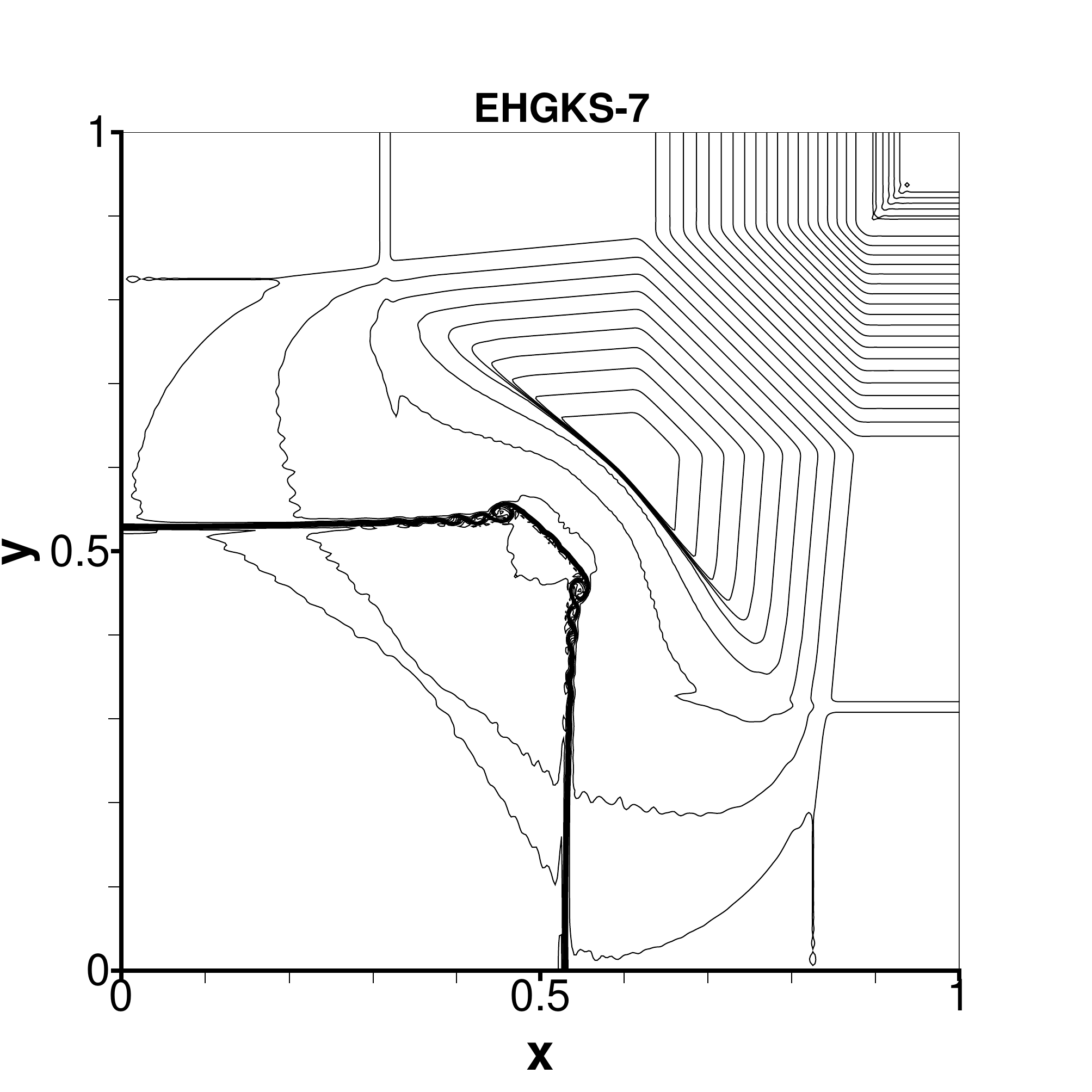}}
\end{minipage}
\begin{minipage}{0.49\linewidth}
  \centerline{\includegraphics[width=\linewidth]{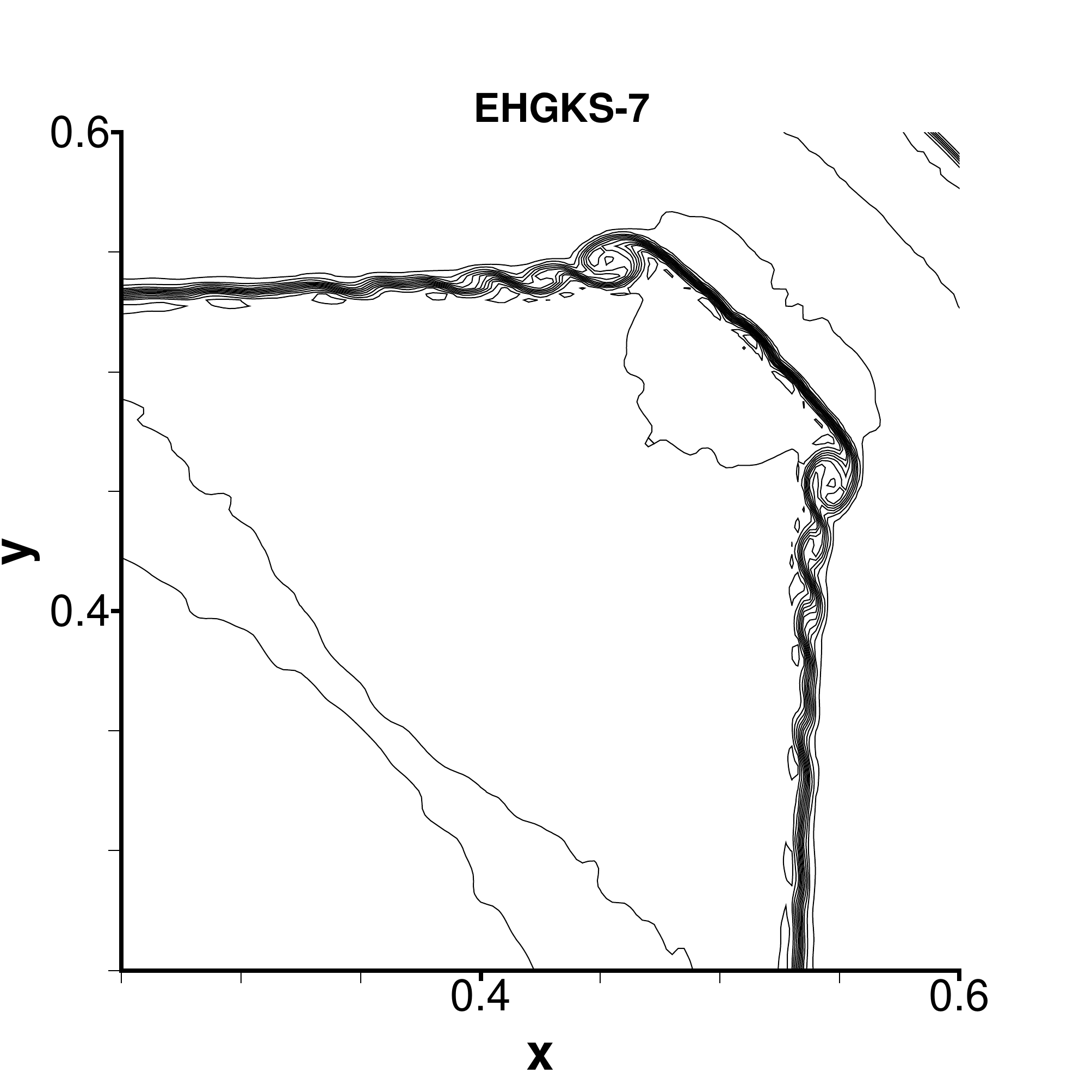}}
\end{minipage}
\caption{\label{fig:2drm2} The density distribution of the first 2D Riemann problem at $t=0.3$ with $400 \times 400$ uniform cells. 30 contours are drawn from 0.3 to 1.}
\end{figure*}

The first case is the interaction of the rarefaction waves and the vortex sheets \cite{pan2016,lax1998}.
The initial conditions are
\begin{eqnarray}
\left( \rho ,U,V,p \right)=
\left\{ \begin{array}{l}
 \left(1,0.1,0.1,1\right),x \ge 0.5,y \ge 0.5, \\
 \left(0.5197,- 0.6259,0.1,0.4\right),x < 0.5,y \ge 0.5, \\
 \left(0.8,0.1,0.1,0.4\right),x < 0.5,y < 0.5, \\
 \left(0.5197,0.1,- 0.6259,0.4\right),x \ge 0.5,y < 0.5. \\
 \end{array} \right.\nonumber
\label{eq:rp40}
\end{eqnarray}
The computational domain is $[0, 1] \times [0, 1]$, and the non-reflecting boundary conditions are used at all boundaries.
The computational domain is divided by $400 \times 400$ uniform cells.
The output time $t= 0.3$.
The density distributions in Fig.~\ref{fig:2drm2} show the roll-up is well captured by the current high-order scheme and confirms its high accuracy.
The same conclusion can be drawn as in the previous tests that the higher-order accuracy of EHGKS, the better resolution on the flow details.

\begin{figure*}[htb!]
\centering
\begin{minipage}{0.49\linewidth}
  \centerline{\includegraphics[width=\linewidth]{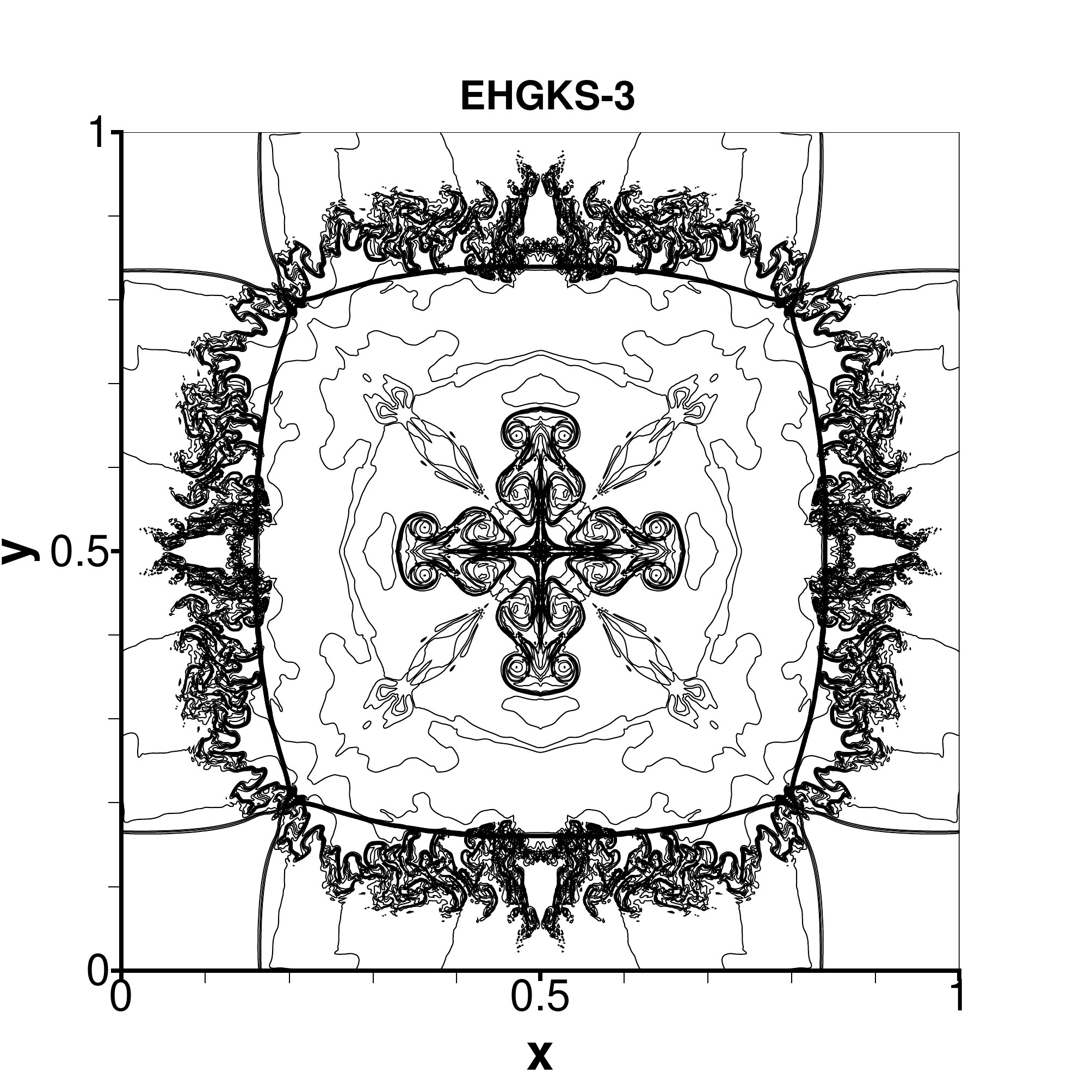}}
\end{minipage}
\begin{minipage}{0.49\linewidth}
  \centerline{\includegraphics[width=\linewidth]{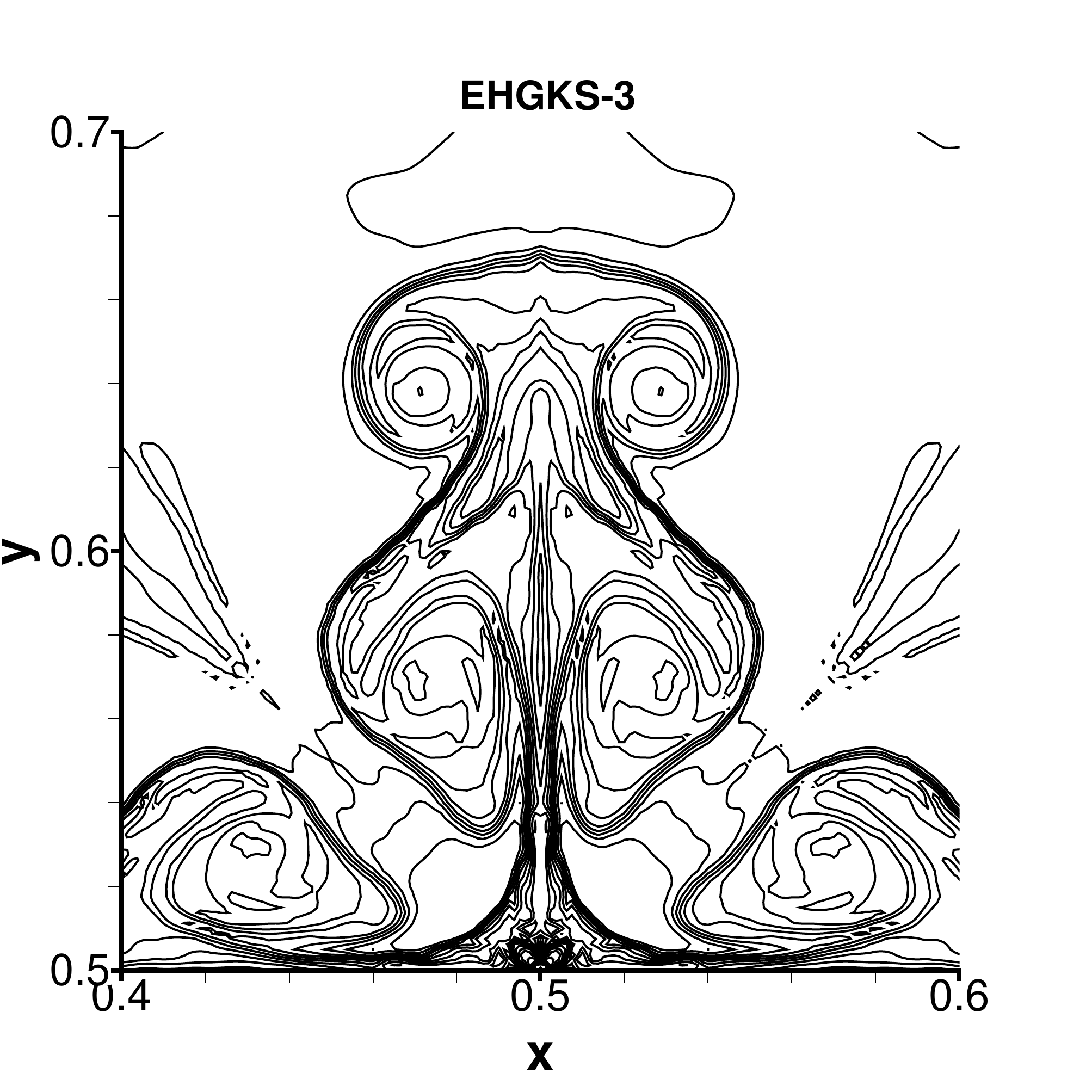}}
\end{minipage}
\begin{minipage}{0.49\linewidth}
  \centerline{\includegraphics[width=\linewidth]{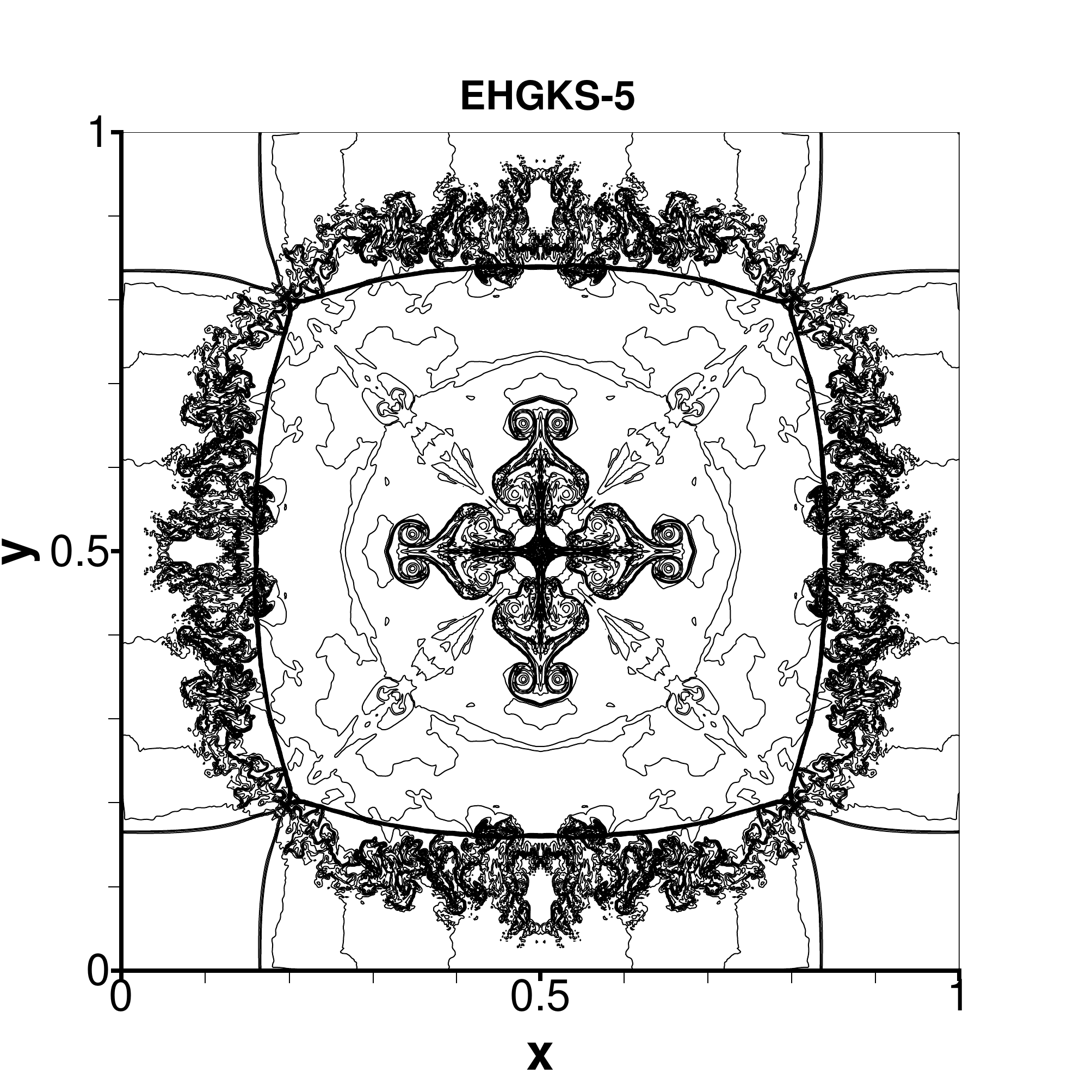}}
\end{minipage}
\begin{minipage}{0.49\linewidth}
  \centerline{\includegraphics[width=\linewidth]{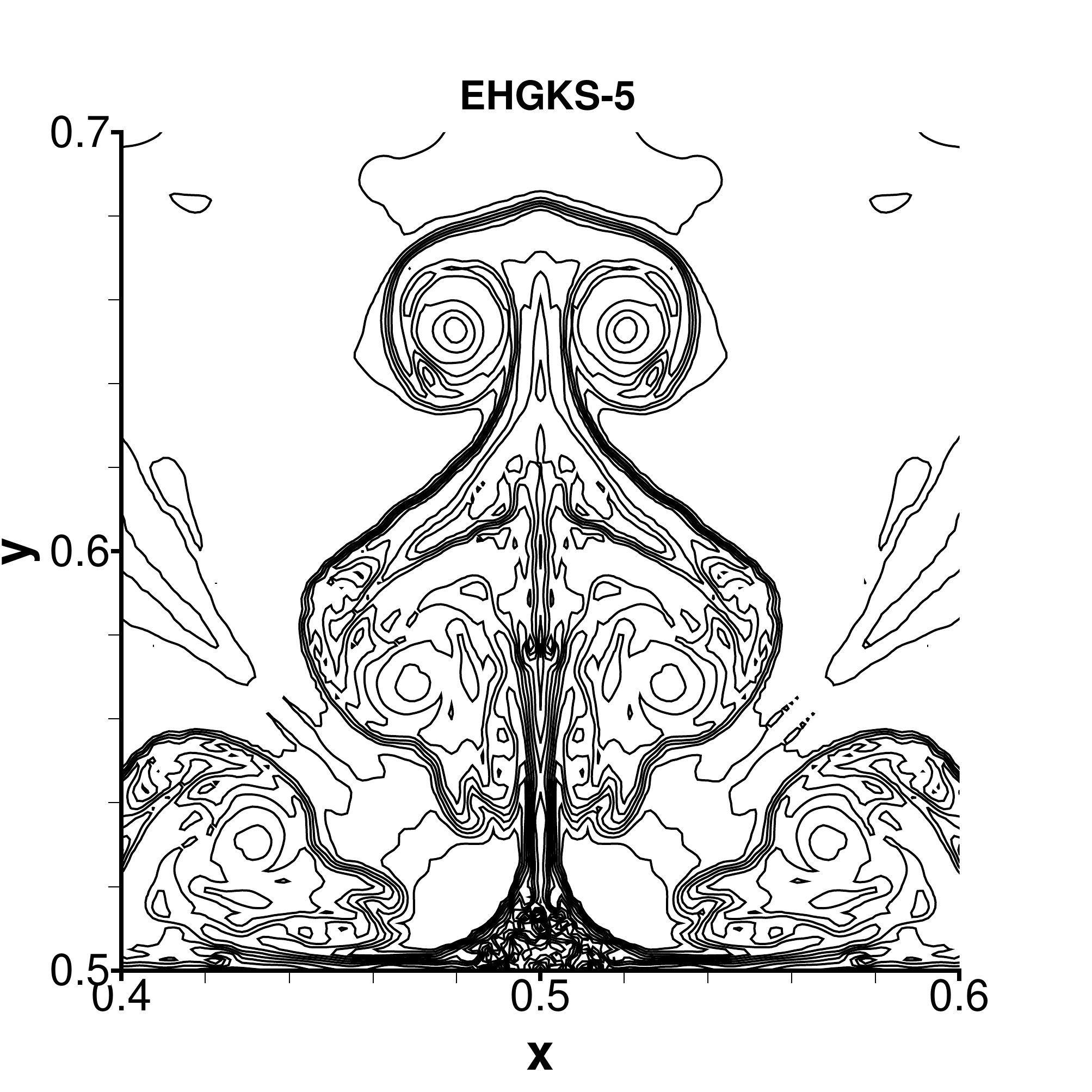}}
\end{minipage}
\begin{minipage}{0.49\linewidth}
  \centerline{\includegraphics[width=\linewidth]{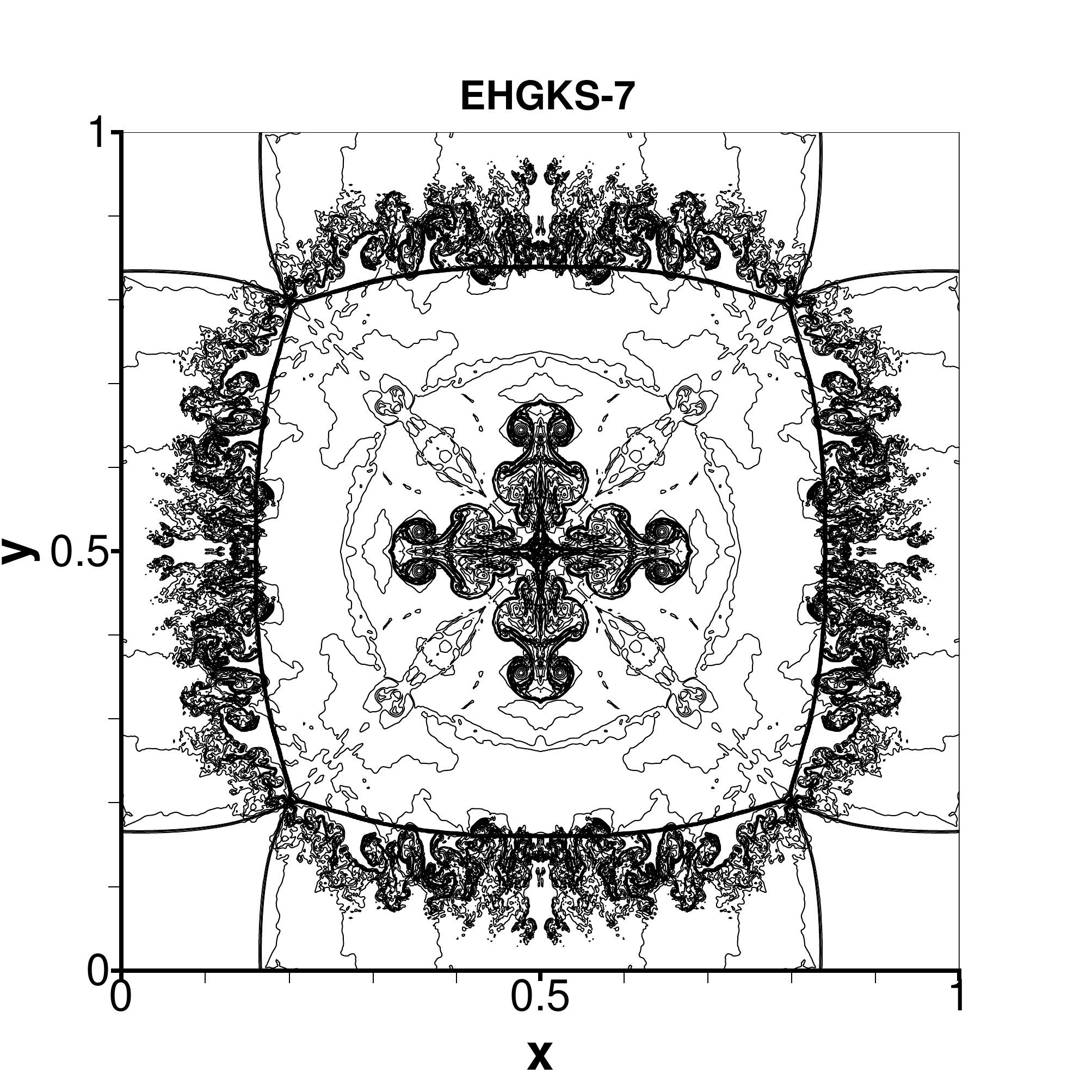}}
\end{minipage}
\begin{minipage}{0.49\linewidth}
  \centerline{\includegraphics[width=\linewidth]{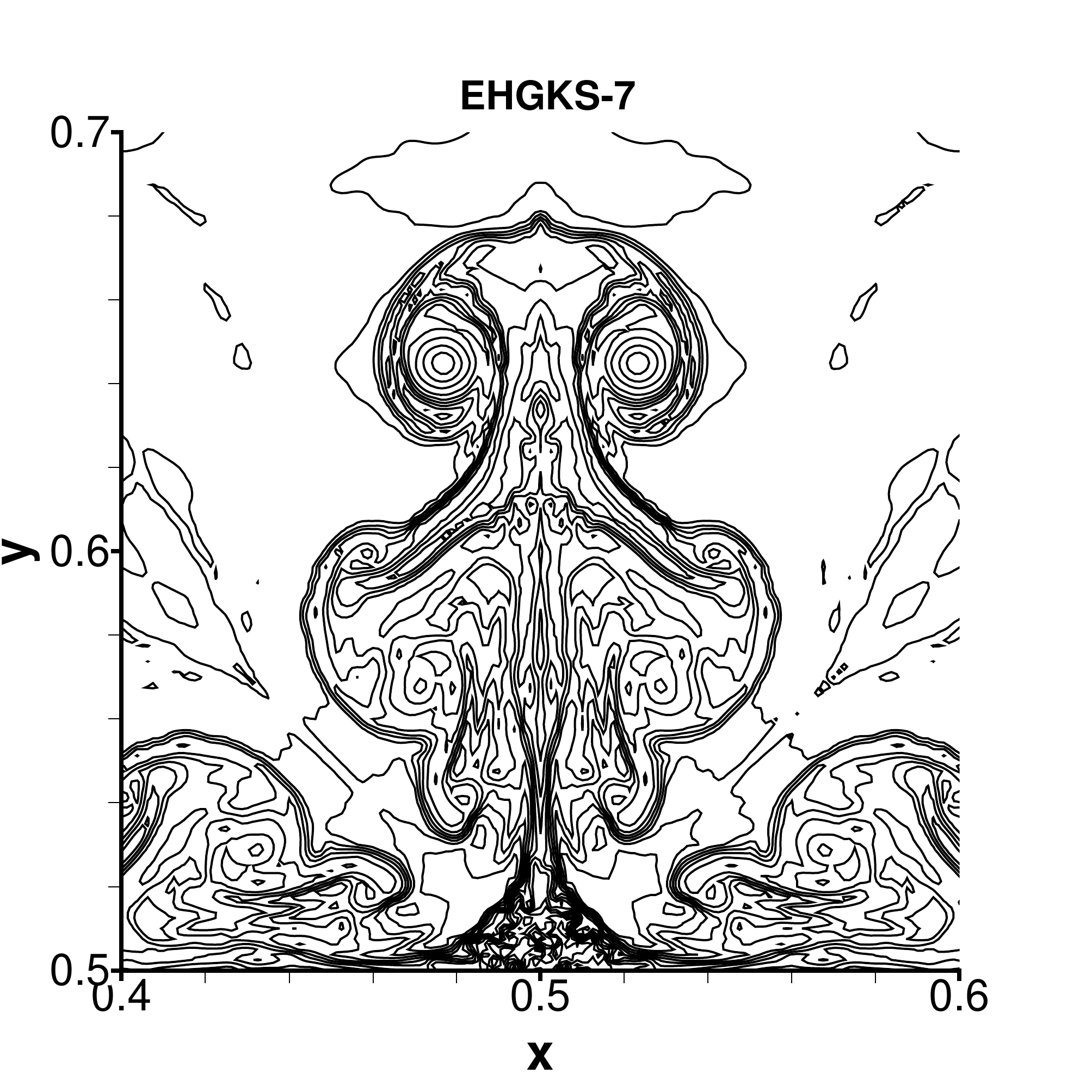}}
\end{minipage}
\caption{\label{fig:2drm} The density distribution of the second 2D Riemann problem at $t=1$ with $800 \times 800$ uniform cells. 30 contours are drawn from 0.1 to 0.6.}
\end{figure*}

In the second case \cite{cyb2016}, the initial conditions are
\begin{eqnarray}
\left( \rho ,U,V,p \right)=
\left\{ \begin{array}{l}
 \left(     1, 0,0, 1   \right), \sqrt{\left(x-0.5\right)^2
                                      +\left(y-0.5\right)^2 } \le 0.3, \\
 \left( 0.125, 0,0, 0.1 \right),\text{else}.
 \end{array} \right.\nonumber
\label{eq:rp0}
\end{eqnarray}
The computational domain is $[0, 1] \times [0, 1]$ divided by $800 \times 800$ uniform cells.
The boundary conditions are all reflective conditions.
The output time $t = 1$ when the shock waves have already hit the boundaries and interacted with the reflected waves \cite{cyb2016}.
Fig.~\ref{fig:2drm} shows more complex structures are captured by the higher-order EHGKS.

\subsection{Double Mach reflection problem}

This case has been extensively adopted to test the performance of numerical schemes in the compressible flows with strong shocks \cite{woodward1984}.
A right-moving shock of $\rma=10$ is initially positioned at $(x, y) =(1/6, 0)$ with $60 ^{\circ}$ to the wall.
The pre-shock conditions are
\begin{eqnarray}
\left( \rho,U,V,p \right) = \left(8,4.125\sqrt 3,- 4.125,116.5 \right),\nonumber
\label{eq:dm0}
\end{eqnarray}
and post-shock conditions are
\begin{eqnarray}
\left( \rho,U,V,p \right) = \left( 1.4,0,0,1 \right).\nonumber
\label{eq:dm1}
\end{eqnarray}
The computational domain is $[0, 3] \times [0, 0.75]$ divided by uniform cells with cell size $\Delta x=\Delta y=1/240$ and $ 1/480$.
The reflective boundary condition is used at the wall.
The pre-shock and post-shock conditions are imposed at the rest boundaries to describe the exact motion of the shock.
$CFL= 0.4$ is used in this case to preserve the stability of EHGKS.
The output time $t=0.2$.

\begin{figure*}[htb!]
\centering
\begin{minipage}{\linewidth}
  \centerline{\includegraphics[width=\linewidth,trim=0 0 0 300,clip]{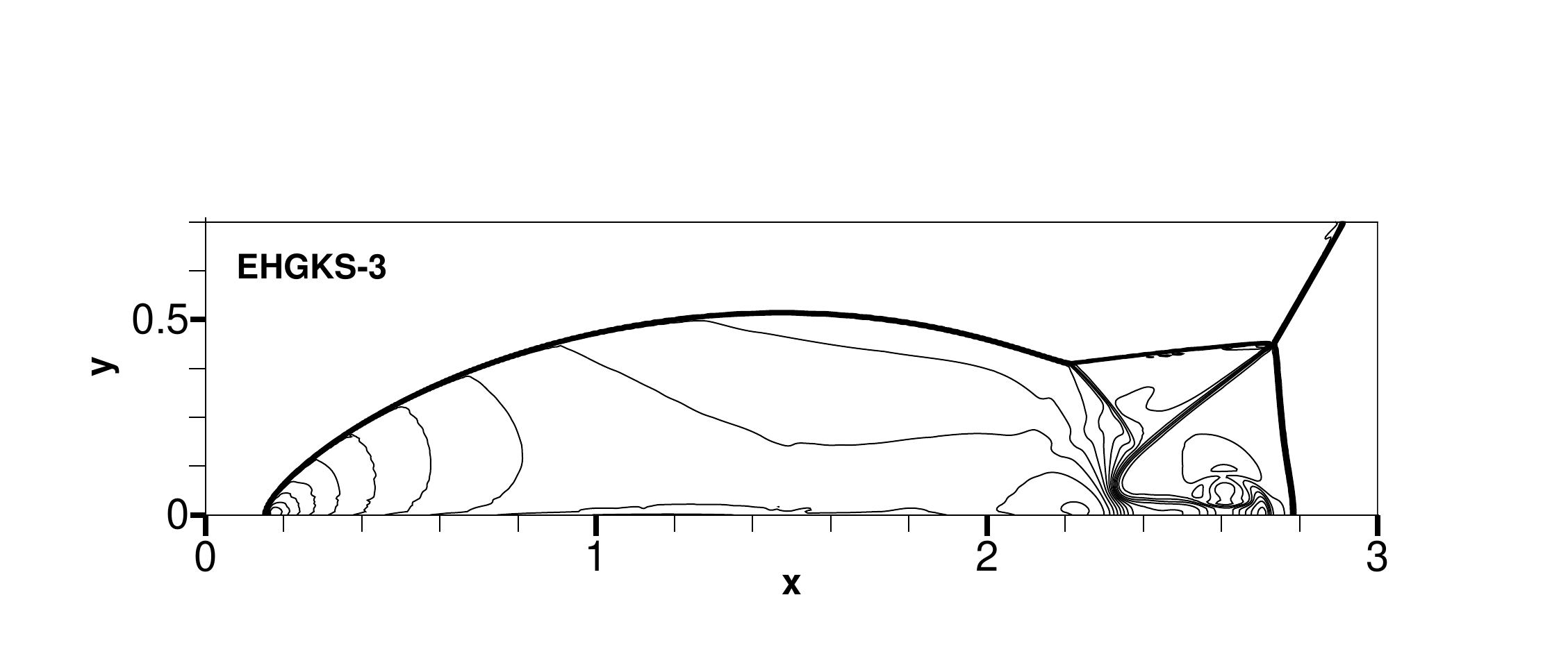}}
\end{minipage}
\begin{minipage}{\linewidth}
  \centerline{\includegraphics[width=\linewidth, trim=0 0 0 300,clip]{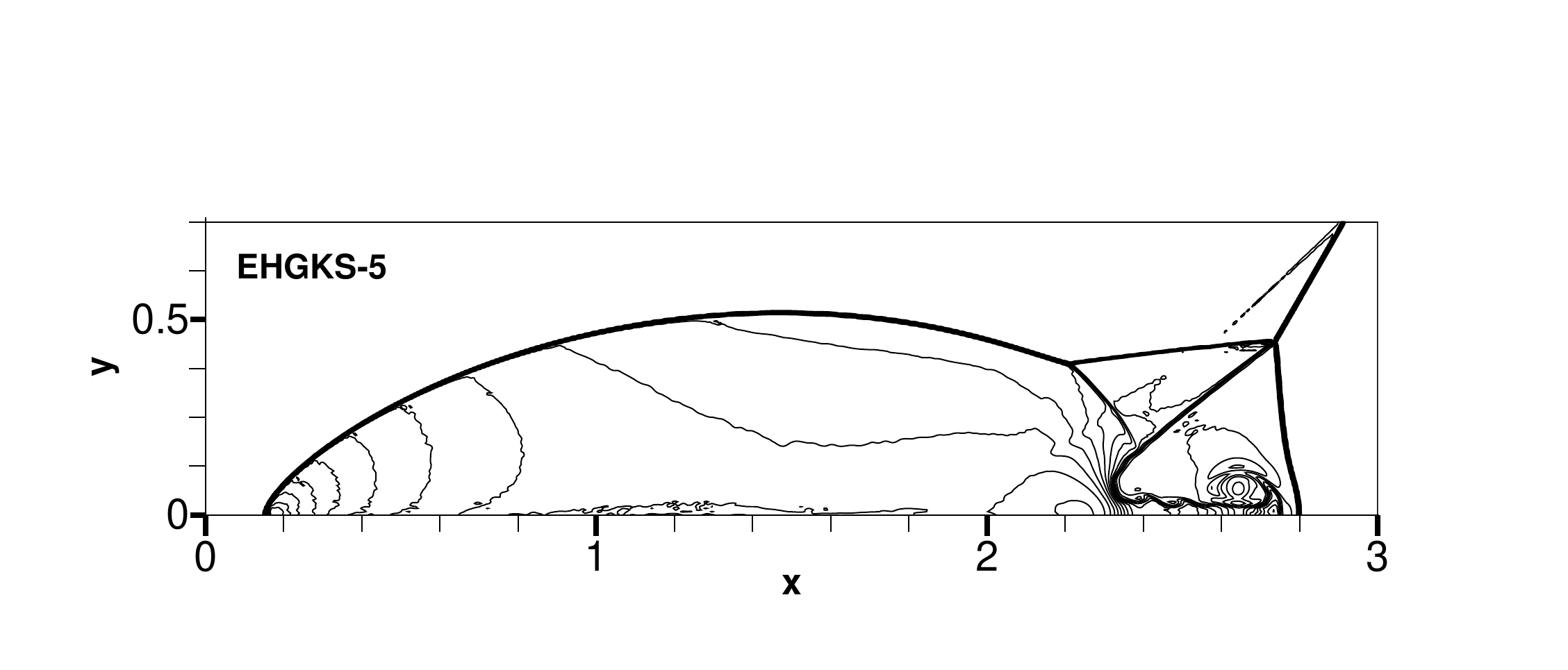}}
\end{minipage}
\begin{minipage}{\linewidth}
  \centerline{\includegraphics[width=\linewidth,trim=0 0 0 300,clip]{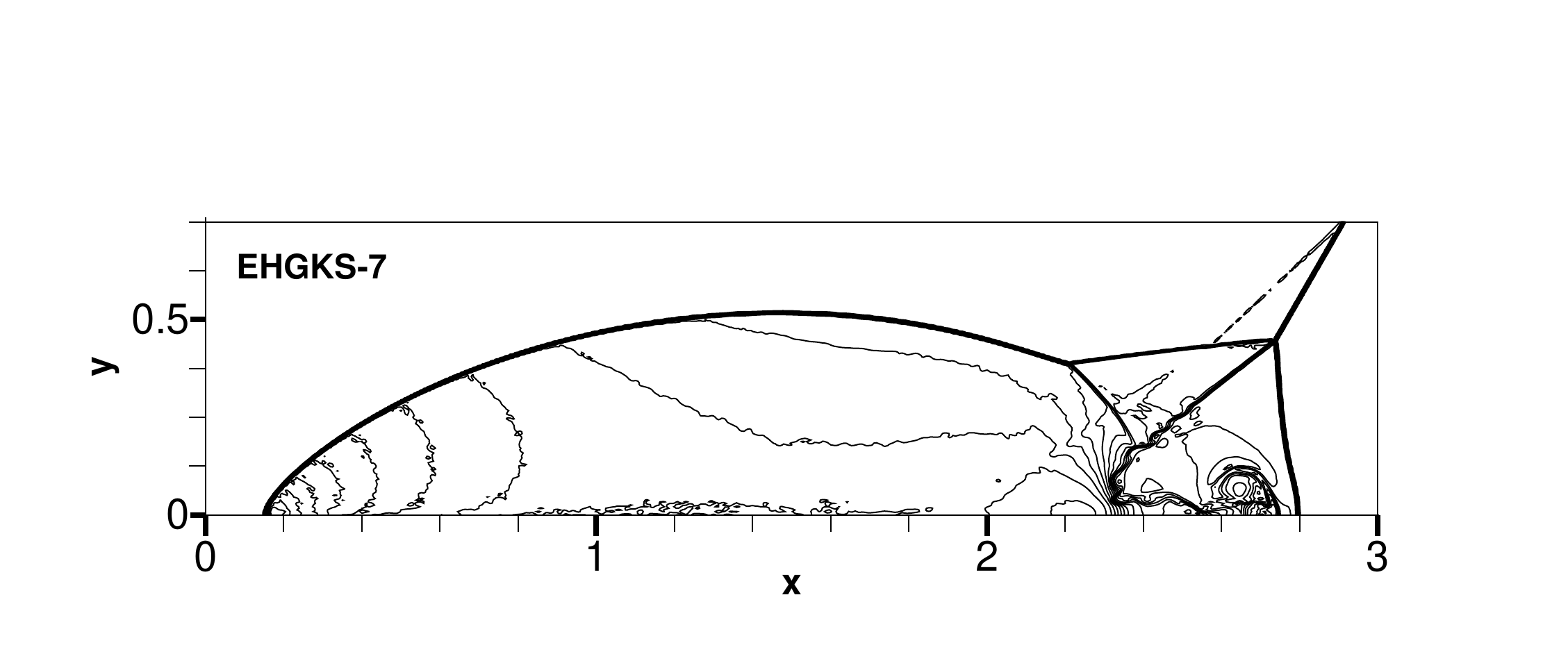}}
\end{minipage}
\caption{\label{fig:doublemach1} The density distribution of the double Mach reflection problem at $t=0.2$ with $\Delta x=\Delta y=1/240$. 30 contours are drawn from 1.731 to 20.92.}
\end{figure*}

\begin{figure*}[htb!]
\centering
\begin{minipage}{0.49\linewidth}
  \centerline{\includegraphics[width=\linewidth,trim=20 20 100 150,clip]{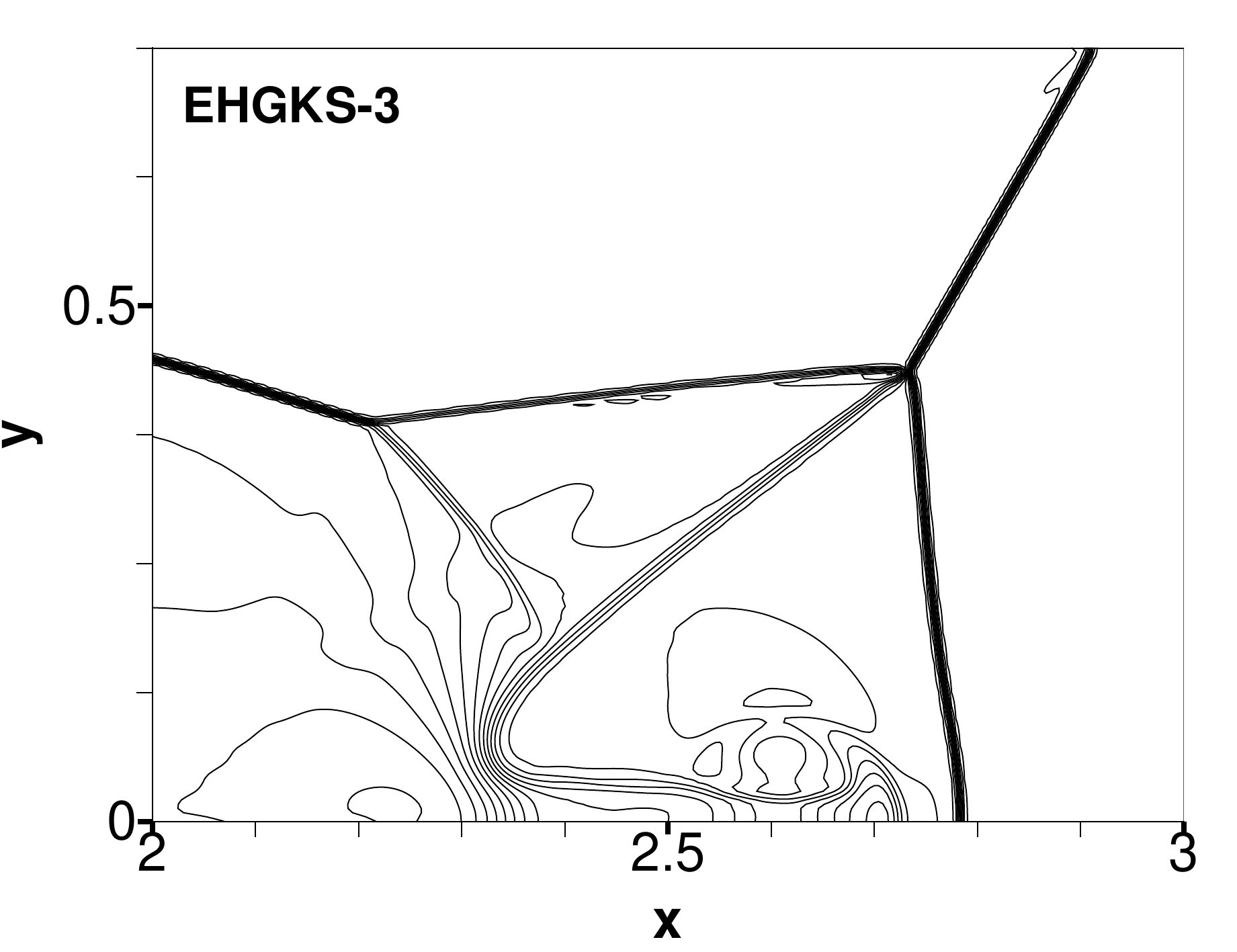}}
\end{minipage}
\begin{minipage}{0.49\linewidth}
  \centerline{\includegraphics[width=\linewidth,trim=20 20 100 150,clip]{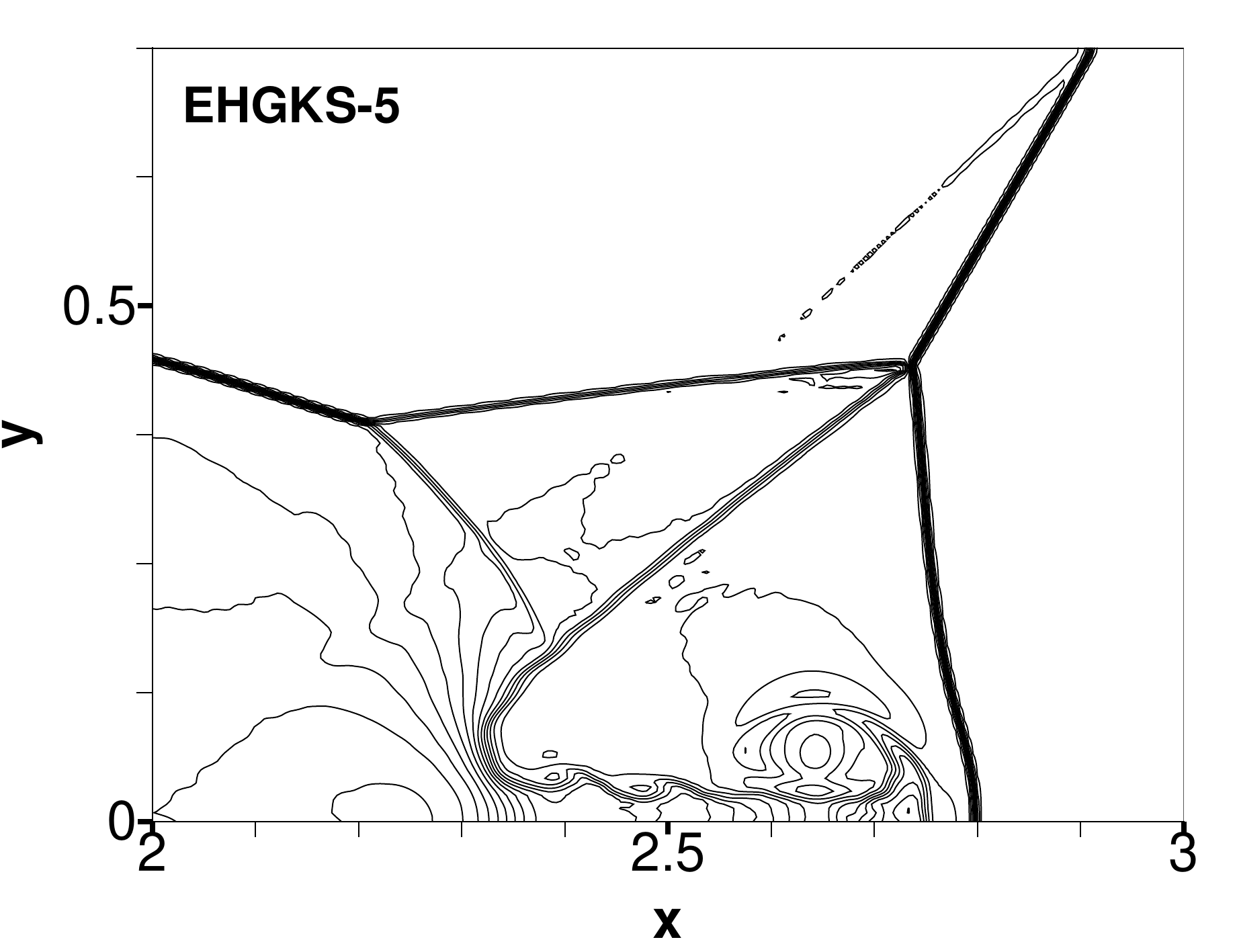}}
\end{minipage}
\begin{minipage}{0.49\linewidth}
  \centerline{\includegraphics[width=\linewidth,trim=20 20 100 150,clip]{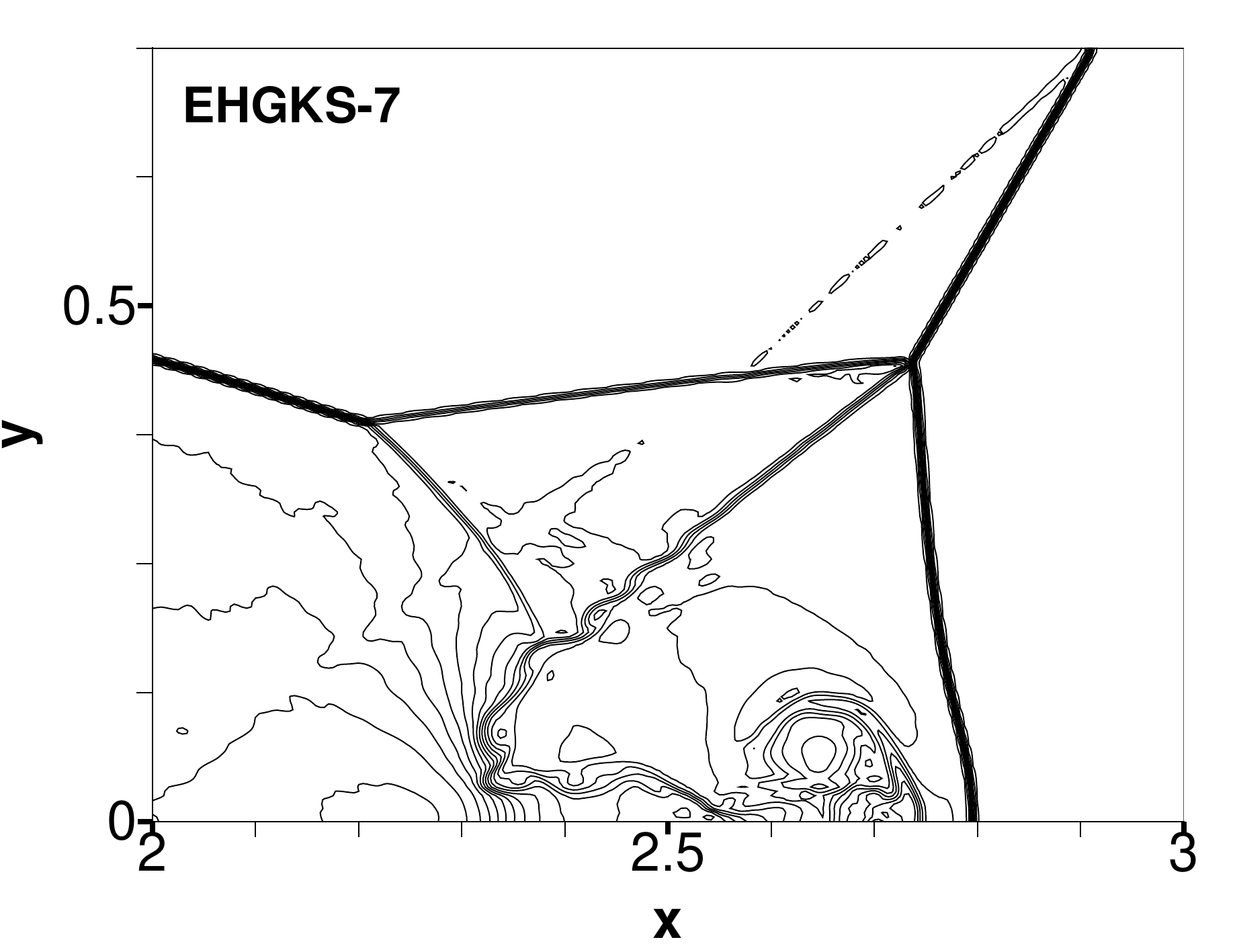}}
\end{minipage}
\caption{\label{fig:doublemach1e} The enlarged density distribution of the double Mach reflection problem at $t=0.2$ with $\Delta x=\Delta y=1/240$. 30 contours are drawn from 1.731 to 20.92.}
\end{figure*}

\begin{figure*}[htb!]
\centering
\begin{minipage}{\linewidth}
  \centerline{\includegraphics[width=\linewidth,trim=0 0 0 300,clip]{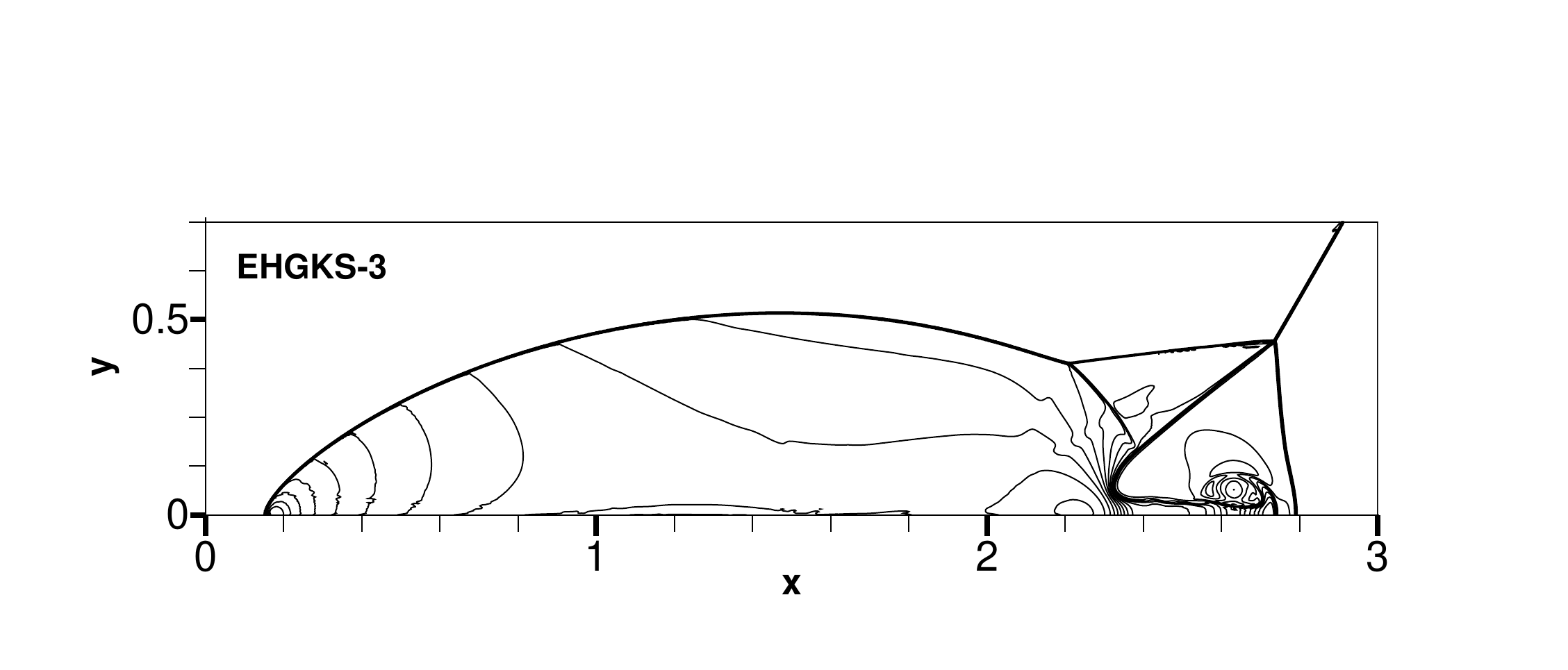}}
\end{minipage}
\begin{minipage}{\linewidth}
  \centerline{\includegraphics[width=\linewidth,trim=0 0 0 300,clip]{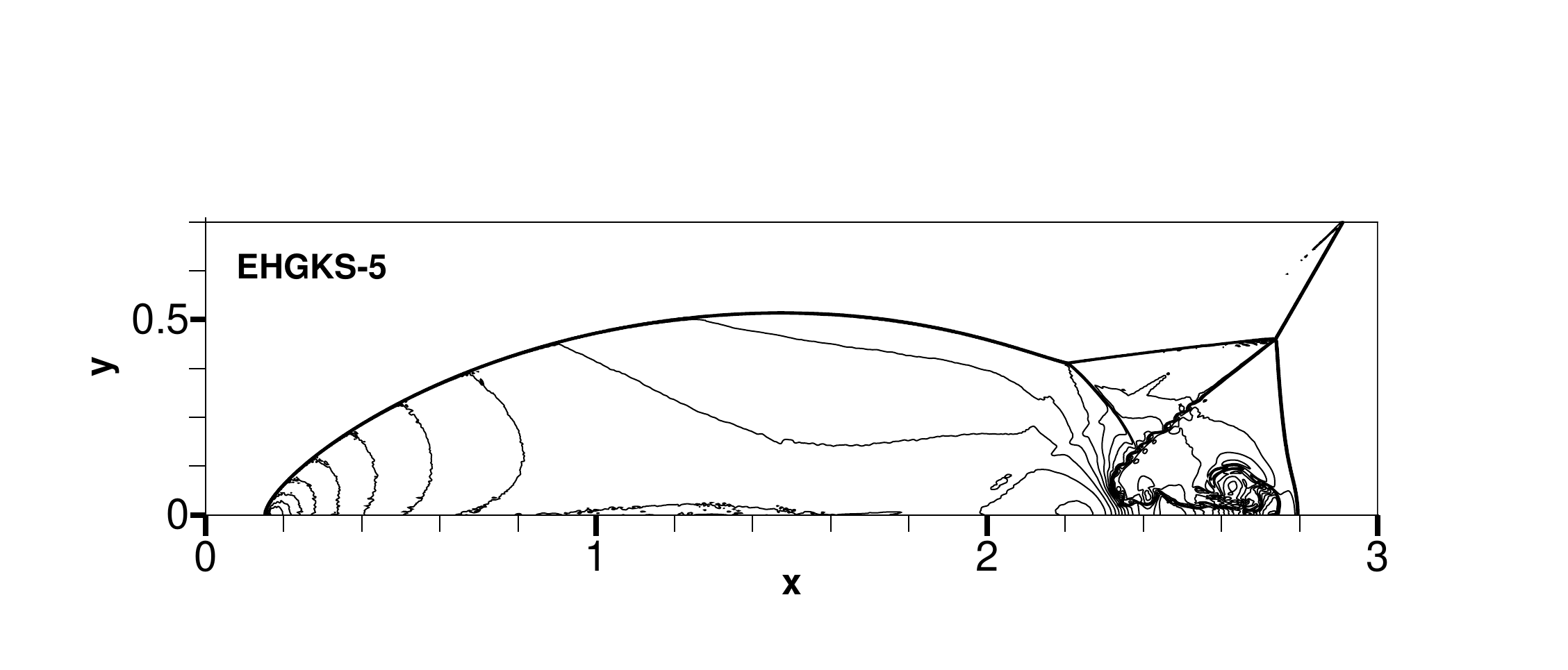}}
\end{minipage}
\begin{minipage}{\linewidth}
  \centerline{\includegraphics[width=\linewidth,trim=0 0 0 300,clip]{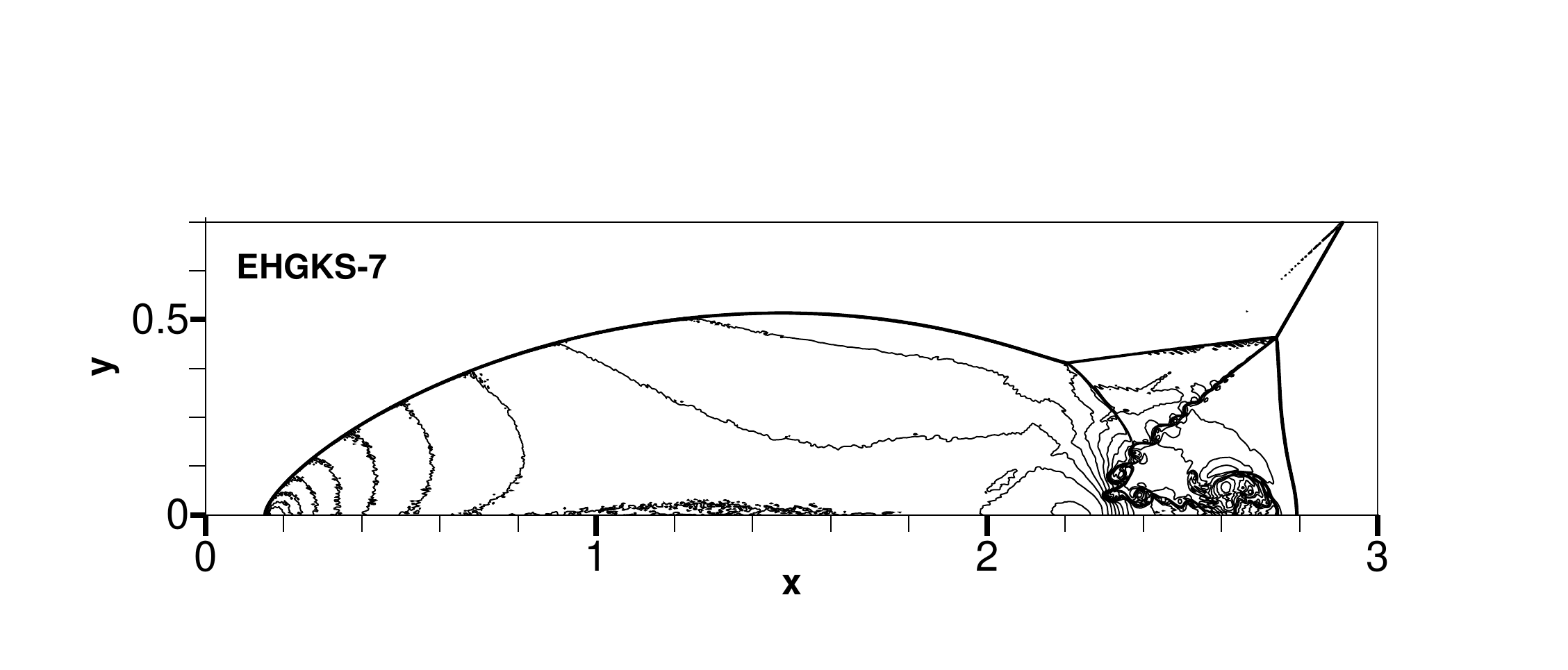}}
\end{minipage}
\caption{\label{fig:doublemach2} The density distribution of the double Mach reflection problem at $t=0.2$ with $\Delta x=\Delta y=1/480$. 30 contours are drawn from 1.731 to 20.92.}
\end{figure*}

\begin{figure*}[htb!]
\centering
\begin{minipage}{0.49\linewidth}
  \centerline{\includegraphics[width=\linewidth,trim=20 20 100 150,clip]{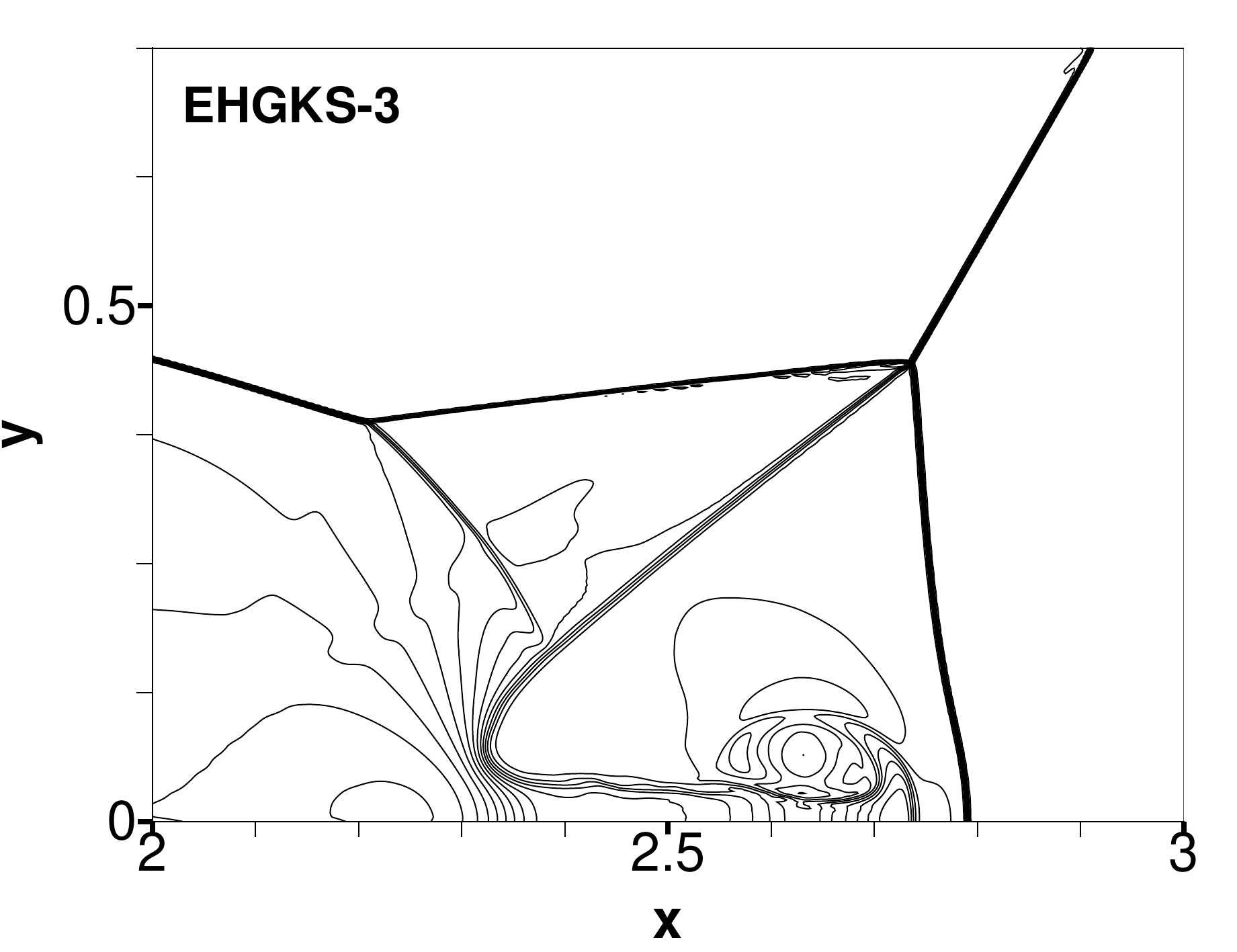}}
\end{minipage}
\begin{minipage}{0.49\linewidth}
  \centerline{\includegraphics[width=\linewidth,trim=20 20 100 150,clip]{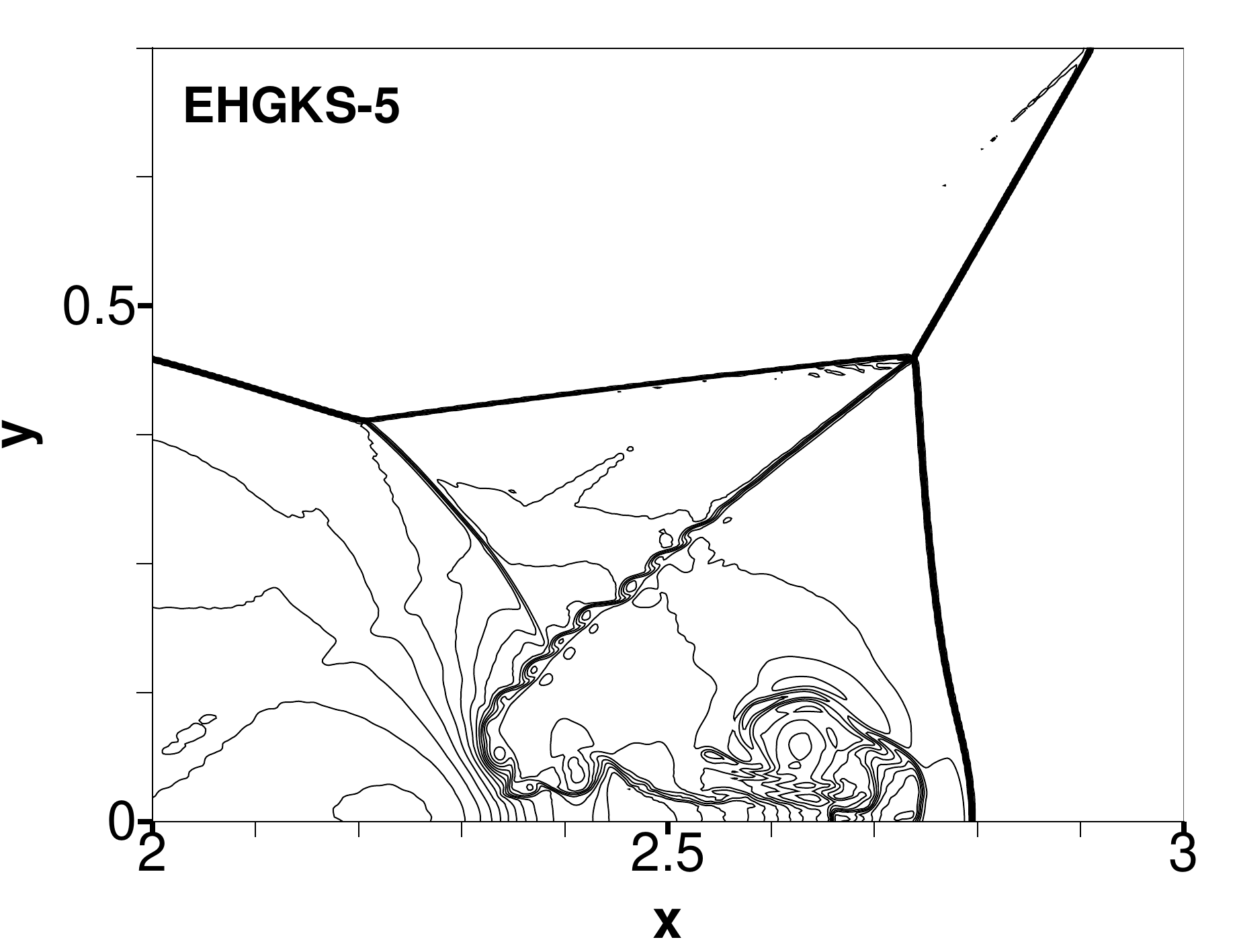}}
\end{minipage}
\begin{minipage}{0.49\linewidth}
  \centerline{\includegraphics[width=\linewidth,trim=20 20 100 150,clip]{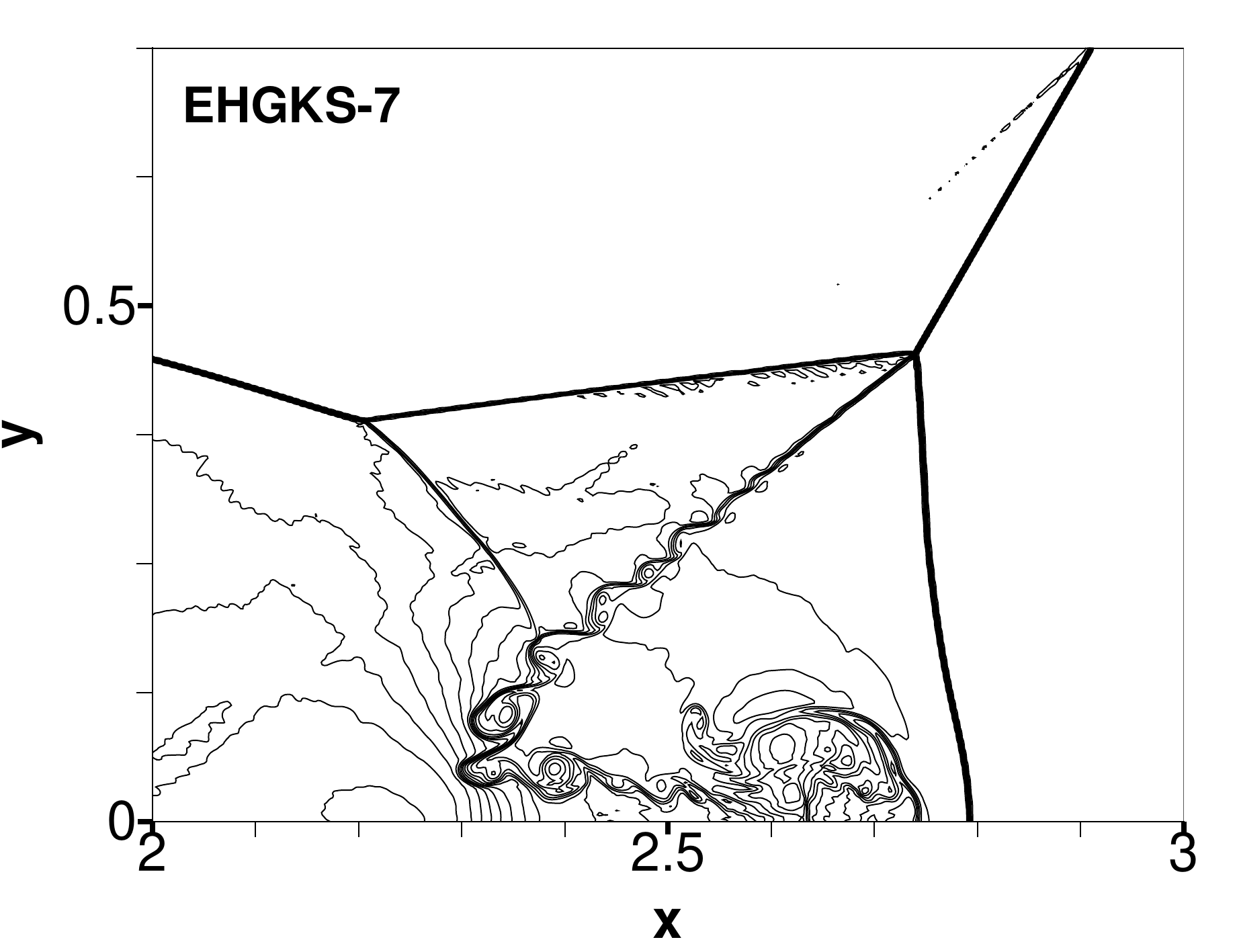}}
\end{minipage}
\caption{\label{fig:doublemach2e} The enlarged density distribution of the double Mach reflection problem at $t=0.2$ with $\Delta x=\Delta y=1/480$. 30 contours are drawn from 1.731 to 20.92.}
\end{figure*}

The density distributions and the local enlargement are shown from Fig.~\ref{fig:doublemach1} to Fig.~\ref{fig:doublemach2e}.
It is clearly observed the instability of the contact line from the triple Mach stem is better resolved by the higher-order EHGKS than the lower-order schemes.
It also demonstrates the advantage of high-order schemes in the simulations of complex flows.

\section{\label{sec:s3}Conclusions }

Based on the extensions, simplifications and modifications on the original HGKS evolution model and the flux evaluation, a more efficient gas-kinetic scheme EHGKS is proposed for the Euler equations of compressible flows.
The new EHGKS takes advantage of both the original HGKS to achieve arbitrary high-order accuracy and strong robustness, and the traditional Lax-Wendroff procedure to significantly reduce the complexity and computational costs of the original HGKS.
The main idea to improve the efficiency contains two parts.
Firstly, we simplify the original HGKS evolution model for the Euler equations.
Inspired by Zhou's simplification on the third-order HGKS, we extend the original HGKS to the case with arbitrary high-order accuracy by eliminating the unnecessary high-order dissipation terms.
Secondly, to avoid computing the complex moments of the derivatives of particle distribution functions, we introduce a Lax-Wendroff procedure to compute the high order derivatives of macroscopic quantities directly.
From the mechanism analysis, EHGKS preserves the high-order accuracy of the original HGKS in the smooth regions and its strong robustness with relaxation to the low-order KFVS solver near the discontinuities.

A sequence of classical test cases are carried out to validate the robustness, accuracy and efficiency of EHGKS.
The results computed by EHGKS-3, EHGKS-5 and EHGKS-7 are present, illustrating the advantage in improving the accuracy. Comparisons between the third-order TVD Runge-Kutta-WENO-GKS demonstrate the high time accuracy of EHGKS. In the typical applications, EHGKS gives a good resolution on the discontinuities and complex flow details.
The efficiency of EHGKS is compared with the original HGKS to solve the Euler equations and the third-order TVD Runge-Kutta-WENO-GKS.
In the case of the third-order accuracy, nearly half the computation cost can be saved by EHGKS-3.
In the case of the fifth-order accuracy, the improvement in efficiency by EHGKS-5 is more than one order of magnitude.
As a summary, the high accuracy, efficiency and strong robustness of EHGKS are consequently confirmed.

In the future, we wish to further construct a compact EHGKS by using the techniques of HWENO or DG, and also plan to extend EHGKS to solve the NS equations.

\section*{Acknowledgements }
This work is supported by NSAF (Grant No. U1630247), the China Postdoctoral Science Foundation (Grant No. 2018M641272), and
NSFC (Grant Nos. 11631008, GZ1465, 11571046).

\appendix

\section{\label{app:rec}Reconstruction}

Firstly, the initial polynomials $\bfW_i(x,0)$ over the $i$-th cell is reconstructed to obtain the point-wise values $\bfW_i(x_{i \pm \frac{1}{2}},0)$ and space derivatives $\frac{{\partial^m \bfW_i(x_{i \pm \frac{1}{2}},0) }}{{\partial x^m}}$ for $m \ge 1$.
The characteristic-wise WENO reconstruction technique \cite{eno1996} is applied to obtain the point-wise values ${\bfW}_{i \pm \frac{1}{2}}$:
\begin{eqnarray}
\bfW_i(x_{i + \frac{1}{2}},0)={\bfW}_{i + \frac{1}{2}},\nonumber\\
\bfW_i(x_{i - \frac{1}{2}},0)={\bfW}_{i - \frac{1}{2}}.\nonumber
\label{eq:weno}
\end{eqnarray}
if without specifications.
To evaluate the point-wise space derivatives, the same smooth function for $\bfW_i(x,0)$ as in \cite{cyb2016} is supposed over the $i$-th cell.
The linear reconstruction is implemented to obtain $\frac{{\partial^m \bfW_i(x_{i \pm \frac{1}{2}},0) }}{{\partial x^m}}$ based on the two point-wise values ${\bfW}_{i \pm \frac{1}{2}}$ after the WENO reconstruction and the centremost cell averages $\overline {\bfW} _i,\overline {\bfW} _{i-1},\overline {\bfW} _{i+1}...$, which gives

{Third order:}
\begin{eqnarray}
&&
\frac{{\partial \bfW_i(x_{i + \frac{1}{2}},0) }}
{{\partial x}} = \frac{{2{\bfW}_{i - \frac{1}{2}}  - 6\overline {\bfW} _i  + 4{\bfW}_{i + \frac{1}{2}} }}
{{\Delta x}},\nonumber\\
&&
\frac{{\partial ^2 \bfW_i(x_{i + \frac{1}{2}},0) }}
{{\partial x^2 }} = \frac{{6{\bfW}_{i - \frac{1}{2}}  - 12\overline {\bfW} _i  + 6{\bfW}_{i + \frac{1}{2}} }}
{{\Delta x^2 }}.\nonumber
\label{eq:weno3}
\end{eqnarray}

{Fifth order:}
\begin{eqnarray}
&&
\frac{{\partial \bfW_i(x_{i + \frac{1}{2}},0) }}
{{\partial x}} = \frac{{12{\bfW}_{i - \frac{1}{2}}  - \overline {\bfW} _{i - 1}  - 31\overline {\bfW} _i  + 2\overline {\bfW} _{i + 1}  + 18{\bfW}_{i + \frac{1}{2}} }}
{{6\Delta x}},\nonumber\\
&&
\frac{{\partial ^2 \bfW_i(x_{i + \frac{1}{2}},0) }}
{{\partial x^2 }} = \frac{{6{\bfW}_{i - \frac{1}{2}}  - \overline {\bfW} _{i - 1}  - 4\overline {\bfW} _i  + 5\overline {\bfW} _{i + 1}  - 6{\bfW}_{i + \frac{1}{2}} }}
{{2\Delta x^2 }},\nonumber\\
&&
\frac{{\partial ^3 \bfW_i(x_{i + \frac{1}{2}},0) }}
{{\partial x^3 }} = \frac{{ - 24{\bfW}_{i - \frac{1}{2}}  + 2\overline {\bfW} _{i - 1}  + 50\overline {\bfW} _i  + 8\overline {\bfW} _{i + 1}  - 36{\bfW}_{i + \frac{1}{2}} }}
{{\Delta x^3 }},\nonumber\\
&&
\frac{{\partial ^4 \bfW_i(x_{i + \frac{1}{2}},0) }}
{{\partial x^4 }} = \frac{{ - 60{\bfW}_{i - \frac{1}{2}}  + 10\overline {\bfW} _{i - 1}  + 100\overline {\bfW} _i  + 10\overline {\bfW} _{i + 1}  - 60{\bfW}_{i + \frac{1}{2}} }}
{{\Delta x^4 }}.\nonumber
\label{eq:weno5}
\end{eqnarray}

{Seventh order:}
\begin{eqnarray}
&&
\frac{{\partial \bfW_i(x_{i + \frac{1}{2}},0) }}
{{\partial x}} =  - \frac{{{{87}}\overline {\bfW} _{i - 1}  - {{3}}\overline {\bfW} _{i - 2}  - {{883}}\overline {\bfW} _i  - {{43}}\overline {\bfW} _{i + 1}  + {{2}}\overline {\bfW} _{i + 2}  + {{480}}{\bfW}_{i - \frac{1}{2}}  + {{360}}{\bfW}_{i + \frac{1}{2}} }}
{{{{180}}\Delta x}},\nonumber\\
&&
\frac{{\partial ^2 \bfW_i(x_{i + \frac{1}{2}},0) }}
{{\partial x^2 }} = \frac{{409\overline {\bfW} _{i - 1}  - 11\overline {\bfW} _{i - 2}  + 89\overline {\bfW} _i  - 71\overline {\bfW} _{i + 1}  + 4\overline {\bfW} _{i + 2}  - 660{\bfW}_{i - \frac{1}{2}}  + 240{\bfW}_{i + \frac{1}{2}} }}
{{{{120}}\Delta x^2 }},\nonumber\\
&&
\frac{{\partial ^3 \bfW_i(x_{i + \frac{1}{2}},0) }}
{{\partial x^3 }} =  - \frac{{51\overline {\bfW} _{i - 1}  + 347\overline {\bfW} _i  + 23\overline {\bfW} _{i + 1}  - \overline {\bfW} _{i + 2}  - 240{\bfW}_{i - \frac{1}{2}}  - 180{\bfW}_{i + \frac{1}{2}} }}
{{{{6}}\Delta x^3 }},\nonumber\\
&&
\frac{{\partial ^4 \bfW_i(x_{i + \frac{1}{2}},0) }}
{{\partial x^4 }} =  - \frac{{5\left( {7\overline {\bfW} _{i - 1}  - 2\overline {\bfW} _{i - 2}  - 73\overline {\bfW} _i  - 17\overline {\bfW} _{i + 1}  + \overline {\bfW} _{i + 2}  + 24{\bfW}_{i - \frac{1}{2}}  + 60{\bfW}_{i + \frac{1}{2}} } \right)}}
{{{{6}}\Delta x^4 }},\nonumber\\
&&
\frac{{\partial ^5 \bfW_i(x_{i + \frac{1}{2}},0) }}
{{\partial x^5 }} =  - \frac{{6\overline {\bfW} _{i - 2}  - 69\overline {\bfW} _{i - 1}  - 329\overline {\bfW} _i  - 29\overline {\bfW} _{i + 1}  + \overline {\bfW} _{i + 2}  + 240{\bfW}_{i - \frac{1}{2}}  + 180{\bfW}_{i + \frac{1}{2}} }}
{{\Delta x^5 }},\nonumber\\
&&
  \frac{{\partial ^6 \bfW_i(x_{i + \frac{1}{2}},0) }}
{{\partial x^6 }} = \frac{{7\left( {\overline {\bfW} _{i - 2}  - 14\overline {\bfW} _{i - 1}  - 94\overline {\bfW} _i  - 14\overline {\bfW} _{i + 1}  + \overline {\bfW} _{i + 2}  + 60{\bfW}_{i - \frac{1}{2}}  + 60{\bfW}_{i + \frac{1}{2}} } \right)}}
{{\Delta x^6 }}.\nonumber
\label{eq:weno7}
\end{eqnarray}
The reconstruction to $\frac{{\partial^m \bfW_i(x_{i - \frac{1}{2}},0) }}{{\partial x^m}}$ is mirror symmetric of the above expressions.
After the reconstruction, the cell interface values $\bfW^{\rm L}$, $\bfW^{\rm R}$ and their space derivatives at $x_{i+\frac{1}{2}}$ are given as
\begin{eqnarray}
&&\bfW^{\rm L}=\bfW_i(x_{i + \frac{1}{2}},0),
\frac{{\partial^m \bfW^{\rm L} }}{{\partial x^m}}=
\frac{{\partial^m \bfW_i(x_{i + \frac{1}{2}},0) }}{{\partial x^m}},\nonumber\\
&&\bfW^{\rm R}=\bfW_{i+1}(x_{i + \frac{1}{2}},0),
\frac{{\partial^m \bfW^{\rm R} }}{{\partial x^m}}=
\frac{{\partial^m \bfW_{i+1}(x_{i + \frac{1}{2}},0) }}{{\partial x^m}}.\nonumber
\label{eq:recon}
\end{eqnarray}

Then, to obtain the space derivatives of $\bfW ^{e}$, the continuous flow distribution hypothesis is adopted \cite{pan2016,ji2017,lqb2010,luo2013}.
In this work, considering the small variations between ${\bfW}^{\rm{L}}$, ${\bfW}^{\rm{R}}$ and ${\bfW}^e$ in the smooth regions, a simple weighting is implemented based on the reconstructed space derivatives of ${\bfW}^{\rm{L}}$ and ${\bfW}^{\rm{R}}$.
The weight $\omega ^e = $\emph{erfc}$ (-\sqrt{\lambda ^e}U^e) /2$ is adopted with both the central and upwind character \cite{estivalezes1996}
\begin{eqnarray}
\frac{{\partial ^m \bfW^e }}{{\partial x}^m} =
\omega ^e \frac{{\partial ^m \bfW^{\rm{L}} }}{{\partial x}^m}
+\left( 1-\omega ^e \right)
\frac{{\partial ^m \bfW^{\rm{R}} }}{{\partial x}^m},\nonumber
\label{eq:wex}
\end{eqnarray}
where \emph{erfc} is th complementary error function.

Other reconstruction techniques are also available, such as the limiters for ${\bfW}^{\rm{L}}$ and ${\bfW}^{\rm{R}}$, the linear reconstruction on $\bfW^e$ \cite{lqb2010,liu2014}.
But it is not the focus of this work.
For simplicity, all the reconstructions are reduced to the zeroth order when ${\bfW}^{\rm{L}}$ or ${\bfW}^{\rm{R}}$ is non-physical with negative density or pressure, and if the pressure difference is with one-order of magnitude among the nearest-layer stencils in the first two steps of all the simulations.
Specifically in the accuracy tests, the linear reconstruction is implemented for ${\bfW}^{\rm{L}}$, ${\bfW}^{\rm{R}}$ and ${\bfW}^e$.
No smoothness indicators are included in the reconstructions for ${\bfW}^{\rm{L}}$ and ${\bfW}^{\rm{R}}$ there.

\bibliography{ehgks}

\end{document}